\RequirePackage{fix-cm}
\documentclass[smallextended]{svjour3}       
\smartqed  

\usepackage[colorinlistoftodos]{todonotes}
\usepackage[colorlinks=true, allcolors=blue]{hyperref}
\usepackage{graphicx}
\usepackage{natbib}
\usepackage{mathptmx}      
\usepackage{latexsym}
\usepackage{amssymb}
\newcommand{\enzo}{\texttt{ENZO }}
\newcommand{\gadget}{\texttt{GADGET }}
\newcommand{\arepo}{\texttt{AREPO }}

\begin{document}

\title{Magnetic Field Amplification in Galaxy Clusters and its Simulation}

\author{Donnert J. \and Vazza F. \and Br\"{u}ggen M. \and ZuHone J.}


\institute{J. Donnert \at
              via P.Gobetti 101 \\
              40129 Bologna, Italy \\
              \email{donnert@ira.inaf.it}           
           \and
           F. Vazza \at
              Dipartimento di Fisica e Astronomia\\
              via Gobetti 93/2\\
              40129 Bologna, Italy\\
              \email{franco.vazza2@unibo.it}     
             \and 
           M. Br\"{u}ggen \at
              Hamburg Observatory, \\
              Gojenbergsweg 112, \\
              21029 Hamburg, Germany, \\
              \email{mbrueggen@hs.uni-hamburg.de}
              \and
           J. ZuHone \at
           	  Smithsonian Astrophysical Observatory\\
              60 Garden St.\\
              Cambridge, MA 02138 USA\\
              \email{john.zuhone@cfa.harvard.edu}
}

\date{Received: date / Accepted: date}

\maketitle

\begin{abstract}
We review the present theoretical and numerical understanding of magnetic field amplification in cosmic large-scale structure, on length scales of galaxy clusters and beyond. Structure formation drives compression and turbulence, which amplify tiny magnetic seed fields to the microGauss values that are observed in the intracluster medium.  This process is intimately connected to the properties of turbulence and the microphysics of the intra-cluster medium. Additional roles are played by merger induced shocks that sweep through the intra-cluster medium and motions induced by sloshing cool cores. The accurate simulation of magnetic field amplification in clusters still poses a serious challenge for simulations of cosmological structure formation. We review the current literature on cosmological simulations that include magnetic fields and outline theoretical as well as numerical challenges.
\end{abstract}

\section{Introduction} \label{sect.intro}

Magnetic fields permeate our Universe, which  is filled with ionized gas from the scales of our solar system up to filaments and voids in the large-scale structure \citep{2015gimf.book.....K}. While magnetic fields are usually not dynamically important, their presence shapes the physical properties of the Baryonic medium \citep{2007mhet.book...85S}. On the largest scales, radio observations remain our most important tool to estimate magnetic fields today (see e.g. van Weeren, this volume). Recent and upcoming advances in instrumentation enable the observation of radio emission on scales of a few kpc at the cluster outskirts and will soon provide three-dimensional magnetic field  distributions in the inter-cluster-medium (ICM) through Faraday tomography \citep{2014skao.rept.....G}. \par
Connecting these new observations to theoretical expectations is a major challenge for the community, due to the complexity of the non-thermal physics in the cosmological context. In the framework of cold Dark Matter, structure formation is dominated by gravitational forces and proceeds from the bottom up: smaller DM halos form first \citep{2016mssf.book...93P}, and baryons flow into the resulting potential well. Through cooling, stars and galaxies form and evolve into larger structures (groups, clusters, filaments), by infall and merging \citep{2010gfe..book.....M}. These processes drive turbulent gas motions and a magnetic dynamo that amplifies some form of seed field to $\mu\mathrm{G}$ values in the center of galaxy clusters. Galaxy feedback injects magnetic fields and relativistic particles (cosmic-ray protons and electrons) into the large-scale structure that interact with shocks and turbulence, get (re-)accelerated and finally become observable at radio frequencies and potentially in the  $\gamma$-ray regime \citep{2002cra..book.....S,2012SSRv..173..557L,bj14}. \par
In the past decade significant progress has been made in the simulation of galaxy formation, with an emphasis on physical models for feedback \citep[e.g.][]{2017ARA&A..55...59N}. Unfortunately, the same is not true for the simulation of turbulence, magnetic fields and cosmic-ray evolution - nearly every step in the chain of non-thermal processes remains open today: \\
What is the origin of the magnetic seed fields and the contributions of various astrophysical sources? What are the properties of turbulence and the magnetic dynamo in the ICM, filaments, and voids? What is the distribution and topology of magnetic fields? What is the spatial distribution of radio dark cosmic-ray electrons in clusters? Where are the cosmic-ray protons? What are their sources? What physics governs particle acceleration in shocks that leads to radio relics? How does turbulence couple to cosmic-rays in radio halos? What are the physical properties (viscosity, resistivity, effective collisional scales) of the diffuse plasma in the ICM, filaments and voids?  \par
Answers have proven themselves difficult to obtain, in part because turbulence is a demanding numerical problem, but also because the physics is different enough from galaxy formation to make some powerful numerical approaches like density adaptivity rather ineffective. Today, JVLA and LOFAR observations have achieved an unprecedented spatial and spectral detail in the observation of magnetic phenomena in cluster outskirts
  \citep[e.g.][]{2014ApJ...794...24O,2017MNRAS.471.1107H,2018ApJ...852...65R}, thereby challenging simulations to increase their level of spatial and physical detail. The gap will likely widen in the next years as SKA precursors like ASKAP see first light \citep{2010AAS...21547013G} and results from the LOFAR survey key science project become available \citep{2017A&A...598A.104S}. 

Here we review the current status on astrophysical and cosmological simulations of magnetic field amplification in structure formation through compression, shocks, turbulence and cosmic-rays. Such a review will naturally emphasize galaxy clusters, simply because there is only weak observational evidence for magnetic fields in filaments and voids. We will also touch on ideal MHD as a model for intergalactic plasmas and introduce fundamental concepts of turbulence and the MHD dynamo. We are putting an emphasis on numerical simulations because they are our most powerful tool to study the interplay of non-thermal physics. This must  also include some details on common algorithms for MHD and their limitations. Today, these algorithms and their implementation limit our ability to model shocks, turbulence and the MHD dynamo in a cosmological framework.  \par

We exclude from this review topics that are not directly related to simulations of the cosmic magnetic dynamo. While we shortly introduce turbulence and dynamo theory, we do not attempt to go into detail, several reviews are available \citep[e.g.][ for an introduction]{2007mhet.book...85S}. We also do not review models for particle acceleration in clusters \citep[][]{bj14} or observations  \citep[see][and van Weeren et al., this volume]{2008SSRv..134...93F}. We also do not discuss in detail the seeding of magnetic fields \citep[see][for recent exhaustive reviews on the topic]{wi11,2012SSRv..166....1R,sub16}, nor the amplification of magnetic fields in the interstellar medium \citep[see][for a recent review]{2016JPlPh..82f5301F} or in galaxies \citep[e.g.][for theoretical reviews]{2010A&A...522A.115S,2012MNRAS.422.2152B,2018MNRAS.tmp.1551M}.

\subsection{Overview}

Galaxy clusters form through the gravitational collapse and subsequent merging of virialized structures into haloes, containing about $80\%$ Dark Matter and $20\%$ Baryons \citep{2002mpgc.book....1S,2005RvMP...77..207V,2015SSRv..188...93P}. From X-ray observations we know that the diffuse thermal gas in the center of haloes with masses $> 10^{14} \,M_\odot$ is completely ionized, with temperatures of $T = 10^8\,\mathrm{K}$ and number densities of $n_\mathrm{th} \approx 10^{-3} \,\mathrm{cm}^{-3}$, \citep[e.g.][]{SA88.1,borgani08}. The speed of sound is then $c_\mathrm{s} = \sqrt{\gamma P/\rho} \approx 1200 \,\mathrm{km}/\mathrm{s}$, where $\gamma = 5/3$ is the adiabatic index at density $\rho$ and pressure $P$. \par
The ideal equation of state for a monoatomic gas is applicable in such a hot under-dense medium, even though the ICM contains $\approx 25 \%$ helium and heavier elements as well \citep[e.g.][]{2010A&ARv..18..127B}. In fact, the intracluster medium is one of the most ideal plasmas known, with a plasma parameter of $g \approx 10^{-15}$ and a Debye length of $\lambda_\mathrm{D} \approx 10^5 \,\mathrm{cm}$ that still contains $\approx 10^{12}$ protons and electrons. In contrast, the mean free path for Coulomb collisions is in the kpc regime (eq. \ref{eq.class_lmfp}). Clearly, electromagnetic particle interactions dominate over two-body Coulomb collisions and plasma waves shape the properties of the medium on small scales \citep[e.g.][table 8.1]{2002cra..book.....S}. \par
Cluster magnetic fields of $1 \,\mu\mathrm{G}$ were first estimated from upper limits on the diffuse synchrotron emission of intergalactic material in a $1 \,\mathrm{Mpc}^3$ volume by \citet{1958ApJ...128....1B}. With the discovery of the Coma radio halo by \citet{1970MNRAS.151....1W}, this was confirmed using equipartition arguments between the  cosmic-ray electron energy density and magnetic energy density \citep[e.g.][]{2005AN....326..414B}. Later estimates based on the rotation measure of background sources to the Coma cluster obtain central magnetic fields of $3-7 \mu\mathrm{G}$ scaling with ICM thermal density with an exponent of $0.5-1$ \citep[e.g.][]{bo10}. Hence the ICM is a high $\beta = n_\mathrm{th} k_\mathrm{B} T / B^2 \approx 100$ plasma, where thermal pressure dominates magnetic pressure.  \par
Based on above estimates, one may hope that magneto-hydrodynamics (MHD) is applicable on large enough scales in clusters (sections \ref{sect.spitzer} and \ref{sect.wcICM}). Then the magnetic field $\vec{B}$ evolves with the flow velocity $\vec{v}$ according to the induction equation \citep{1961AmJPh..29..647L}: 
\begin{eqnarray}
\frac{\partial \vec{B}}{\partial t}  &=& - \vec{v} \cdot \nabla \vec{B} + \vec{B} \cdot \vec{\nabla} \vec{v} - \vec{B} \vec{\nabla} \cdot \vec{v} - \eta \Delta \vec{B}, \label{eq.induction}
 \label{eq:induction1}
\end{eqnarray}
where the first term accounts for the advection of field lines, the second one for stretching, the third term for compression and the fourth term for the magnetic field dissipation with the diffusivity $\eta = c_\mathrm{s} / 4\pi\sigma$ and the conductivity $\sigma$. Because the ICM is a nearly perfect plasma ($\beta_\mathrm{pl} \gg 1$), conductivity is very high, diffusivity likely very low ($\eta \approx 0$). Then the induction equation \ref{eq.induction} predicts that magnetic fields are frozen into the plasma and advected with the bulk motions of the medium \citep{2006PhT....59a..58K}. Because equation \ref{eq.induction} is a conservation equation for magnetic flux,  magnetic fields cannot be created in the MHD framework, but have to be seeded by some mechanism, also at high redshift (section \ref{sect.seed}). However, current upper limits on large-scale magnetic fields exclude large-scale seed fields above $\sim \mathrm{nG}$ \citep{PLANCK2015}, 
and back-of-the-envelope calculations show that pure compression cannot produce $\mu\mathrm{G}$ in clusters from such initial values (section \ref{sect.compr}). X-ray observations have revealed substantial turbulent velocities in a few clusters \citep{sc04,2014Natur.515...85Z,2016Natur.535..117H}. These are in agreement with estimates from rotation measurements \citep{2003A&A...412..373V,2011A&A...529A..13K} that can also be used to constrain magnetic field power spectra \citep{2012A&A...540A..38V,2016A&A...591A..13V,2017A&A...603A.122G}. \par
It is reasonable to assume some form of turbulent dynamo in the clusters and possibly filaments  \citep{1980ApJ...241..925J,1981A&A....93..407R,1989MNRAS.241....1R,1992ApJ...386..464D,1993ApJ...411..518G,1997ApJ...480..481K,1998Ap&SS.263...87S,su06,2006A&A...453..447E}, but it is necessary to consider plasma-physical arguments to understand the fast growth of seed fields by many orders of magnitude \citep{sch05,2007mhet.book...85S}. There are clear theoretical predictions for idealized MHD dynamos \citep[e.g.][]{2004ApJ...612..276S,po15}, which show that magnetic fields are amplified though an inverse cascade at the growing \emph{Alfv\'{e}n scale}, where the field starts back-reacting on the flow. This is called the \emph{small-scale dynamo} (section \ref{sect.ssd}). However, the astrophysical situation differs significantly from these idealized models: structure formation drives turbulence localized, episodic and multi-scale in the presence of a strong gravitational potential in galaxy clusters (section \ref{sect.physturb}), and the magneto-hydrodynamical properties of the medium are far from clear \citep{2009ApJS..182..310S}. Shocks and cosmic-rays amplify magnetic fields as well and are very difficult to model (section \ref{sect.shocks}). \par

With JVLA, LOFAR, ASKAP and the SKA, the Alfv\'{e}n scale comes within the range of radio observations: radio relics are now spatially resolved to a few kpc in polarization; low-frequency surveys are expected to find hundreds of radio halos and mini-halos; Faraday tomography will allow to map magnetic field structure also along the line of sight (see van Weeren et al., this volume). Future X-ray missions will put stringent bounds on turbulent velocities in clusters and constrain magnetic field amplification by draping and sloshing in cold fronts  (section \ref{sect.slosh}). 

\section{Magnetic seeding processes} \label{sect.seed}

Let us begin with a short overview of proposed seeding mechanisms; a detailed review can be found e.g. in \citep{sub16}. It is very likely that more than one of these mechanisms contributes to the magnetization of the large-scale structure. Hence an important question for simulations of magnetic field amplification is the influence of these seeding mechanisms on the final magnetic field.
 
\subsection{Primordial mechanisms}

Several mechanisms for the initial seed field have been suggested to start the dynamo amplification process within galaxies and galaxy clusters. Some of the proposed scenarios involve the generation of currents during inflation, phase transitions and baryogenesis \citep[e.g.][]{1973MNRAS.165..185H,2010PhRvD..82h3005K,wi11,2011ApJ...726...78K,2013A&ARv..21...62D,sub16,2016PhyS...91j4008K}. These primordial seed fields may either produce small \citep[$\leq ~\rm Mpc$, e.g.][]{1967SvA....10..634C} or large \citep[e.g.][]{1970SvA....13..608Z,1988PhRvD..37.2743T} coherence lengths, whose structure may still persist until today \citep[e.g.][]{2018arXiv180302629H}, in the emptiest cosmic regions, possibly also carrying information on the generation of primordial helicity \citep[e.g.][]{2005A&A...433L..53S,2009IJMPD..18.1395C,2016PhyS...91j4008K}.

Owing to uncertainties in the physics of high energy regimes in the early Universe, the uncertainty in the outcome of most of the above scenarios is rather large and fields in the range of $\sim ~10^{-34}-10^{-10}$ G are still possible.

The presence of magnetic fields with rms values larger than a few co-moving $\sim ~\rm nG$ on $\leq ~\rm Mpc$ scales at $z \approx 1100$ is presently excluded by the analysis of the CMB angular power spectrum by Planck \citep{PLANCK2015,2014PhRvD..89d3523T}, while higher limits are derived for primordial fields with much larger coherence length \citep[][]{1997PhRvL..78.3610B}. Conversely, the lack of detected Inverse Compton cascade around high redshift blazars was used to set lower limits {\footnote{See however \citet{2012ApJ...752...22B} for a different interpretation.}} on cosmological seed fields of $\geq 10^{-16} ~ \rm G$ on $\sim \rm ~Mpc$ \citep[][]{2009ApJ...703.1078D,2010Sci...328...73N,2011ApJ...727L...4D,2014ApJ...796...18A,2015PhRvD..91l3514C,2015PhRvL.115u1103C}.  

\subsection{Seeding from Galactic Outflows}

At lower redshift ($z \leq 6$) galactic feedback can transport magnetic fields from galactic to more rarefied scales such as galaxy clusters. In lower mass haloes, star formation drives winds of magnetized plasma into the circum-galactic medium \citep[e.g.][]{Kronberg..1999ApJ,Volk&Atoyan..ApJ.2000,donn09,2006MNRAS.370..319B,sam17} and into voids \citep[][]{beck13}. At the high mass end, active galactic nuclei (AGN) can magnetize the central volume of clusters through jets \citep[e.g.][]{2008A&A...482L..13D,2009ApJ...698L..14X,donn09} and even the intergalactic medium during their violent quasar phase \citep[][]{2001ApJ...556..619F}. Just taking into account the magnetization from dwarf galaxies in voids, a lower limit of the magnetic field in voids has been derived as $\sim 10^{-15} ~\rm G$ \citep[][]{beck13,sam17}.

If magnetic fields have been released by processes triggered during galaxy formation, they might have affected the transport of heat, entropy, metals and cosmic rays in forming cosmic structures \citep[e.g.][]{2016mssf.book...93P,2008PhRvL.100h1301S}. \par
Additional processes such as the ``Biermann-battery'' mechanism \citep[][]{1997ApJ...480..481K}, aperiodic plasma fluctuations in the inter-galactic plasma \citep[][]{2012ApJ...758..102S},  resistive mechanisms \citep[][]{mb11} or ionization fronts around the first stars  \citep[][]{2005A&A...443..367L} might provide additional amplification to the primordial fields starting from $z \leq 10^3$, i.e. after recombination. 

\section{Magnetic Field Amplification in the Intra-Cluster Medium}\label{sect.ampl}

\subsection{Amplification by Compression} \label{sect.compr}

\begin{figure*}
	\centering
	\includegraphics[width=0.9\textwidth]{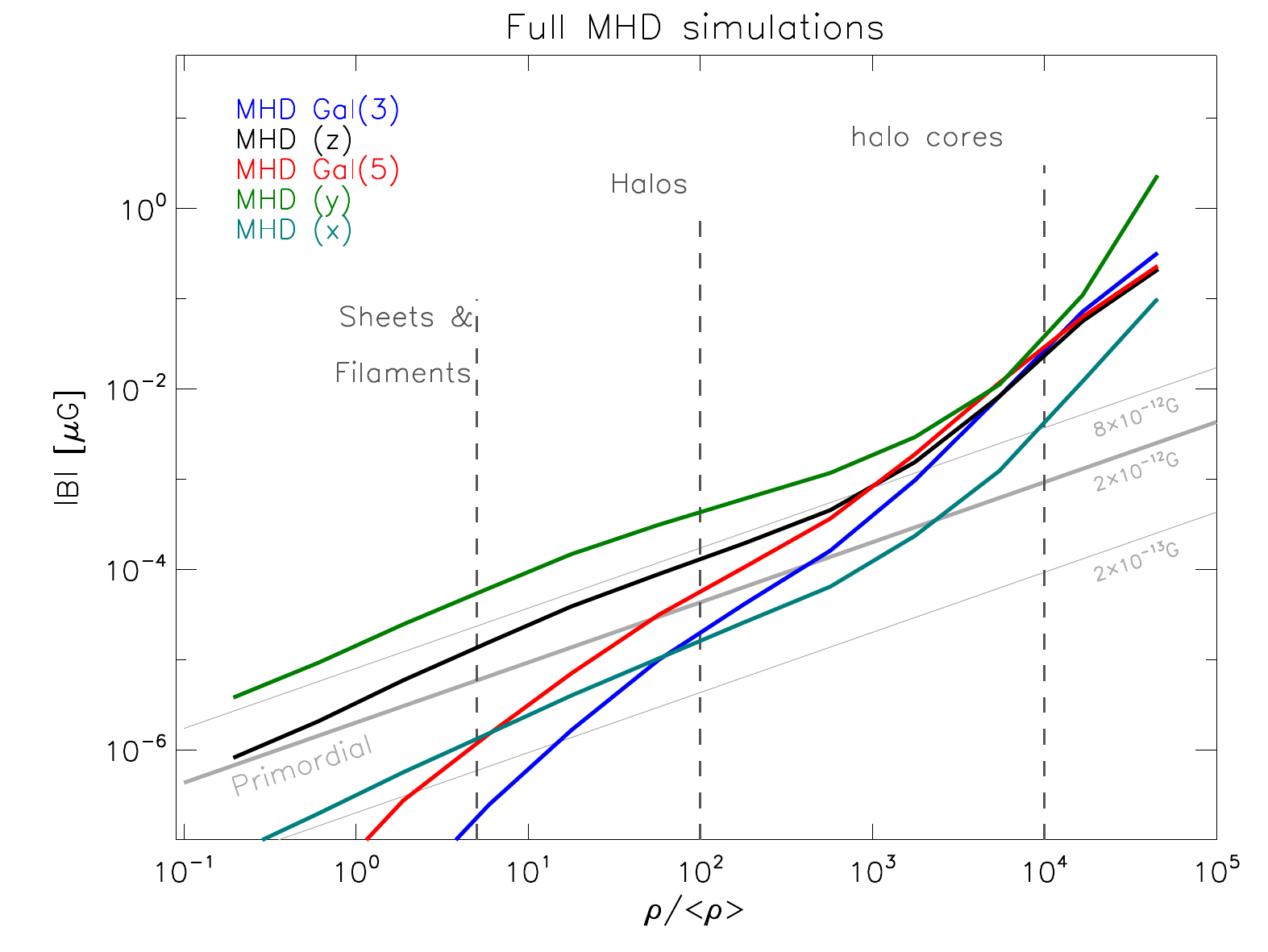}
    \caption{Magnetic field strength as a function of over density in cosmological SPH simulations. Starting from 3 different cosmological seed field strengths: $2\times 10^{-13} \rm G$ (dark green), $2\times 10^{-12} \rm G$ (black), $8\times 10^{-12} \rm G$ (green) \citep{2008SSRv..134..311D,2005JCAP...01..009D}. Adiabatic evolution solely by compression in grey. Runs with galactic seeds are in red and blue.} \label{fig:ComprAmpl}
\end{figure*}

From the third term in the induction equation (Eq. \ref{eq.induction})) we find that a positive divergence of the velocity field $\nabla \cdot \vec{v}$, i.e. a net inflow, results in the growth of the magnetic field \citep{2012MNRAS.423.3148S}. Indeed, it is a basic result of MHD that magnetic flux $\Phi$ is conserved \citep[e.g.][]{2006PhT....59a..58K} leading to the scaling of the magnetic field with density: 
\begin{eqnarray}
B(\rho) &\propto& B(z_\star) \left( \frac{\rho}{\left<\rho\right>}\right)^{2/3}. \label{eq.Brho}
\end{eqnarray}
For a galaxy cluster with an average over-density of $\Delta = \rho/\left<\rho\right> \approx 100$ this means that adiabatic compression can amplify the seed field by up to a factor of $\sim 20$ within the virial radius (or $\sim 180$ within the cluster core, where the density can be $\approx 2500$ the mean density). This refers to the average magnetic field inside a radius of the cluster. The peak density and magnetic field can be much higher. However, depending on the redshift and environment of the seed fields, the expectation from adiabatic amplification can be lower. Nonetheless, observations find a scaling exponent of magnetic field strength with cluster density of $0.5-1$, which is compatible with amplification by compression. \par
In figure \ref{fig:ComprAmpl} we reproduce a central result from early cosmological SPMHD (smooth particle magneto-hydrodynamics) simulations \citep{2005JCAP...01..009D,2008SSRv..134..311D}. They show the magnetic field strength over density in a cosmological simulation with cosmological seed fields of $B(z_\star) = 2\times 10^{-13}\,\mathrm{G}$ (dark green), $B(z_\star) = 2\times 10^{-12}\,\mathrm{G}$ (black), $B(z_\star) = 8\times 10^{-12}\,\mathrm{G}$ (dark green) co-moving, seeded at $z_\star = 20$ alongside the analytical expectation from equation \ref{eq.Brho}. Runs with galactic seeding in blue and red. At central cluster over-densities ($\rho/\left<\rho\right> > 1000$), all but one simulations reach $\mu$G field strengths. Thus different seeding models are indistinguishable here. Differences to galactic field seeding appear only at lower densities. \par
In runs with a cosmological seed field, amplification is mostly caused by compression below over-densities of 1000. At larger over-densities, a dynamo caused by velocity gradients along the field lines in the first term of the induction equation \ref{eq.induction} operates and leads to much higher field strengths. This is characteristic for turbulence in structure formation, which we will discuss next. Simulations of the cosmic dynamo and their limitations will be covered later in section \ref{sect.DynSim}. \par

\subsection{A Brief Introduction to Turbulence}\label{sect.turb}

Let us first introduce a few key concepts of turbulence used throughout the review. For a more detailed exposure, we refer the reader to the vast literature available on Astrophysical turbulence \citep[e.g.][]{1966hydr.book.....L,2006PhT....59a..58K,2009SSRv..143..387L}. \par
A key idea of the Kolmogorov picture of turbulence is that random fluid motions with \emph{velocity dispersion}\footnote{Note that velocity and velocity dispersion (i.e. root-mean-square of the power-spectrum at scale $k$) are used somewhat interchangeably in the literature. Similarly we denote the velocity dispersion with $v$ as the distinction is usually clear by context.} $v$ of size or scale $l$ (''eddies'') break up into two eddies of half the size due to the convective $\vec{v} \cdot \nabla \vec{v}$ term in the fluid equations. This process constitutes a local energy transfer from large to small scales at a rate $k v$, where $k= 2\pi/l$ is the wave vector. This process continues at each smaller length scale which leads to a cascade of velocity fluctuations down to smaller scales with decreasing kinetic energy. At an \emph{inner scale} $k_\nu$, the local kinetic energy becomes comparable to viscous forces, which  dissipate the motion into thermal energy or, in case of a dynamo, also magnetic energy via the Lorentz force. At each scale, the cascading time scale is the \emph{eddy turnover time} $\tau_l = l/v_l$ and for continuous injection of velocity fluctuations at the \emph{outer scale} a steady state is reached. If the kinetic energy density of these fluctuations is $1/2 \rho v^2 = \rho/2 \int I(k) \,\mathrm{d}k$ (assuming isotropy), then it can be shown that the velocity power spectrum $I(k)$ is \citep{1941DoSSR..30..301K,1991RSPSA.434....9K}:
\begin{eqnarray}
	I(k) &\propto& v_0^2 \frac{k_0^{2/3}}{k^{5/3}},\label{eq.cascade}
\end{eqnarray}
where $v_0$ is the velocity dispersion of the largest eddy at scale $k_0$. We note that $v_0$ is a velocity fluctuation on top of the mean. This dispersion of the associated random velocity field then scales as $v^2 \propto l^{2/3}$. It follows that the energy of turbulence is dominated by the largest scales and that viscous forces are important close to the dissipative inner scale, where motions are slowest. The range of scales where equation \ref{eq.cascade} is valid is called the \emph{inertial range}, and the \emph{Reynolds number} is defined as:
\begin{eqnarray}
	R_e &=& \frac{v_0}{k_0\nu} \label{eq.Re}\\
    	&\propto& \left( \frac{l_0}{l_\nu} \right)^{4/3} \label{eq.Re}
\end{eqnarray}
with the \emph{kinematic viscosity} $\nu$. 
The role of small scales is universal in the sense that the cascading does not depend on the driving scale or velocity  (assuming homogeneity, scale invariance, isotropy and locality of interactions) \citep{2007mhet.book...85S}. We note that turbulence is not limited to velocity fluctuations around a mean caused by a superposition of velocity eddies. The velocity field causes density and pressure fluctuations as well, because these are coupled via the fluid equations. For sub-sonic turbulence the fluctuations will be adiabatic. This has been used to place an upper limit on the kinematic viscosity in the Coma cluster of $\nu < 3\times10^{29} \,\mathrm{cm}^2/\mathrm{s}$  on scales of 90 kpc using X-ray data \citep{2004A&A...426..387S}.  \par

Whether or not the stage of the dynamo amplification is reached in an astrophysical system ultimately depends on the magnetic Reynolds number (eq. \ref{eq.Rm}) and on the nature of the turbulent forcing in the ICM \citep[][]{fed14,bm16}. The magnetic Reynolds number is set by the outer scale and the dissipation scale, so it is worth discussing the latter next. For galaxy clusters, these scales are connected to the physics of the ICM plasma.

\subsubsection{The Spitzer Model for the ICM}\label{sect.spitzer}

As noted in the introduction,  most theoretical and numerical studies approximate the ICM plasma as a fluid. However, the MHD equations as a statistical description of the many-body plasma are applicable only, if equilibration processes between ionized particles act on length and time scales much smaller than ''the scales of interest'' of the fluid problem, i.e. if collisional equilibrium among particles (protons, electron, metal ions) is maintained so local particle distributions become Maxwellian and temperature and pressure are well defined \citep{1966hydr.book.....L}. \par
In the ''classic'' physical picture of the ICM, this arises from ion-ion Coulomb scattering, with a viscosity $\nu_{\rm ii}$, over a mean free path $l_{\rm mfp}$ which is given by the Spitzer model for fully ionized plasmas \citep{1956pfig.book.....S}. It can be shown that a \emph{whole cluster} is then ''collisional'' in the sense that $r_\mathrm{vir} \gg l_\mathrm{mfp}$ \citep{SA86.1}, with:
\begin{eqnarray}
    l_{\rm mfp} &\approx& 23  \left(\frac{n_\mathrm{th}}{10^{-3} \,\mathrm{cm}^{-3}}\right)^{-1} \left( \frac{T}{10^8 \,\mathrm{K}}\right)^{2}\,\mathrm{kpc}. \label{eq.class_lmfp}
 \end{eqnarray}
Under these conditions, the Reynolds number (equation \ref{eq.Re}) of the ICM in a cluster during e.g. a major merger is \citep[e.g.][]{2007MNRAS.378..245B}:
\begin{eqnarray}
R_e &=& \frac{L v_L}{\nu_{\rm ii}} \\
&\approx & 52 \,\frac{v_L}{10^3 \,\mathrm{km/s} } 
\cdot \frac{L}{300\,\mathrm{kpc}} \cdot \frac{n}{10^{-3} \,\mathrm{cm}^{-3}} \cdot \left(\frac{T}{8 \mathrm{keV}}\right)^{-5/2} \cdot \left( \frac{\log \Lambda}{40} \right)
\end{eqnarray}
where $L$ is a typical eddy size (ideally the injection scale of turbulence), $\log \Lambda$ is the Coulomb logarithm \citep{2011hea..book.....L} and  $v_L$ is the rms velocity within the scale $L$.  Thus based on typical values of the ICM, the Reynolds number would hardly reach $R_e \sim 10^2$ in most conditions. \par
In contrast, rotation measures inferred from observations of radio galaxies have demonstrated field reversals on kpc scales, implying much larger Reynolds numbers \citep{2008MNRAS.391..521L,2010A&A...522A.105G,bo13,2011A&A...529A..13K,2012A&A...540A..38V}. Turbulent gas motions from AGN feedback have been observed directly  with the Hitomi satellite in the Perseus cluster \citep{2016Natur.535..117H} showing velocity dispersions of $\sim 200$ km/s on scales of $< 60$ kpc. This is not compatible with a medium based solely on Coulomb collisions. \par
Thus it is unavoidable to consider a more complex prescription of  the ICM plasma.  In the future, stronger constraints on the velocity structure of gas motions in galaxy clusters will be provided by the XIFU instrument on the Athena satellite \citep[][]{2013arXiv1306.2322E,2018arXiv180502577R}.\par
We note that modern numerical simulations of galaxy clusters reach and exceed spatial resolutions of the Spitzer collisional mean free path. It follows that other processes than Coulomb scattering have to maintain collisionality on smaller scales for these simulations to be valid at all. Just adding a magnetic field to the Spitzer model, i.e. Coulomb scattering plus a Lorentz force, does not suffice to make the ICM collisional on kpc scales. In a microphysical sense the MHD magnetic field is a mean magnetic field that arises \emph{after} averaging over micro-physical quantities \citep[adiabatic invariants][]{2002cra..book.....S}. 

\subsubsection{Turbulence and the Weakly-Collisional ICM} \label{sect.wcICM}

In the MHD limit, turbulence can excite three MHD waves, of which two have compressive nature (fast and slow modes, similar to sound waves) and one is solenoidal (Alfv\'{e}n mode). The Alfv\'{e}n speed is given by \cite{1942Natur.150..405A}:
\begin{eqnarray}
v_\mathrm{A} &=& \frac{B}{\sqrt{4 \pi \rho}} \label{eq.va}\\
             &=& 69 \frac{B}{1\,\mu\mathrm{G}} \left(\frac{n_\mathrm{th}}{10^{-3}\,\mathrm{cm}^{-3}}\right)^{-1/2} \,\mathrm{km}/\mathrm{s},
\end{eqnarray}
with the number density of (thermal) ions $n_\mathrm{th}$.  \par
Numerical simulations of cluster formation find turbulent velocities at the outer scale of several hundred km/s \citep{2014ApJ...782...21M}, which means that ICM turbulence starts off \emph{super-Alfv\'{e}nic} on the largest scales. Thus the magnetic field is dynamically not important near the outer scale and field topology is shaped by fluid motion. \par
Integrating equation \ref{eq.cascade} over $k$, we find that $v_l \propto l^{1/3}$ and with equation \ref{eq.va} the \emph{Alfv\'{e}n scale}, where the magnetic field back-reacts on turbulent motions \citep{2007MNRAS.378..245B}:
\begin{eqnarray}
    l_A \approx 100 \left( \frac{B}{\mu\mathrm{G}} \right)^3 \left( \frac{L_0}{300 \,\mathrm{kpc}} \right) \left( \frac{V_\mathrm{L}}{10^3 \, \mathrm{km}/\mathrm{s}}\right)^{-3} \left( \frac{n_\mathrm{th}}{10^{-3} \,\mathrm{cm}^{-3}} \right)^{\frac{3}{2}} \,\mathrm{pc}, \label{eq.lA}
\end{eqnarray}
which is already smaller than the classical mean free path derived before and leads to a Reynolds number of a few 1000. As we will see, this scale is crucial to numerically resolve magnetic field growth by turbulence. \par
In principle, one has to consider three separate turbulent cascades, whose interplay changes close around Alfv\'{e}n scale \citep[see][ and ref. therein]{bl11}. Here the character of turbulence dramatically changes. The Lorentz force introduces strong anisotropy to fluid motions, viscosity and turbulent eddies become anisotropic and non-local interactions between modes in the turbulent cascade start to be important.  See \citet{1997ApJ...485..680G,2007mhet.book...85S} for a more detailed picture of these processes.  \par
That leaves us to ask, what is it that keeps the ICM collisional on scales much smaller than the Alfv\'{e}n scale, so MHD is applicable at all? \citet{sch05, 2006ApJ...640L.175B, 2007mhet.book...85S, 2008PhRvL.100h1301S} propose that due to the large Spitzer mean free path, the non-ideal MHD equations are not sufficient to estimate viscosity and obtain a Reynolds number for the ICM. Kinetic calculations reveal that particle motions perpendicular to the magnetic field are suppressed and motions parallel to the field can exist and excite firehose and mirror instabilities. The instabilities inject MHD waves, which act as scattering agents (magnetic mirrors). Scattering off these self-exited modes isotropizes particle motions on very small length and time scales. This picture is confirmed also by hybrid-kinetic simulations \citep{2014PhRvL.112t5003K}.  \par
Under these conditions, a lower limit of the viscous scale of the ICM is given by the mobility of thermal protons in a magnetic field, which is the Larmor radius \citep[e.g.][]{sch05,bm16,bl11}:
\begin{eqnarray}
l_{\rm mfp} &=& r_\mathrm{Lamor} \\
            &\approx& 3 \cdot  10^{-12} \rm kpc  \left( \frac{T}{10 \,\mathrm{keV}} \right) \left( \frac{B}{\mu \,\mathrm{G}} \right)^{-1}. \label{eq.lmfp}
\end{eqnarray}
In this case, the effective Reynolds number of the ICM becomes:
\begin{equation}
R_{e,\mathrm{eff}} = \left( \frac{l_0}{l_\nu} \right)^{4/3} \sim 10^{19}, \label{eq.Retrue}
\end{equation}
This estimate predicts a highly turbulent ICM down to non-astrophysical scales and establishes collisionality on scales of tens of thousands of kilometers. This is good news for simulators, because the fluid approximation is well motivated in galaxy clusters and probably valid down to scales forever out of reach of simulations \citep{2017MNRAS.465.4866S,2014ApJ...781...84S}. \par
The bad news is that the physics of the medium is complicated, so that e.g.  transport properties of the ICM are dominated by scales out of reach for simulations and observations. One example is heat conduction, where some estimates from kinetic theory predict no conduction in the weakly-collisional limit \citep{2008PhRvL.100h1301S,2011MNRAS.417..602K}. Indeed, only an upper limit was found by comparing observations with simulations \citep{2015ApJ...798...90Z}. Thus, the properties of the medium cannot be constrained any further. Additionally, the likely presence of cosmic-ray protons makes the picture of generation and damping of compressive and Alfv\'{e}n modes/turbulence even more involved (figure \ref{fig:turbGraph}) \citep{2002cra..book.....S,2011MNRAS.412..817B,2004MNRAS.350.1174B}.\par

Now that we have established that MHD is very likely applicable down to sub-pc scales, we can discuss how (large-scale) magnetic fields can be amplified by turbulence in the MHD limit.

\subsection{The Small-scale Dynamo} \label{sect.ssd}

\begin{figure}
\centering
\includegraphics[width=0.49\textwidth]{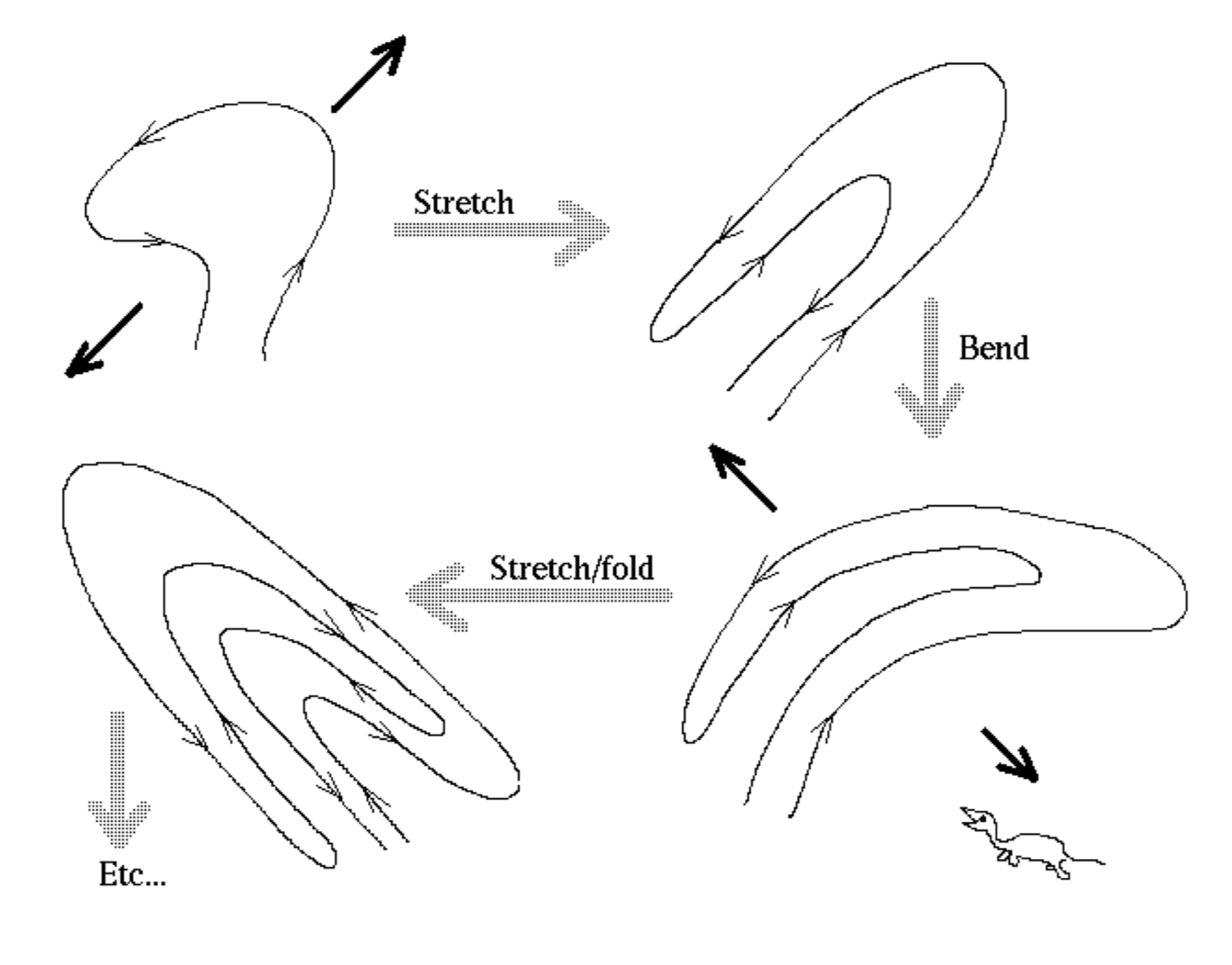}
\includegraphics[width=0.49\textwidth]{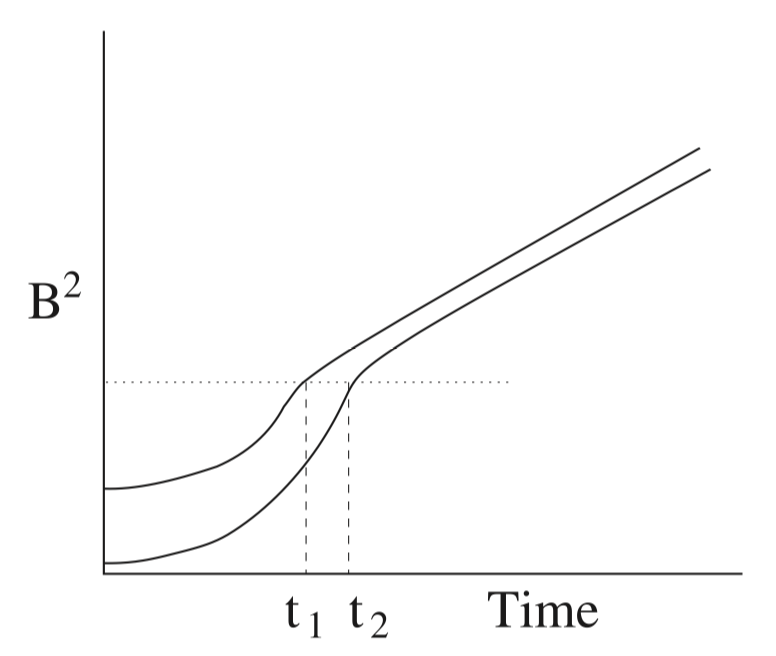}
\caption{Left: Cartoon illustrating the stretching and folding for magnetic field lines on small scales from \citet{2002ApJ...576..806S}. Right: Cartoon from \citet{2009ApJ...693.1449C} depicting the growth of magnetic energy in driven turbulence simulations with very weak initial magnetic field. The initial seed field sets the timescale for the end of the kinematic dynamo and the beginning of the non-linear dynamo.  }
\label{fig:dynamo}
\end{figure}
If a magnetic field is present in a turbulent flow, the properties of turbulence can change significantly  due to the back-reaction of the field on the turbulent motions \citep{1967PhFl...10..859K,1997ApJ...485..680G}. In a magnetic dynamo, the kinetic energy of turbulence is transformed into magnetic energy, which is a non-trivial theoretical problem. The dissipation of magnetic energy into heat occurs at the \emph{resistive scale} $l_\eta$ and the \emph{magnetic Reynolds number} is defined as:
\begin{eqnarray}
	R_{m} &=& \frac{v_0}{k_0 \eta} \label{eq.Rm}\\
    	&\propto& \left( \frac{l_\eta}{l_0} \right)^{4/3}
\end{eqnarray}
The magnetic Prandtl number relates resistive with diffusive scales  eq. \ref{eq.Retrue}.
\begin{eqnarray}
	P_{m} &=& \frac{\nu}{\eta}  \label{eq.Pm} \\
          &=&  \frac{R_m}{R_e}  = \left( \frac{l_\nu}{\l_\eta} \right)^{4/3} \nonumber
\end{eqnarray}
For a theoretical framework of the gyrokinetics on small scales, including a discussion on cluster turbulence, we refer the reader to \citet{2009ApJS..182..310S}.\par
In a simplified picture, magnetic field amplification by turbulence is a consequence of the stretching and folding of pre-existing field lines by the random velocity field of turbulence, which amplifies the field locally due to flux conservation (figure \ref{fig:dynamo}, left) \citep{1950RSPSA.201..405B,1951PhRv...82..863B}. If a flux tube of radius $r_1$ and length $l_1$ with magnetic field strength $B_1$ is stretched to length $l_2$ and radius $r_2$, mass conservation leads to:
\begin{eqnarray}
 \frac{r_2}{r_1} &=& \sqrt{\frac{l_1}{l_2}}.
\end{eqnarray}
The magnetic flux $S_1 = \pi r_1^2 B$ is conserved in the high-$\beta$ regime, so for an incompressible fluid: 
\begin{eqnarray}
	B_2 &=& B_1 \frac{l_2}{l_1}
\end{eqnarray}
By e.g. folding or shear (figure \ref{fig:dynamo}, left) the field can be efficiently amplified \citep{1972SvPhU..15..159V,2002PhRvE..65a6305S}. Repeating this process leads to an exponential increase in magnetic energy, if the field does not back-react on the fluid motion (figure \ref{fig:dynamo}, right). In a turbulent flow the folding occurs on a time scale of the smallest eddy turnover time, i.e. close to the viscous scale. Flux tubes are tangled and merged, and their geometry/curvature is set by the resistive and viscous scales of the flow. The energy available for magnetic field growth is the rate of strain $\delta v/l$  \citep{2005mpge.conf...86S}. { We note that due to the universality of scales in turbulence, the dynamo process does not depend on the actual magnetic field strength and time scale of the system. As long as the conditions for a small scale dynamo are satisfied, field amplification will proceed as shown in figure \ref{fig:dynamo}, right.} \par
\begin{figure}
\centering
\includegraphics[width=0.45\textwidth]{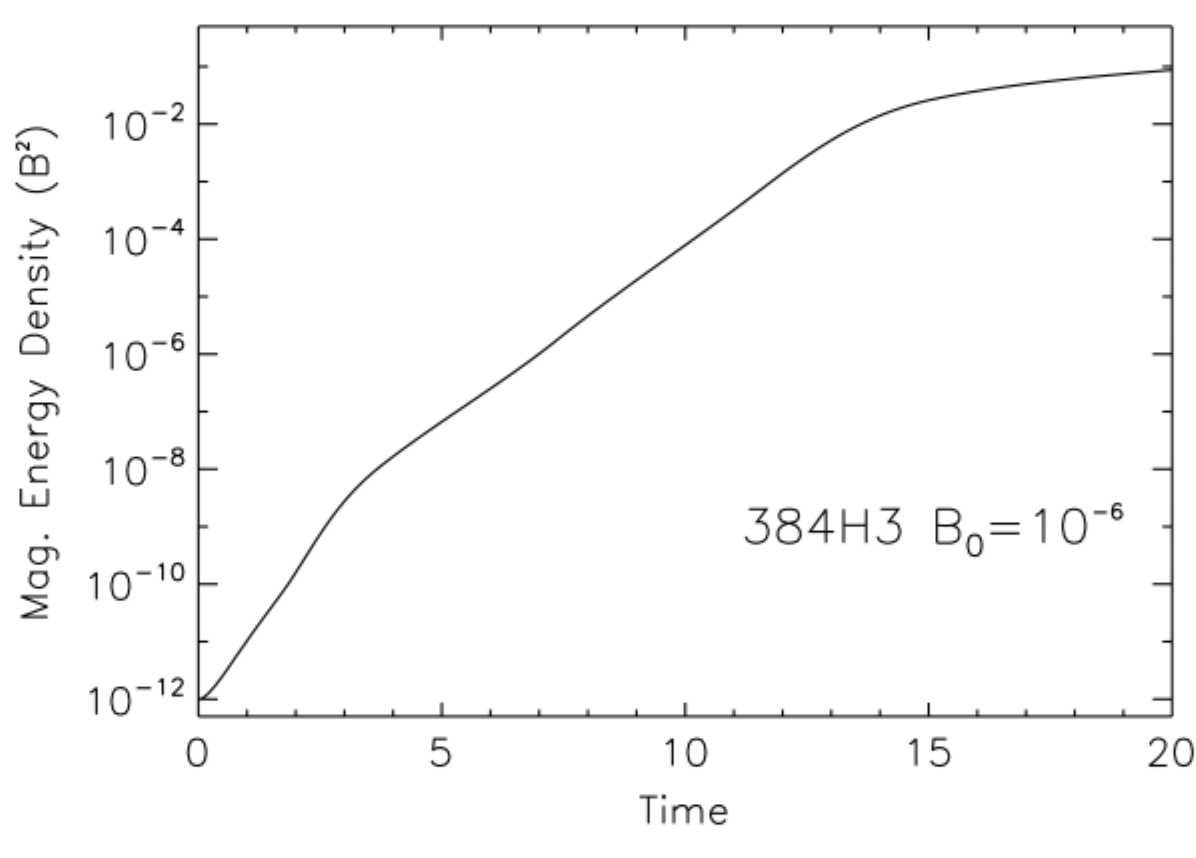}
\includegraphics[width=0.45\textwidth]{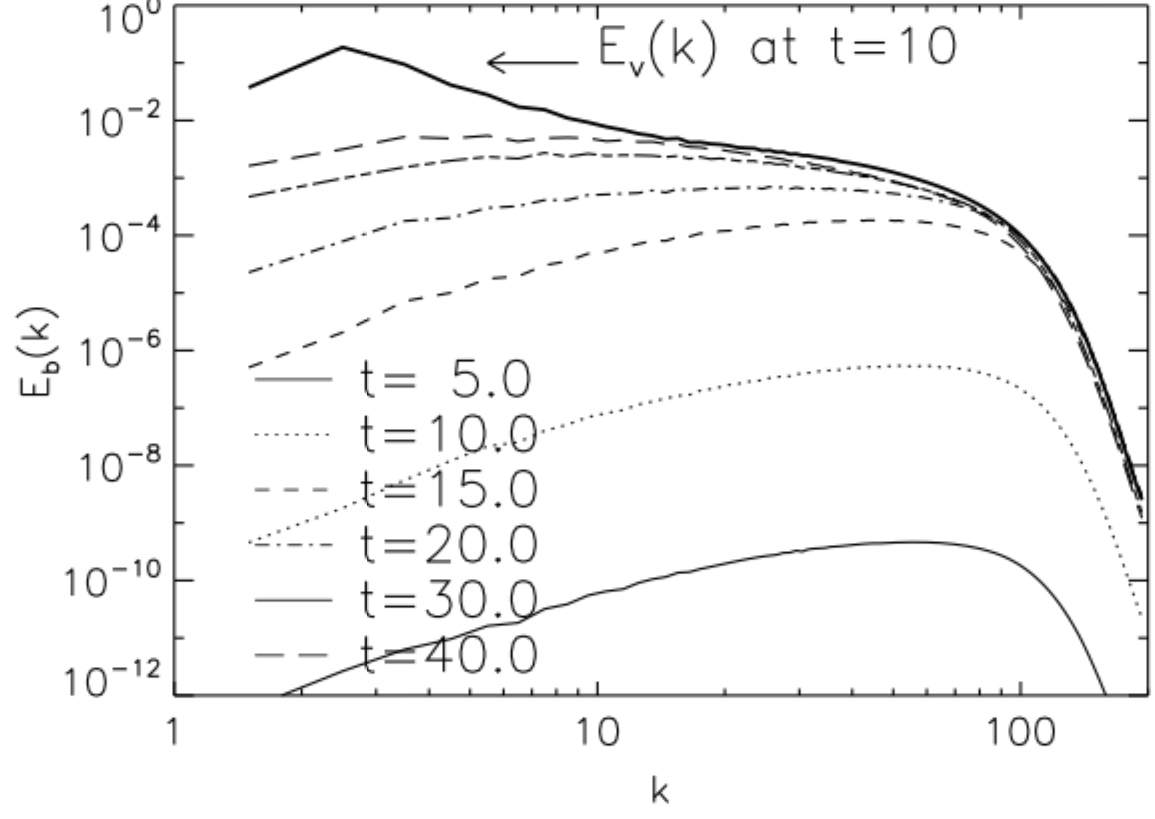}
\caption{Left: Evolution of magnetic energy over time in a simulation of driven turbulence. The transition from kinematic to non-linear dynamo occurs at $t=15$. Right: Magnetic field energy spectra over wave number for different times of the same run. Both figures by \citet{2009ApJ...693.1449C}. }
\label{fig:dynamo2}
\end{figure}

For a small ($10^{-13}\,\mathrm{G}$) initial seed field in galaxy environments or proto-clusters (see section \ref{sect.seed}), back-reaction is negligible, $P_m$ is very large and a \emph{small-scale dynamo} (SSD) operates in the \emph{kinematic regime} of exponential amplification without back-reaction \citep{1992ApJ...396..606K}. The SSD proceeds from small to large scales in an inverse cascade starting at the resistive scale.  A rigorous treatment of this process based on Gaussian random fields in the absence of helicity was first presented by \citet{1968JETP...26.1031K}, for an instructive application to proto-clusters see e.g. \citet[][]{2011ApJ...731...62F}, \citet{schober13}, and \citet{2013MNRAS.432..668L}. For a unique experimental perspective on the kinematic dynamo see \citet{2015PNAS..112.8211M}. In figure \ref{fig:dynamo2}, we reproduce the time evolution of magnetic energy (left) and of the magnetic and kinetic power spectra (right) from an idealized simulation of the MHD dynamo \citep{2009ApJ...693.1449C}. Here $k_\nu = 1/l_nu \approx 100$, and the kinematic dynamo proceeds until $t = 15$. An instructive numerical presentation can be also found in \citep{po15}, { a detailed exposure is presented in \citet{2004ApJ...612..276S}.} \par
The exponential growth of the kinematic dynamo is stifled quickly \citep{2011ApJ...741...92B}, once the magnetic field starts to back-react on the turbulent flow. The dynamo then enters the \emph{non-linear regime} and turbulence  grows a steep inverse cascade with an outer magnetic scale $l_\mathrm{B}$. In figure \ref{fig:dynamo2}, this occurs for $t > 15$ and $k_\mathrm{B} = 1/l_\mathrm{B} \approx 10$ at $t = 40$.  In principle, growth will continue until equipartition with the turbulent kinetic energy is attained \citep{2004PhRvE..70c6408H,2005PhR...417....1B,2009ApJ...693.1449C,po15,bm16}. \par
What does this mean for galaxy clusters? Above we had motivated a lower limit for the viscous scale in proto-clusters of around 1000 km (eq. \ref{eq.lmfp}) and Reynolds numbers of up to $10^{19}$. The resistive scale is highly uncertain, but likely small enough for an SSD to occur. The large Reynolds number leads to a growth timescale of the kinematic dynamo of $\tau \approx 1000 \, \mathrm{yrs}$ \citep{2002ApJ...576..806S,2004ApJ...612..276S,bm16}. It is clear that this exponential growth is so fast that it will complete in large haloes before galaxy clusters start forming at redshifts 2-1. The kinematic dynamo efficiently amplifies even smallest seed fields until back-reaction plays a role, i.e. the Alfv\'{e}n scale approaches the viscous scale. \par
Depending on the physics of the seeding mechanism, the kinematic phase will take place in the environment of high redshift galaxies that is polluted by jets and outflows, in proto-clusters or, in the case of a cosmological seed field, in all collapsing over-dense environments  at high redshift \citep{1983flma....3.....Z, 1992ApJ...396..606K,1997ApJ...480..481K, 2013MNRAS.432..668L}.\par 
 However, contrary to the idealized turbulence simulations shown in figure \ref{fig:dynamo2}, turbulent driving in clusters occurs highly episodic and at multiple scales at once (section \ref{sect.physturb}), so the equipartition regime is never reached. Instead, the magnetic field strength and topology will depend on the driving history of the gas parcel under consideration. It is also immediately clear that as opposed to amplification by isotropic compression, this dynamo erases all imprint of the initial seed field. Thus we cannot hope to constrain seeding processes from magnetic fields in galaxy clusters, but instead have to look to filaments and voids, where the dynamo may not be driven efficiently.

\subsection{Cosmic-ray Driven Amplification and Plasma Effects}

Magnetic fields can be amplified by a range of effects caused by cosmic rays. Current-driven instabilities, e.g. \citep{2004MNRAS.353..550B}, have been shown to amplify magnetic fields by considerable factors \citep{2010ApJ...717.1054R}. The electric current that drives this instability comes from the drift of CRs. The return electric current of the  plasma leads to a transverse force that can amplify transverse perturbations in the magnetic field.
\cite{2004MNRAS.353..550B} pointed out that the fastest instability is caused by the return background plasma current that compensates the current produced by CRs streaming upstream of the shock. It is important to note that this instability is non-resonant and can be treated using ideal MHD. The Bell or non-resonant streaming (NRS) instability has been tested in various numerical studies using a range of methods ranging from pure MHD \citep{2008ApJ...678..939Z}, full PIC \citep{2011ApJ...733...63R},  hybrid \citep{2014ApJ...783...91C,2014ApJ...783...91C,2014ApJ...794...46C} to Vlasov or PIC-MHD \citep{2013MNRAS.430.2873R,2008MNRAS.386..509R,2015ApJ...809...55B}; see \citet{2016RPPh...79d6901M} for a review. { In strong SNR shocks, a non-resonant long-wavelength instability can amplify magnetic fields as well \citep{2009AstL...35..555B,2011MNRAS.410...39B}, but this has not been confirmed by simulations. A full non-linear calculation is needed to take into account the feedback of the CRs on the shock structure that may lead to a significant modification of the shocks structure \citep[e.g.][]{md01,2006ApJ...652.1246V,2014ApJ...789..137B}. Recent $\gamma$-ray observations of SNR challenge this picture, so CR spectra might be steeper than the test-particle prediction \citep{2012JCAP...07..038C,2014ApJ...783...33S}. }  All the aforementioned effects operate on length scales comparable to the gyro-radius of protons. 

Filamentation instabilities can act on larger scales, as do models where CRs drive a turbulent dynamo \citep{2012MNRAS.427.2308D, 2013MNRAS.436..294B}. In the latter case, the turbulence is caused by the cosmic-ray pressure gradient in the upstream region which exerts a force on the upstream fluid that is not proportional to the gas density. Density fluctuations then lead to fluctuations in the acceleration which, in turn, produce further density fluctuations. CRs are also able to generate strong magnetic fields at shock fronts which is invoked to explain the high magnetic field strengths in several historical supernova remnants. This was first studied in the context of the high magnetic field strengths deduced from X-ray observations of supernova remnants. In fast shocks, the streaming of CRs into the upstream region triggers a class of plasma instabilities that can grow fast enough to produce very strong magnetic fields \citep{2000MNRAS.314...65L}. \par

\begin{figure}
\begin{center}
\includegraphics[width=0.95\textwidth]{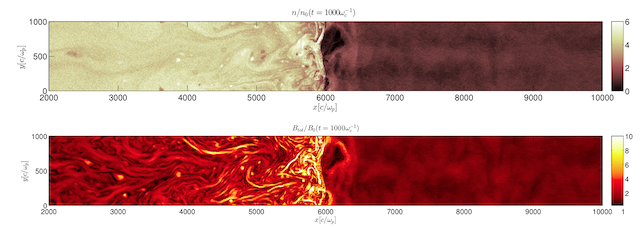}
\caption{Ion number density (top) and magnetic field strength (bottom) for a parallel shock wave with Mach number $M=20$ at 1000 $\omega_c^{-1}=mc/eB_0$ from \citep{2014ApJ...794...46C}.\label{fig:caprioli}}
\end{center}
\end{figure}

More recently, \cite{2013MNRAS.430.2873R} have studied a CR-driven filamentation instability that also results from CR streaming, but contrary to the Bell-instability generates long-wavelength perturbations. \citet{2014ApJ...794...46C} have investigated CR-driven filamentation instabilities using a di-hybrid method where electrons are treated as a fluid and protons as kinetic particles. While progress in this field has grown substantially over the past years, very few PIC simulations for weak shocks in high-$\beta$ plasmas have been done \citep[e.g.][]{2016ApJ...818L...9G}.\par

In analytical work \citep[e.g.][]{2016MNRAS.459.2701M}, it has been shown that microphysical plasma instabilities can produce a more efficient small-scale dynamo than its MHD counterpart described above. In this picture, shearing motions drive pressure anisotropies that excite mirror or firehose fluctuations \citep[as seen in direct numerical simulations of collisionless dynamo; see][]{2016PNAS..113.3950R}. These fluctuations lead to anomalous particle scattering that lead to field growth. 
As shown in \cite{2014MNRAS.440.3226M}, these scatterings can decrease the effective viscosity of the plasma thereby allowing the turbulence to cascade down to smaller scales and thus develop greater rates of strain and amplify the field faster. Within a number of large eddy turn-over times, this process can result in magnetic fields that saturate near equipartition with the kinetic energy of the ICM.\par

While the total budget of cosmic ray protons stored in clusters is now constrained to $\leq 1 \%$ (on average) for the thermal gas energy by the latest collection of Fermi-LAT data \citep[][]{fermi14}, it cannot be excluded that a larger fraction of cosmic rays may exist close to shocks in the intra-cluster medium. At present, the limits that can be derived from $\gamma$-rays are of $\leq 15\%$, at least in the case of the (nearby) relics in Coma \citep[][]{2014MNRAS.440..663Z}.

\subsection{Processes that drive turbulence in clusters}\label{sect.physturb}

\begin{figure*}
	\centering
	\includegraphics[width=0.95\textwidth]{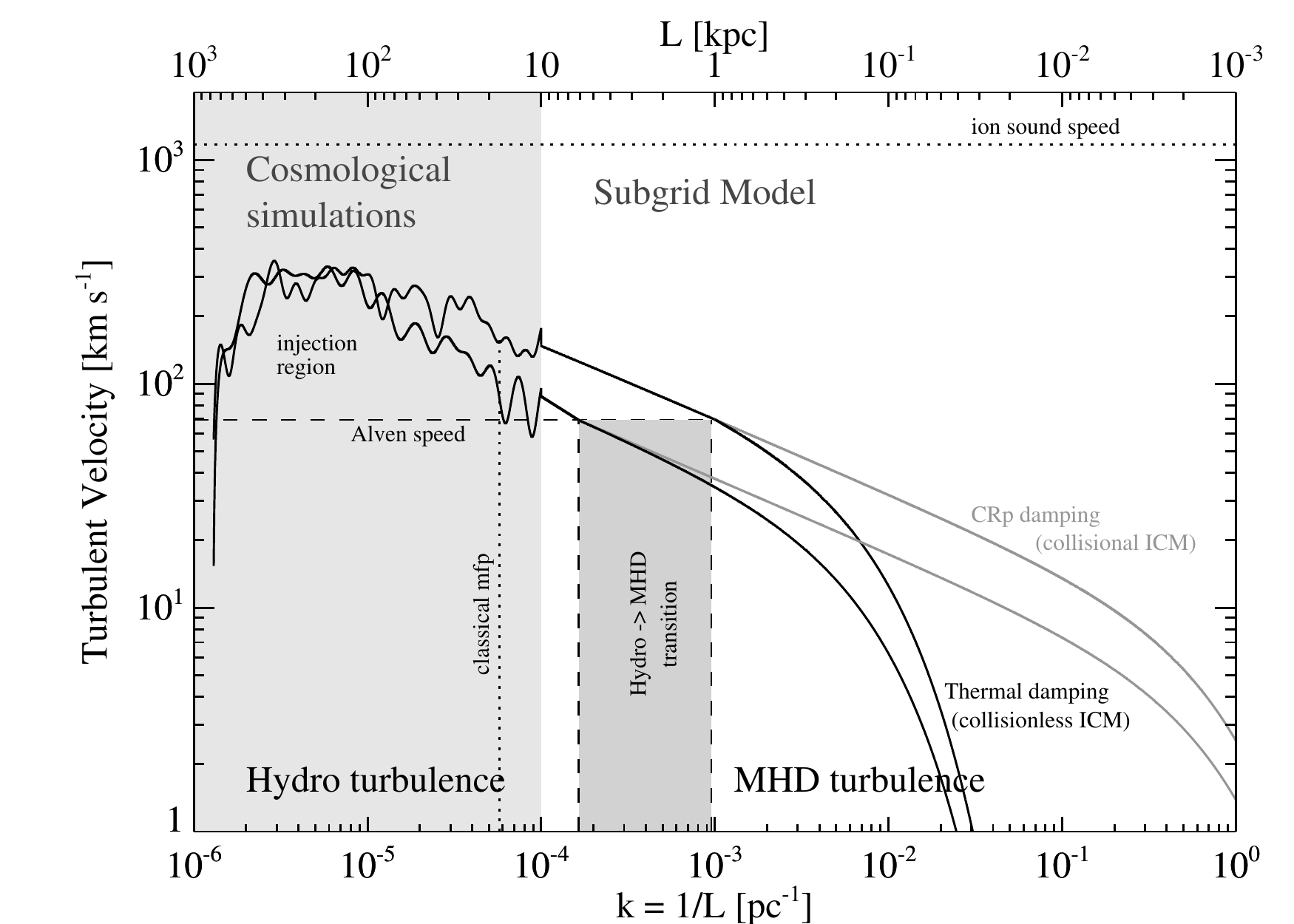}
    \caption{Cartoon depicting the cascade of only compressive turbulence over length scale in galaxy clusters, considering damping from thermal ions and cosmic-ray protons \citep{2014MNRAS.443.3564D}.} \label{fig:turbGraph}
\end{figure*}

The accretion of gas and Dark Matter subunits is a main driver of turbulence in clusters. During infall, gas gets shock-heated around the virial radius (Mach numbers $\sim 10$). In major mergers, the displacement of the ICM creates an eddy on the scale of the cluster core radii \citep[e.g.][]{2014MNRAS.443.3564D}. Shear flows generated by in-falling substructure inject turbulence through Kelvin-Helmholtz (K-H) and Rayleigh-Taylor (R-T) instabilities \citep[e.g.][]{su06,yysu17b,2016MNRAS.463..655K}. Feedback from central AGN activity, radio galaxies and galactic winds inject turbulence on even smaller scales \citep[e.g.][]{chu04,2005ApJ...628..153B,gaspari11a}.  As a result of this complex interplay of episodic driving motions on scales of half a Mpc to less than a kpc, the intra-cluster medium is expected to include weak-to-moderately-strong shocks ($\mathcal{M} \leq 5$) and hydrodynamic shear, leading to a turbulent cascade down to the dissipation scale. \par

The solenoidal component (Alfv\'{e}n waves) of the cascade will drive a turbulent dynamo, while the compressive component (fast \& slow modes) produces weak shocks and adiabatic compression waves, which can in turn generate further small-scale solenoidal motions \citep[e.g.,][]{po15,va17turb}. The relative contributions from both components will depend on the turbulent forcing and its intensity \citep{2011ApJ...731...62F,po15}. Compressive and solenoidal components of the turbulent energy are also expected to accelerate cosmic-ray protons and electrons via second-order Fermi processes, which again alters the properties of turbulence on small scales \citep[see ][ for a review]{bj14}. In figure \ref{fig:turbGraph} we reproduce a cartoon plot of the compressive cascade in clusters from \citet{2014MNRAS.443.3564D} that depicts the relevant scales: the classical mean free path (eq. \ref{eq.class_lmfp}), the Alf\'{v}en scale (eq. \ref{eq.lA}) and the dissipation scales, if the cascade is damped by thermal protons ($k > 10^{-2}$) or CR protons ($k \approx 1$). The graph also includes the sound speed and the Alfv\'{e}n speed and marks the regions accessible by current cosmological simulations. A more involved graph can be found in \citet{{bj14}}. \par
 
Note that the simple ''Kolmogorov'' picture of turbulence (section \ref{sect.turb}) with a single well defined injection scale, an inertial range and a single dissipation scale is oversimplified in galaxy clusters. As argued above, structure formation leads to an increase of the outer/driving scale with time and injection concurrently takes place at many smaller scales and can be highly intermittent.  Thus a strictly-defined inertial range does probably not exist and turbulence may be more loosely defined in clusters than in other fields of astrophysics. Cosmological simulations can be used to capture the complexity of these processes.

\section{Simulations of Turbulence and the Small Scale Dynamo in Clusters} \label{sect.DynSim}

\subsection{Simulations of Cluster Turbulence}\label{sect.cosmoturb}

\begin{figure}
\centering
\includegraphics[width=0.4\textwidth]{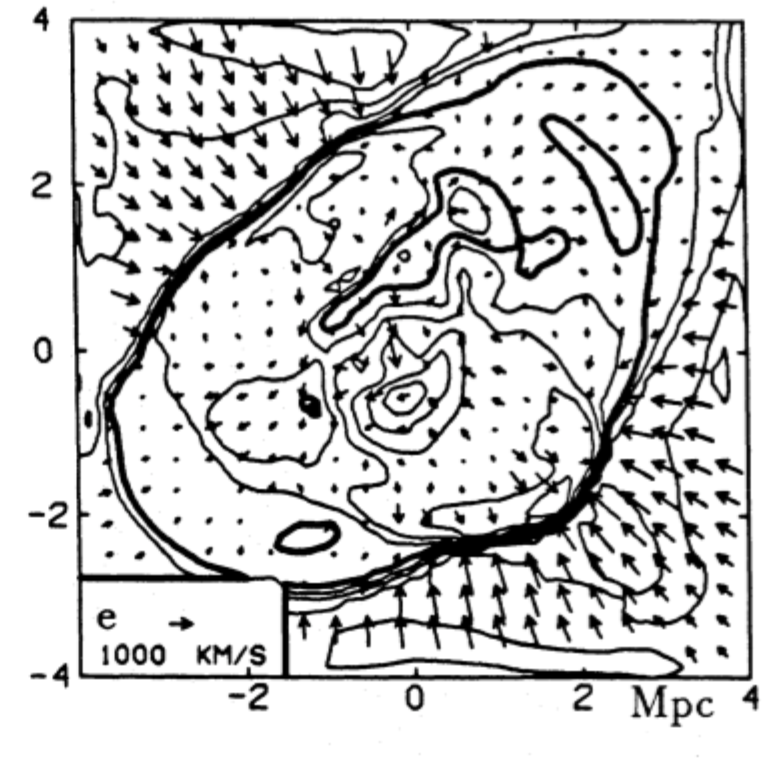}
\includegraphics[width=0.59\textwidth]{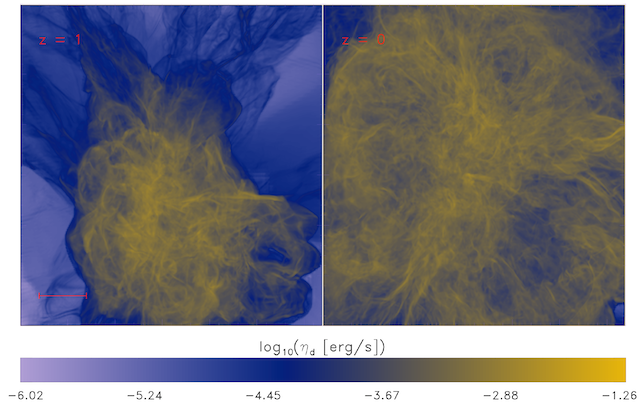}
\caption{Left: Gas motions in the first Eulerian simulation of a merging cluster \citep{1993A&A...272..137S}. Central and Right panel: projected enstrophy energy flux for a state-of-the-art Eulerian simulation with {\enzo} at $z=1$ and $z=0$, taken from \citep[][]{wi17}.}
\label{fig:simcomp}
\end{figure}

 Simulations of merging clusters have been pioneered by \citet{1990ApJ...363..349E,1992MNRAS.257...11T}, who reported a shock traveling outward during a merger. \citet{1993A&A...272..137S} for the first time used a Eulerian PPM scheme with $60^3$ zones to follow the gas dynamics in an idealized merger (figure \ref{fig:simcomp}, left) \citep[see also][]{1993ApJ...407L..53R,1997ApJS..109..307R}. Using idealized adaptive mesh refinement Eulerian merger simulations, \citet{2001ApJ...561..621R} for the first time report ram pressure stripping and turbulence, with eddy sizes of ''several hundred kpc'' [..] ''pumped by DM driven oscillations of the gravitational potential''. \citet{2005ApJ...629..791T,2004ApJ...606L.105A} used a TVD scheme to study the driving of shocks and turbulence by substructure in idealized cluster simulations. They focused on the injection of instabilities and gas stripping (see section \ref{sect.slosh}). \par

In cosmological simulations, turbulence was first studied by \citet{2005MNRAS.364..753D} using SPH with a low viscosity scheme \citep[for shocks see][]{2000ApJ...542..608M}. They find subsonic velocity dispersions of $400-800$ km/s on scales of 20 to 140 kpc, with turbulent energy fractions of 5-30 per cent and a trend for higher turbulent energies in higher mass clusters. Turbulent energy spectra from their simulations were flatter than the Kolmogorov expectation, but might have been limited by numerics (see section \ref{subsec:numerics}). Their work was extended to a sample of 21 clusters by \citet{2006MNRAS.369L..14V}, who provided scaling laws for the turbulent energy over cluster mass, see \citet{2011A&A...526A.158V} for a later study. \par

In a seminal contribution, \citet{ry08} studied the generation and evolution of turbulence in a Eulerian cosmological cluster simulation. They showed that turbulence is largely solenoidal, not compressive, with subsonic velocities in clusters and trans-sonic velocities in filaments. In agreement with prior SPH simulations, they find a clear trend of rms velocity dispersion with cluster mass and turbulent energy fractions/pressures of $10-30\%$. They also propose a vorticity based dynamo model, which we will discuss in section \ref{sect.cosmosimdyn}.\par
The influence of turbulent pressure support on cluster scaling relations was studied by \citet{2007ApJ...668....1N,lau09,2010ApJ...725.1452S,2010ApJ...721.1105B,2012ApJ...758...74B,2014ApJ...792...25N,2017MNRAS.470..142S}. Consistently, turbulent pressure increases with radius in simulated clusters, which is related to the increased thermal pressure caused by the central potential of the main DM halo. An analytic model for non-thermal pressure support was presented by \citep{2014MNRAS.442..521S}, and also validated by numerical simulations \citep{2015MNRAS.448.1020S, 2016MNRAS.455.2936S}. First power spectra of turbulence in Eulerian cosmological cluster simulations were presented by \citet{2009ApJ...698L..14X,2009A&A...504...33V}. Their kinetic spectra roughly follow the Kolmogorov scaling. The simulations reach an ''injection region'' of turbulence larger than 100 kpc, an inertial range between 100 kpc and 10 kpc and a dissipation scale below 10 kpc. Thus their Reynolds number was 10-100. \par
The next years saw improvements in resolution of cluster simulations, due to the inevitable growth in computing power. Increasingly higher Reynolds numbers could be reached and/or additional physics could be implemented usually with adaptive mesh refinement (AMR). \citet{va11turbo} studied a sample of simulated clusters with Reynolds number of up to 1000. They also developed new filtering techniques to estimate turbulent energy locally. They showed that the turbulent energy in relaxed clusters reach only a few percent. \citet{2009ApJ...707...40M,iapichino11} added a subgrid-scale model for unresolved turbulence to their simulations and studied the evolution of turbulent energy. They found that peak turbulent energies are reached at the formation redshift of the underlying halo. Their subgrid model shows that unresolved pressure support is usually not a problem in cluster simulations, and that half of the simulated ICM shows large vorticity.  \citet{2011ApJ...726...17P} simulated a sample of merging clusters and found a scaling of turbulent energy with cluster mass as $\propto M^{5/3}$, consistent with earlier SPH results \citep[][]{2006MNRAS.369L..14V}. The influence of minor mergers on the injection of turbulence in a idealized scenario of a cool core cluster was simulated with anisotropic thermal conduction by \citet{2011MNRAS.414.1493R}. They found that long-term galaxy motions excite subsonic turbulence with velocities of 100-200 km/s and give a detailed theoretical model for the connection between vorticity and magnetic fields. \par
\citet{va12filter} used an improved local filter to estimate the turbulent diffusivity in their simulations as $D_\mathrm{turb}\approx 10^{29-30} \,\mathrm{cm}^2/\mathrm{s}$ and identify accretion and major mergers as dominant drivers of cluster turbulence.\par
\begin{figure}
\centering
\includegraphics[width=0.9\textwidth]{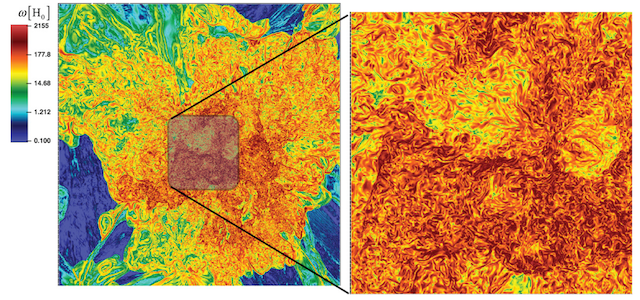}
\caption{Vorticity map for the innermost regions of a simulated $\sim 10^{15} \rm M_{\odot}$ galaxy cluster at high resolution, in the ''Matrioska run'' by \citet{2014ApJ...782...21M}.} \label{fig:miniati}
\end{figure}
In a series of papers, \citet{2014ApJ...782...21M,2015ApJ...800...60M} introduced static Eulerian mesh refinement simulations to the field. They reach a peak resolution of $\approx 10 \,\mathrm{kpc}$ covering the entire virial radius of a massive galaxy cluster with a PPM method. Consistent with previous studies they find that shocks generate 60\% of the vorticity in clusters. Their adiabatic simulations show turbulent velocity dispersions above 700 km/s, regardless of merger state. The analysis using structure functions reveals that solenoidal/incompressible turbulence with a Kolmogorov spectrum dominates the cluster, while compressive turbulence with a Burgers slope \citep{Burgers1939} become more important towards the outskirts. They propose that a hierarchy of energy components exists in clusters, where gravitational energy is mostly dissipated into thermal energy, then turbulent energy and finally magnetic energy with a constant efficiency \citep{2015Natur.523...59M}. { Vorticity maps from their approach are reproduced in figure \ref{fig:miniati}}\par

In the most recent studies, the resolution has been improved to simulate the first early baroclinic injection of vorticity in cluster outskirts \citep[e.g.][]{va17turb,2017MNRAS.469.3641I}  as well as its later amplification via compression/stretching during mergers \citet{wi17b}. 
Using the Hodge-Helmholtz decomposition, high resolution Eulerian simulations measure a very large fraction of turbulence being dissipated into solenoidal motions 
\citep[][]{2014ApJ...782...21M,va17turb,wi17b}. Baroclinic motions inject enstrophy on large scales, while dissipation and stretching terms govern its evolution. \par
Recent simulations using Lagrangian methods focus on including more subgrid physics in the setup to study the influence of magnetic fields on galaxy formation. \citet{2015MNRAS.453.3999M} show that the redshift evolution of the rms velocity fluctuations in the ``Illustris TNG'' galaxy formation simulations is independent of seed magnetic fields. 

\subsection{Cosmological Simulations of Magnetic Fields in Galaxy Clusters}\label{sect.cosmosimdyn}

\begin{figure*}
  \includegraphics[width=0.33\textwidth,height=0.33\textwidth]{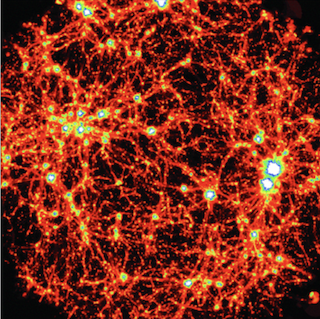}
  \includegraphics[width=0.33\textwidth,height=0.33\textwidth]{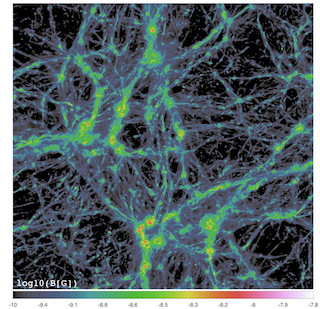}
  \includegraphics[width=0.33\textwidth,height=0.33\textwidth]{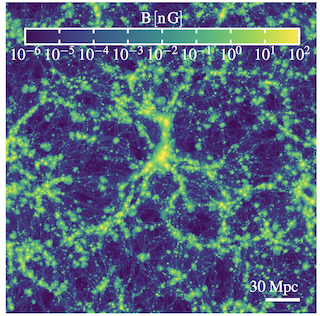}
\caption{Projection of magnetic field strength in three cosmological simulations using different MHD approaches and solvers. Left: non-radiative \gadget SPH simulations with galactic seeding by \citet{donn09}, based on the MHD method by \citet{2009MNRAS.398.1678D}. Middle: non-radiative \enzo MHD simulation on a fixed grid by \citet{va14mhd} using the Dender cleaning \citep[][]{ded02}; Right:  Simulation with full ``Illustris TNG'' galaxy formation model using a Lagrangian finite volume method \citep{ma17}.} \label{fig:cosmo1}  
\end{figure*}

Pioneering studies of magnetic fields in simulated large-scale structures were conducted by \citet{1992ApJ...386..464D,1997ApJ...480..481K,1999ApJ...518..594R}.
First full MHD simulations of cluster magnetic fields from nG cosmological seeds have been presented by \citet{1999A&A...348..351D,2002A&A...387..383D,2011MNRAS.418.2234B}. They found a correlation of the magnetic field strength the ICM gas density with an exponent of $0.9$, using smooth particle hydrodynamics (SPH) \citep{2009MNRAS.398.1678D,2016MNRAS.455.2110B}. This is close to the theoretical expectation for spherical collapse (figure \ref{fig:ComprAmpl}, equation \ref{eq.Brho}) and it is in-line with observations from Faraday rotation measures. In the center of clusters, their simulations obtain a magnetic field strength of $3-6 \,\mu\mathrm{G}$, over a wide range of cluster masses. Subsequently the simulations were used to model giant radio haloes \citep{2000A&A...362..151D,2010MNRAS.401...47D}, the influence of the field on cluster mass estimates \citep{2000A&A...364..491D,2001A&A...369...36D}, the propagation of ultra high energy cosmic-rays \citep{2005JCAP...01..009D} and the distribution of fast radio bursts \citep{2015MNRAS.451.4277D}. \cite{donn09,2013MNRAS.435.3575B} presented models for cluster magnetic fields seeded by galaxy feedback, and established that different seeding models can lead to the same cluster magnetic field. \citet{2012MNRAS.422.2152B} showed  theoretical and numerical models for magnetic field seeding and amplification in galactic haloes. { We reproduce projected magnetic field strengths in cosmological simulations from three different methods, \gadget (SPH), \enzo (Eulerian finite volume) and \arepo (Lagrangian finite volume) in figure \ref{fig:cosmo1}.}\par
\citet{ry08} established the connection between shock driven vorticity during merger events and magnetic field amplification in clusters using Eulerian cosmological simulations. They applied a semi-analytic model of the small scale dynamo coupled to the turbulent energy to derive $\mu\mathrm{G}$ fields in clusters \citep[see also][]{bm16}. \par

\begin{figure*}
  \includegraphics[width=0.45\textwidth,height=0.4\textwidth]{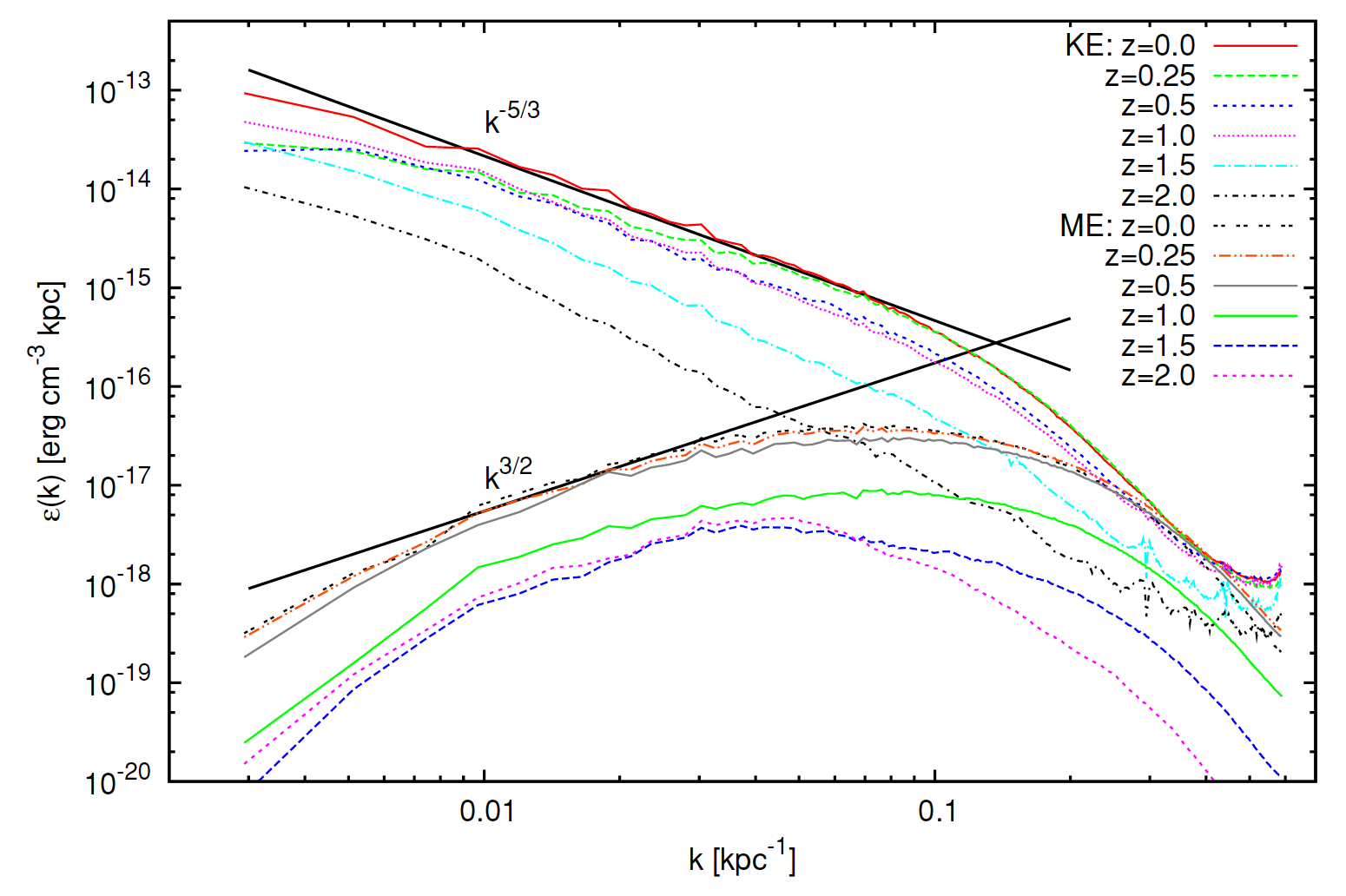}
  \includegraphics[width=0.45\textwidth,height=0.4\textwidth]{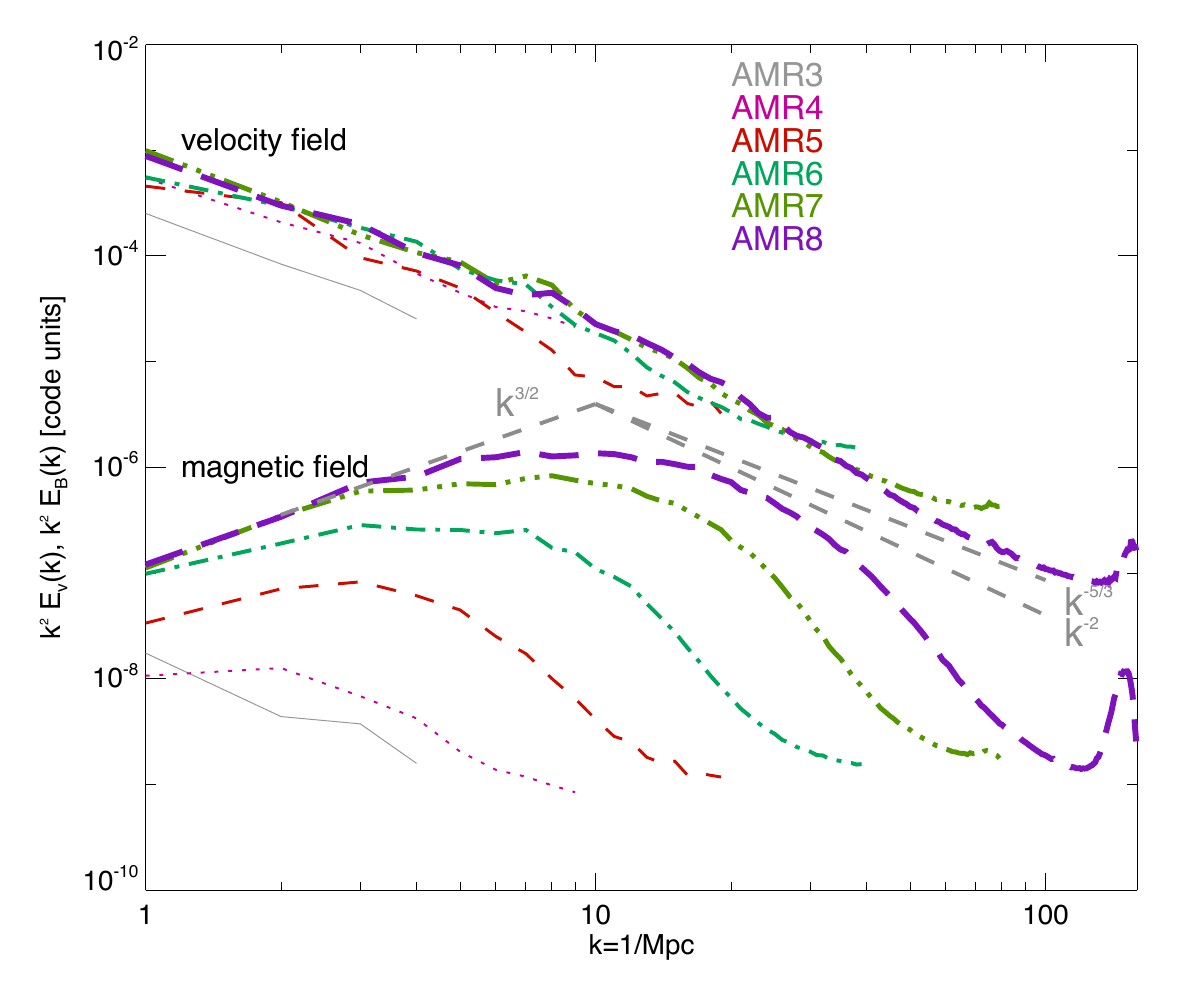}
\caption{3-dimensional kinetic and magnetic power spectra in \enzo  MHD simulations by \citet{2009ApJ...698L..14X} (assuming a seeding of magnetic fields by AGN) and by \citet{va18mhd}, assuming a primordial magnetic field of $0.1$ nG (comoving), as a function of resolution.}
\label{fig:dynamo_sim}
\end{figure*}

\citet{2009ApJ...698L..14X,2011ApJ...739...77X} used AGN seeding in the first direct Eulerian MHD cluster simulations to obtain magnetic field strengths of $1-2 \mu\mathrm{G}$ in clusters with a second order TVD method and constrained transport \citep{2008ApJS..174....1L}. { We reproduce power spectra from this simulation in figure \ref{fig:dynamo_sim}, left}. Considering cosmological seed fields, \citet{va14mhd} used large uniform grids to simulate magnetic field amplification in a massive cluster. \citet{ruszkowski11} presented a simulation of cluster magnetic fields with anisotropic thermal conduction. They find that conduction eliminates the radial bias in  turbulent velocity and magnetic fields that they observe without conduction. \par

\begin{figure*}
 \includegraphics[height=0.45\textwidth]{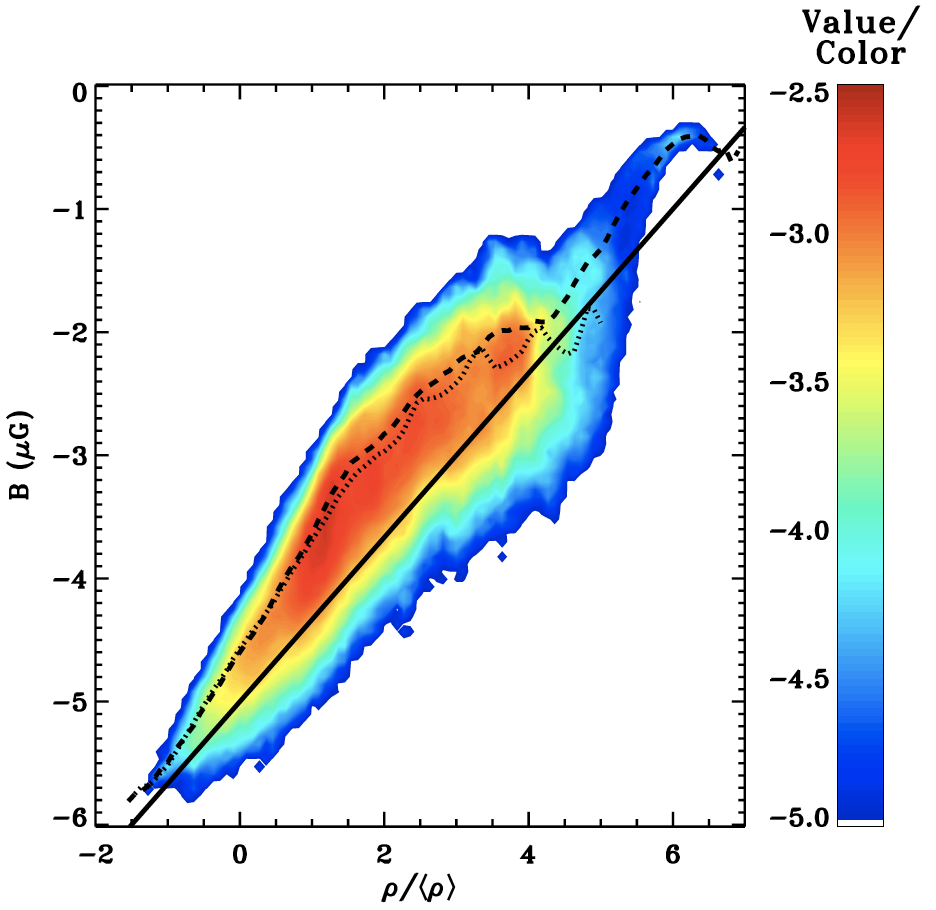}
 \includegraphics[height=0.45\textwidth]{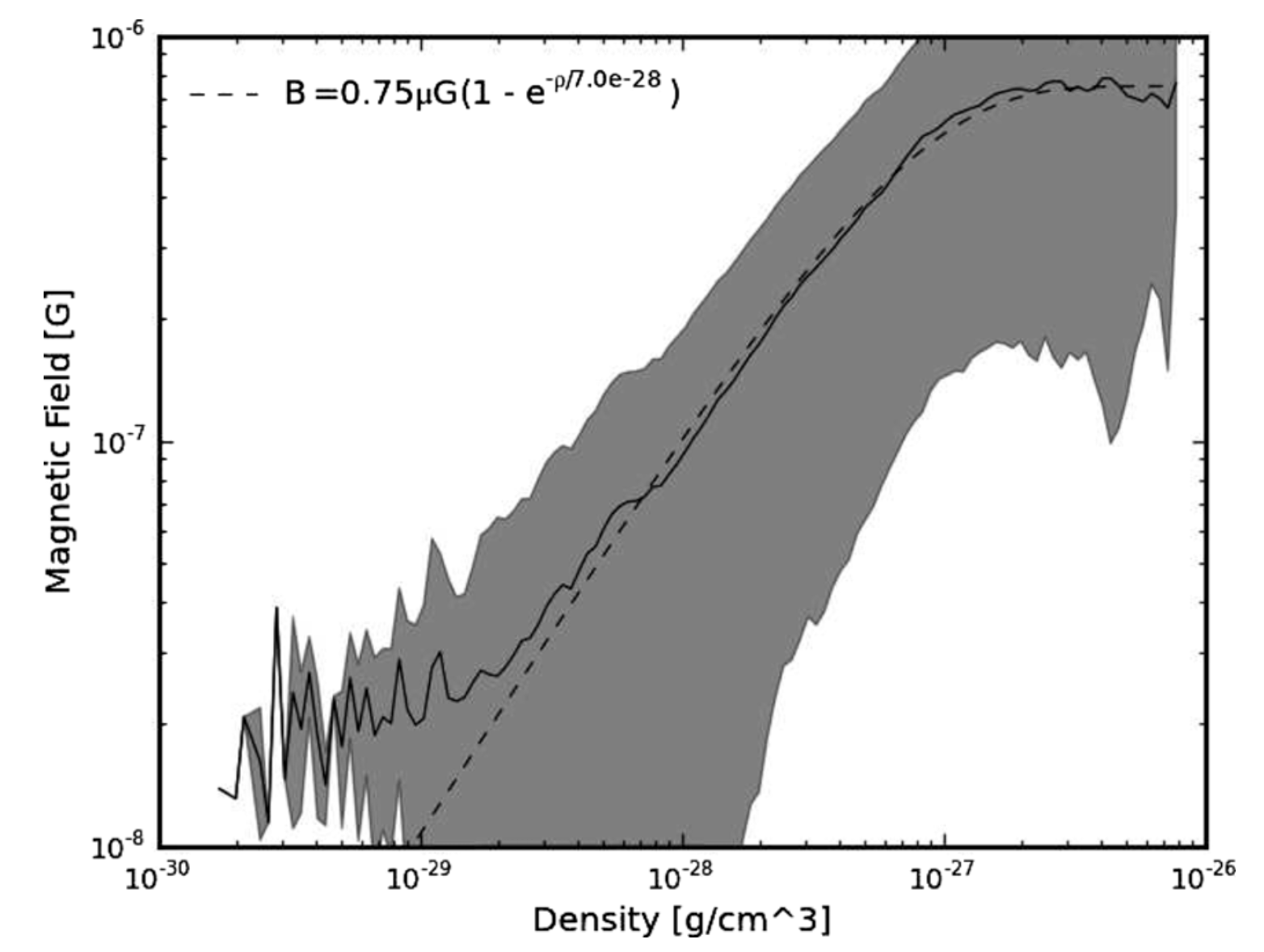}
 \includegraphics[height=0.45\textwidth]{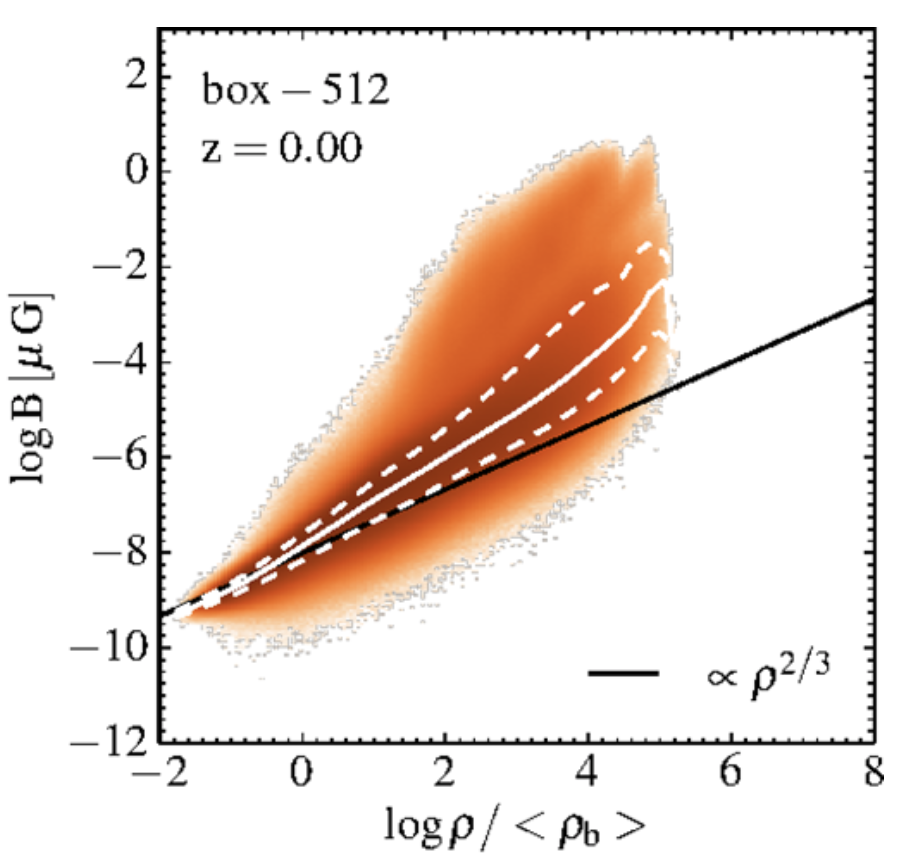}
 \includegraphics[height=0.45\textwidth]{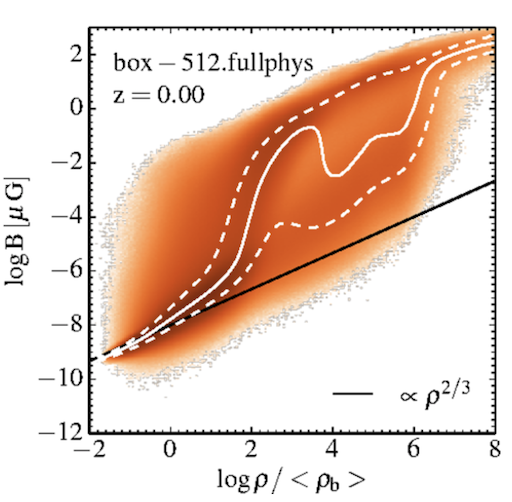}  
\caption{Phase diagrams for different cosmological simulations, like figure \ref{fig:ComprAmpl}. Top left: RAMSES CT simulation of a cooling-flow galaxy cluster \citep[][]{2008A&A...482L..13D,2009MNRAS.399L..49D}. Top right: \enzo CT simulation of a major merger cluster (fields injected by AGN activity) \citep[][]{sk13}. Bottom left, right: \arepo  (Powell scheme) simulation without and with Illustris galaxy formation model, respectively \citep{2015MNRAS.453.3999M}. }
\label{fig:cosmo2}
\end{figure*}

Within the limit of available numerical approaches, modern simulations find that adiabatic compression/rarefaction of magnetic field lines is the dominant mechanism across most of the cosmic volume (see figure \ref{fig:cosmo2}), with increasing departures at high density, $\rho \geq 10^2 \langle \rho \rangle$, when dynamo amplification sets in. Additional scatter in this relation is also found in presence of additional sources of magnetization or dynamo amplification, such as e.g. feedback from AGN, as shown by the comparison between non-radiative and ''full physics'' runs. Using a Lagrangian finite volume method, \citet{2015MNRAS.453.3999M, ma17,2018MNRAS.476.2476M} showed magnetic field seeding and evolution with the ``Illustris'' subgrid model for galaxy formation, also including explicit diffusivity. They obtained $\mu\mathrm{G}$ magnetic fields in clusters when they included seeding from galaxy feedback (figure \ref{fig:cosmo2}, bottom).\par
\begin{figure*}
  \includegraphics[width=0.95\textwidth]{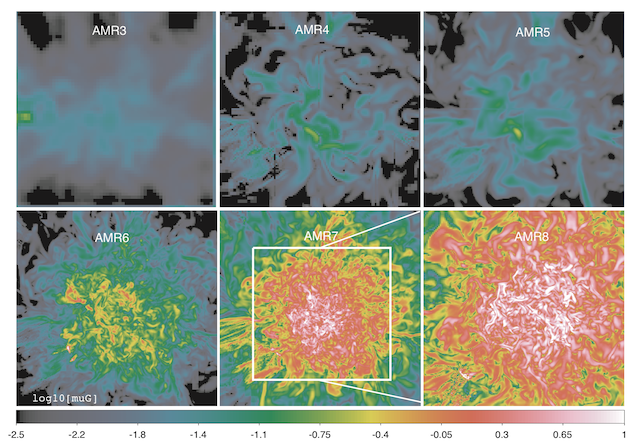}
\caption{Map of projected mean magnetic field strength for re-simulations of a cluster with increasing resolution, for regions of $8.1 \times 8.1 ~\rm Mpc^2$ around the cluster center at $z=0$.  Each panel shows the mass-weighted magnetic field strength (in units of $\log_{\rm 10} [\mu G]$ for a slice of $\approx 250$ kpc along the line of sight. Adapted from \citet[][]{va18mhd}.}
\label{fig:dynamo_francosim}       
\end{figure*}
Recently, \citet[][]{va18mhd} simulated the growth of magnetic field as low as $0.03 ~\rm nG$ up to $\sim 1-2~ \rm \mu G$ using AMR with a piece-wise linear finite volume method. By increasing the maximum spatial resolution in a simulated $\sim 10^{15}M_{\odot}$ cluster, they observed the onset of significant small-scale dynamo for resolutions $\leq 16 ~\rm kpc$, with near-equipartition magnetic fields on $\leq$ 100 kpc scales for the best resolved run ($\approx 4 ~\rm kpc$), see Figure \ref{fig:dynamo_francosim}. They estimated that $\sim 4\%$ turbulent kinetic energy was converted into magnetic energy. The amplified 3D fields show clear spectral, topological and dynamical signatures of the small-scale dynamo in action, with mock Faraday Rotation roughly in-line with observations of the Coma cluster \citep[][]{bo13}. A significant non-Gaussian distribution of field components is consistently found in the final cluster, resulting from the superposition of different amplification patches mixing in the ICM.  \par

\subsection{Cosmological Simulations of Magnetic Fields Outside of Galaxy Clusters}

\begin{figure}
\centering
\includegraphics[width=0.49\textwidth]{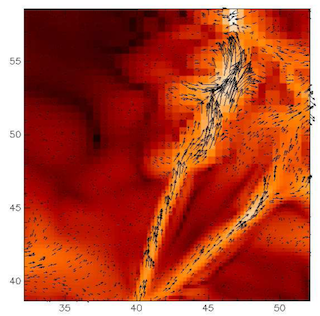}
\includegraphics[width=0.49\textwidth]{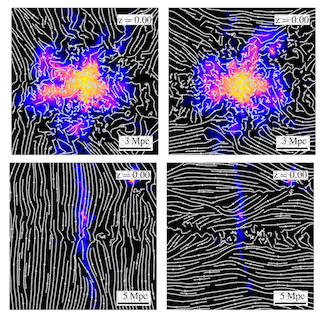}
\caption{Left: magnetic field vectors for a cosmic filaments simulated with AMR using FLASH \citep[][]{br05}. Right: magnetic field vectors around a massive galaxy cluster (top) and a filament (bottom) simulated with AREPO \citep[][]{2015MNRAS.453.3999M} and for two different topologies of uniform seed magnetic fields.}
\label{fig:fila}
\end{figure}

The peripheral regions of simulated galaxy clusters mark the abrupt  transition from supersonic to subsonic accretion flows, and the onset of the virialization process of the 
in-falling gas.  The accreted gas moves supersonically with respect to the warm-hot intergalactic medium in the cluster periphery, which  triggers 
$\mathcal{M} \sim 10-100$ strong shocks in the outer regions
of clusters and in the filaments attached to them  \citep[e.g.][]{ry03,pf06}. Downstream of such strong shocks, supersonic turbulence is injected towards structures, together with a first inject of vorticity by oblique shocks  \citep[e.g.][]{ka07,ry08,wi17}.
In these physical conditions, the SSD is predicted to be less efficient, because of the predominance of compressive forcing of turbulent motions \citep[][]{ry08,2011PhRvL.107k4504F,jones11,2013NJPh...15b3017S,po15}. In this case, the maximum magnetic field arising from SSD amplification in the $10^5 ~\rm K\leq T \leq 10^{7} ~\rm K$ medium of filaments would be $\sim 0.01-0.1 ~\rm \mu G$ \citep[e.g.][]{ry08,va14mhd}.  Direct numerical simulations investigated the small-scale dynamo amplification of primordial fields in cosmic filaments, so far reporting no evidence for dynamo amplification, unlike for galaxy clusters simulated with the same method and at a similar level of spatial detail \citep[][]{va14mhd}. This trend is explained  by the observed predominance of compressive turbulence at all resolutions (unlike in clusters, where turbulence gets increasingly solenoidal as resolution is increased), as well  as by the limited amount of turnover times that infalling gas experiences before being accreted onto clusters \citep[e.g.][]{ry08,va14mhd}. If these results will be confirmed by simulations with even larger resolutions, it has the important implication that the present-day magnetization of filaments should be anchored to the 
seeding events of cosmic magnetic fields, posing a strong case for future radio observations \citep[e.g.][]{2016MNRAS.462..448G,va17cqg}.
In this scenario  the outer regions of galaxy clusters and filaments are expected to retain information also on the {\it topology} of initial seed fields even today, as shown in numerical simulations at high resolution \citep[e.g.][see also figure \ref{fig:fila}]{br05,2015MNRAS.453.3999M}, in case the magnetic fields have a primordial origin.
Conversely, if the fields we observe in galaxy clusters are mostly the result of seeding from active galactic nuclei and galactic activities, the magnetization at the scale of filaments and cluster outskirts is predicted to be low \citep[e.g.][]{donn09,2009ApJ...698L..14X,2015MNRAS.453.3999M}. Future surveys in polarization should have the sensitivity to investigate the outer regions of galaxy clusters down to $\sim 1-10 ~\rm rad/m^2$ \citep[e.g.][]{2015aska.confE.113T,2015aska.confE..95B,2016A&A...591A..13V}, which is enough to discriminate among most extreme alternatives in cluster outskirts \citep[e.g.][]{va17cqg}.

\subsection{Discussion} \label{subsec:numerics}

Simulations of magnetic field amplification in clusters have reproduced key observations for two decades now. Most of the early progress has been achieved with Lagrangian methods originally developed in the galaxy formation context, most notably SPH \citep{2002A&A...387..383D}. These simulations reproduce the magnetic field strength inferred from rotation measures in clusters and have been used extensively to model related astrophysical questions. However, the adaptivity of Lagrangian methods and the particle noise in SPH limits their ability to resolve the structure of the magnetic field, especially in low density environments (cluster outskirts, filaments).  \par
Clear theoretical expectations for the small scale dynamo in clusters have been established \citep{ry08,bm16}, also from idealized simulations \citep[e.g.][]{2004ApJ...612..276S,2009ApJ...693.1449C,po15}. Some of these expectations have been tested in cosmological simulations using Eulerian codes \citep[e.g.][]{va18mhd}. Recent Eulerian simulations approach observed field strengths in clusters, but do not reach field strengths obtained from Lagrangian approaches. \citet{bm16} argued that due to numerical diffusion, Eulerian approaches spend too much time in the exponential/kinetic growth phase, thus the non-linear growth phase is severely truncated. { Following \citet{2004ApJ...612..276S}, a clear indicator for the presence of a dynamo that cannot be produced via compression is the anti-correlation of magnetic field strength and its curvature $\vec{K}$:
\begin{eqnarray}
    \vec{K} &= \frac{ \left( \vec{B} \cdot \nabla \right) \vec{B}}{\vec{B}^2},
\end{eqnarray}
so that $\vec{B}\vec{K}^\frac{1}{2} = const$, where the exponent has to be obtained from the magnetic field distribution. In cluster simulations, only \citet{2018MNRAS.474.1672V} have demonstrated consistent curvature correlations. We note that in galactic dynamos, consistent results  have recently  been achieved with Eulerian and Lagrangian codes \citep{2017ApJ...843..113B,2016MNRAS.457.1722R,2017MNRAS.469.3185P,2018arXiv180809975S}, but only \citet{2018arXiv180809975S} showed a curvature relation. }\par
{ In clusters, }all simulations show an exponential increase in magnetic field strength followed by a non-linear growth phase \citep[e.g.][]{2012MNRAS.422.2152B}. However, the timescale of exponential growth is set by the velocity power/rate of strain at the resolution scale, which in turn is determined by the MHD algorithm (resolution, dissipation/noise). The real kinematic dynamo in primordial haloes is far below the resolution scale of every numerical scheme \citep{bm16} and needs to be treated with an large eddy approach \citet{2005JSP...121..823Y,2009ApJ...693.1449C}. \par
As we have motivated above, dynamo theory predicts that the final structure of cluster magnetic fields is shaped by turbulence near the Alfv\'{e}n scale, because this is where the eddy turnover time is smallest (equation \ref{eq.lA}, a few kpc in a massive cluster merger). Thus an accurate simulation of field topology has to faithfully follow the velocity field and the magnetic field near this scale in the non-linear growth phase, i.e. at least achieve Reynolds numbers (eq. \ref{eq.Re}) of 300-500 at redshifts $z<1$ during a major merger  \citep{2004MNRAS.353..947H,bm16}. For an outer scale of 300 kpc, this implies evolution of turbulence velocity and magnetic field growth at about 1 kpc, \emph{including numeric effects}. \par
 This makes the small scale dynamo in clusters is a very hard problem, because it combines the large dynamical range of scales in cosmological clustering with the evolution of two coupled vector fields (turbulence and magnetic fields) near the resolution scale. Additionally, seeding on smaller scales by galactic outflows may play an important role. Hence, it is likely the numerical dissipation scale that shapes the outcome of MHD simulations in a cosmological context. We now provide a short discussion of effective Reynolds numbers and numerical limitations in current approaches.

\subsubsection{Effective Reynolds Numbers}

From the numerical viewpoint, the Reynolds number of a flow increases with the effective dynamic range reached inside a given volume. Its upper limit is set by the driving scale and the spatial resolution in the volume of interest following equation \ref{eq.Re}. However, in any numerical scheme the effective dynamic range and Reynolds number of the flow are reduced by the cut-off of velocity and magnetic field power near the numerical dissipation scale in Fourier space \citep[e.g.][]{2003PhRvE..68b6304D}. Simply put, numerical error takes away velocity and magnetic field power close the resolution scale in most schemes. The shape of the velocity power spectrum on small scales determines how much velocity power (rate of strain $\delta u/l$, see section \ref{sect.ssd}) is available to fold the magnetic field and drive the small-scale dynamo. Thus a less diffusive (finite volume) code reaches higher effective Reynolds numbers, faster amplification and a more tangled field structure at the same resolution. \par
\begin{figure}
\centering
\includegraphics[width=0.45\textwidth]{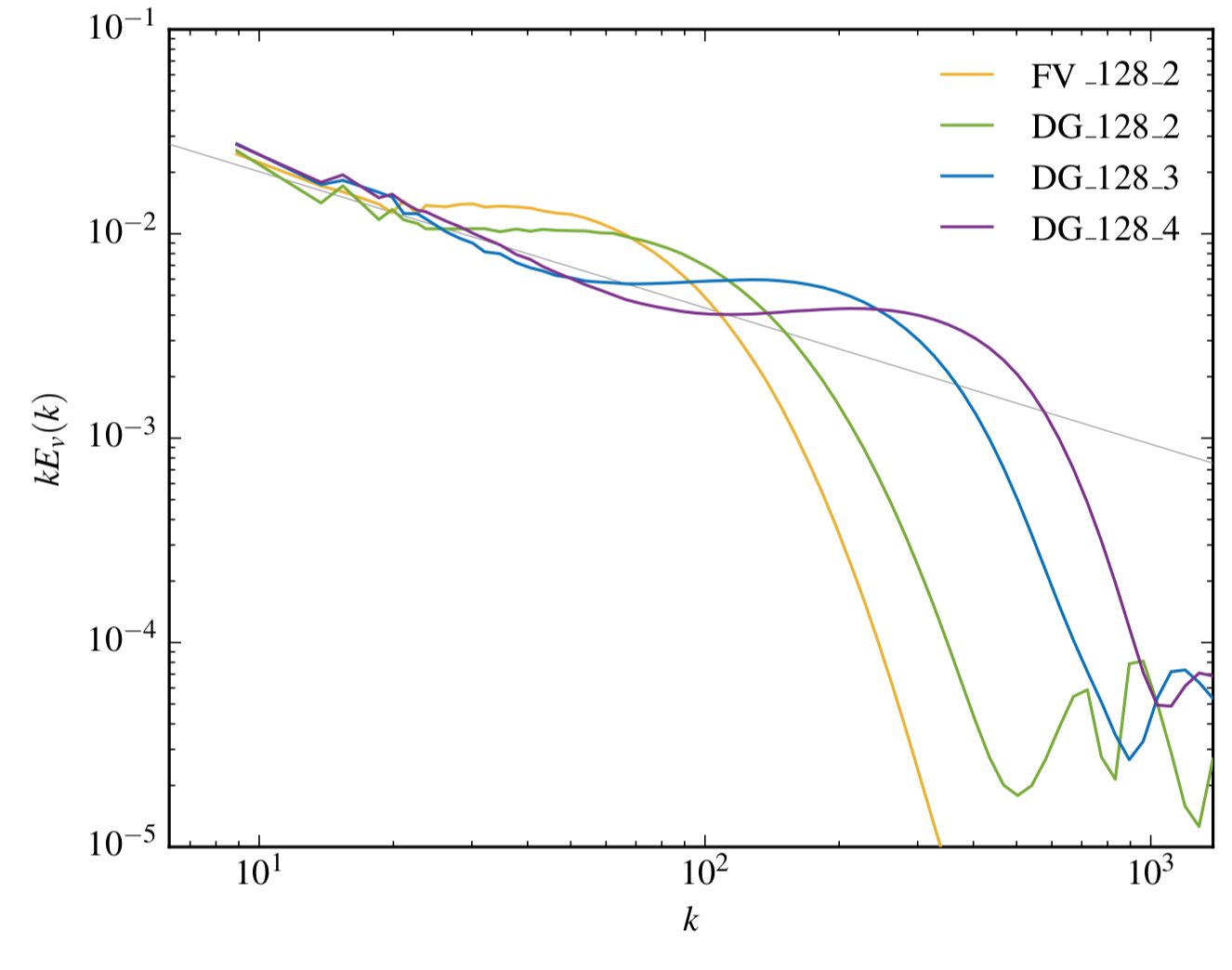}
\includegraphics[width=0.45\textwidth]{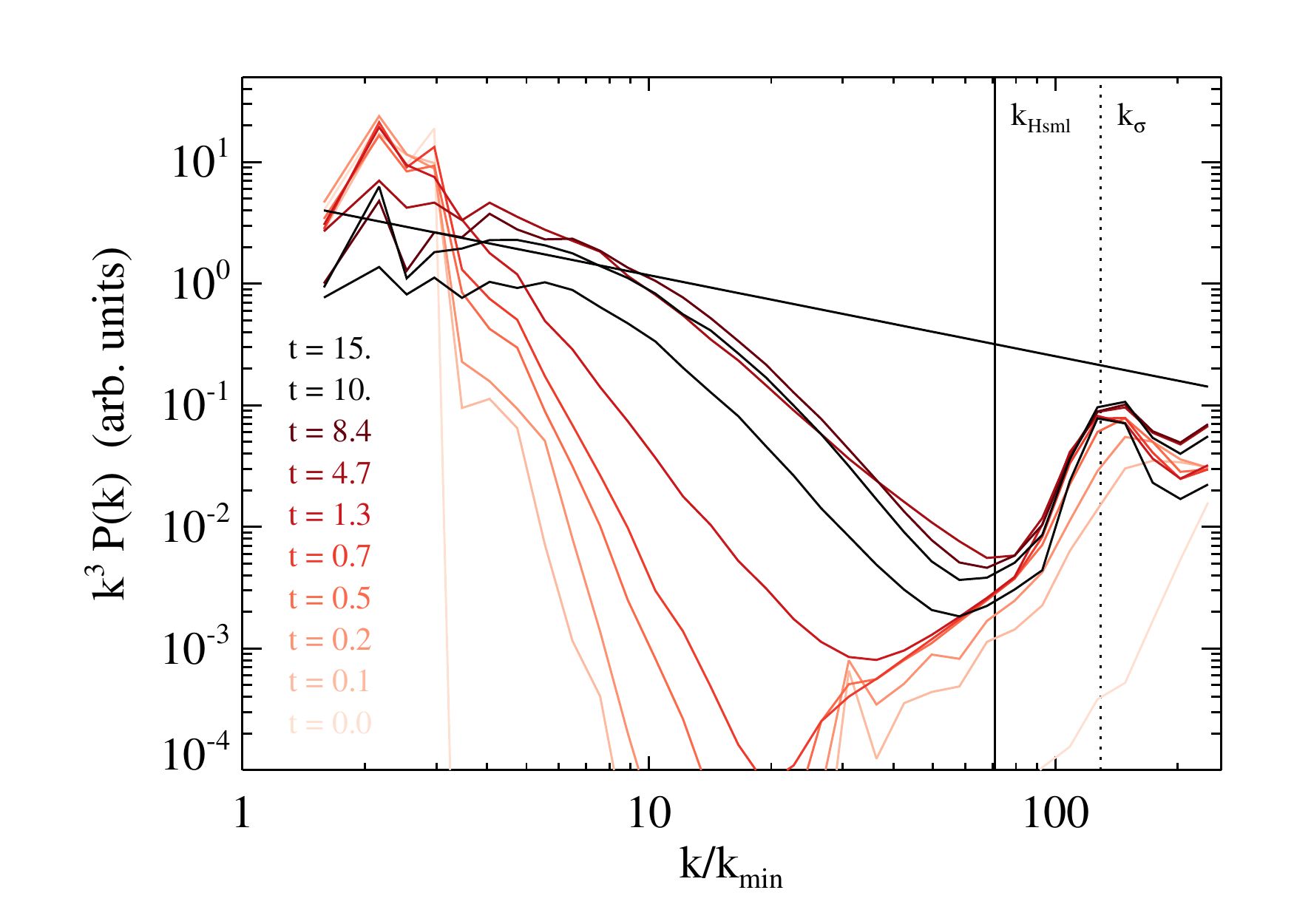}
\caption{Left: Velocity power spectra of driven compressible turbulence with a second order finite volume scheme (yellow) and Discontinuous Galerkin schemes (2nd order: green, 3rd order blue, 4th order purple) \citep{2016arXiv160209079B}. Right: Velocity power spectra of decaying turbulence simulated with modern SPH \citep{2016MNRAS.455.2110B}. The spectra were obtained using wavelet kernel binning to remove aliasing above the kernel scale $k_\mathrm{hsml}$.}
\label{fig:schaal}
\end{figure}

We can quantify this behavior by introducing an \emph{effective Reynolds number} of an MHD simulation of turbulence as:
\begin{eqnarray}
    R_{\rm e,min} & \approx \left( \frac{L}{\epsilon \Delta x} \right)^{4/3},
\label{eq:rme}
\end{eqnarray}
where $\Delta x$ is the resolution element, $\epsilon$ is a factor depending on the diffusivity of the numerical method, and $L$ is the outer scale (in clusters 300-500 kpc, section \ref{sect.physturb}). As a conservative estimate, one may assume in modern SPH codes $\epsilon \ge 10$ (\citet{2012MNRAS.420L..33P}, figure \ref{fig:schaal}), in hybrid codes $\epsilon \approx 10$ \cite{2015MNRAS.450...53H}. For second order finite difference/volume codes one often assumes $\epsilon \approx 7$ (e.g. \citet{2011ApJ...737...13K,2016MNRAS.457.1722R}).
In figure \ref{fig:schaal} left, we reproduce velocity power spectra from a driven compressible turbulence in a box simulation with $128^3$ zones using the finite volume (FV) code \arepo and the discontinuous Galerkin (DG) code \texttt{TENET }  \citep{2016arXiv160209079B}. Second order FV is shown in yellow, while second, third and fourth order DG power spectra are shown in green, blue and purple, respectively. The formal resolution / Nyquist scale remains constant in all runs. However, with increasing order of spatial and time interpolation, viscosity reduces, the effective dissipation scale shrinks, velocity power on small scales increases, the inertial range grows in size, and with it the effective Reynolds number of the simulation (i.e. $\epsilon$ decreases). Note that the DG scheme has more power near the dissipation scale than the FV scheme, even at the same order (green vs. yellow). This indicates that formal convergence order is not sufficient to determine effective Reynolds numbers at a given resolution. $\epsilon$ obviously depends on implementation details and has to be determined empirically with driven turbulence ``in a box'' simulations. For a recent review on high-order finite-volume schemes, see \citet{2017LRCA....3....2B}. \par

\subsubsection{Dynamos in Eulerian Schemes} \label{sect.eul}

In non-adaptive Eulerian cluster simulations the effective Reynolds number is set by the resolution of the grid and the diffusivity of the numerical method \citep[e.g.][]{2011ApJ...737...13K}. \citet{2011ApJ...731...62F} and \citet{2013MNRAS.432..668L} reported that only by resolving the Jeans length of a halo with $\geq 64$ cells the small-scale dynamo can develop (e.g. $R_M \sim 32$ setting $\epsilon=2$ in Eq.\ref{eq:rme}) in a proto-galactic halo of $10^{6}M_{\odot}$ at $z \sim 10$. However, \citet[][]{va14mhd} reported that small-scale amplification can begin before $z=0$ in $\sim 10^{14} M_{\odot}$ galaxy clusters if their virial diameter is resolved with at least $\geq 100$ cells ($R_M \sim 50$), while in order to approach energy equipartition between turbulence and magnetic fields by $z=0$ one needs to resolve the virial diameter with $\geq 1500$ elements ($R_M \sim 750$ in the ideal case). These differences likely arise from the shapes of the numerical dissipative and resistive scales. The underlying Eulerian methods were either second or first order accurate and used CT or Dedner cleaning to constrain magnetic field divergence.\par

 Eulerian structure formation simulations produce flows with supersonic velocities relative to the simulation grid. At the same time, the truncation error of Eulerian methods is inherently velocity dependent \citep{2010MNRAS.401.2463R,2016arXiv160209079B}. It has been shown that these errors do not pose a problem for the simulation of clusters in a cosmological context \citep{2009MNRAS.395..180M}, but they may suppress the growth of instabilities close to the dissipation scale \citep[e.g.][]{2010MNRAS.401..791S} and thus further reduce the effective Reynolds number of the simulation. We note that poorly un-split Eulerian schemes may also affect angular momentum conservation close to the resolution scale and further reduce the accuracy of e.g. galaxy formation simulations, where angular momentum conservation is desirable to produce disc galaxies. \par
\begin{figure}
\centering
\includegraphics[width=0.45\textwidth]{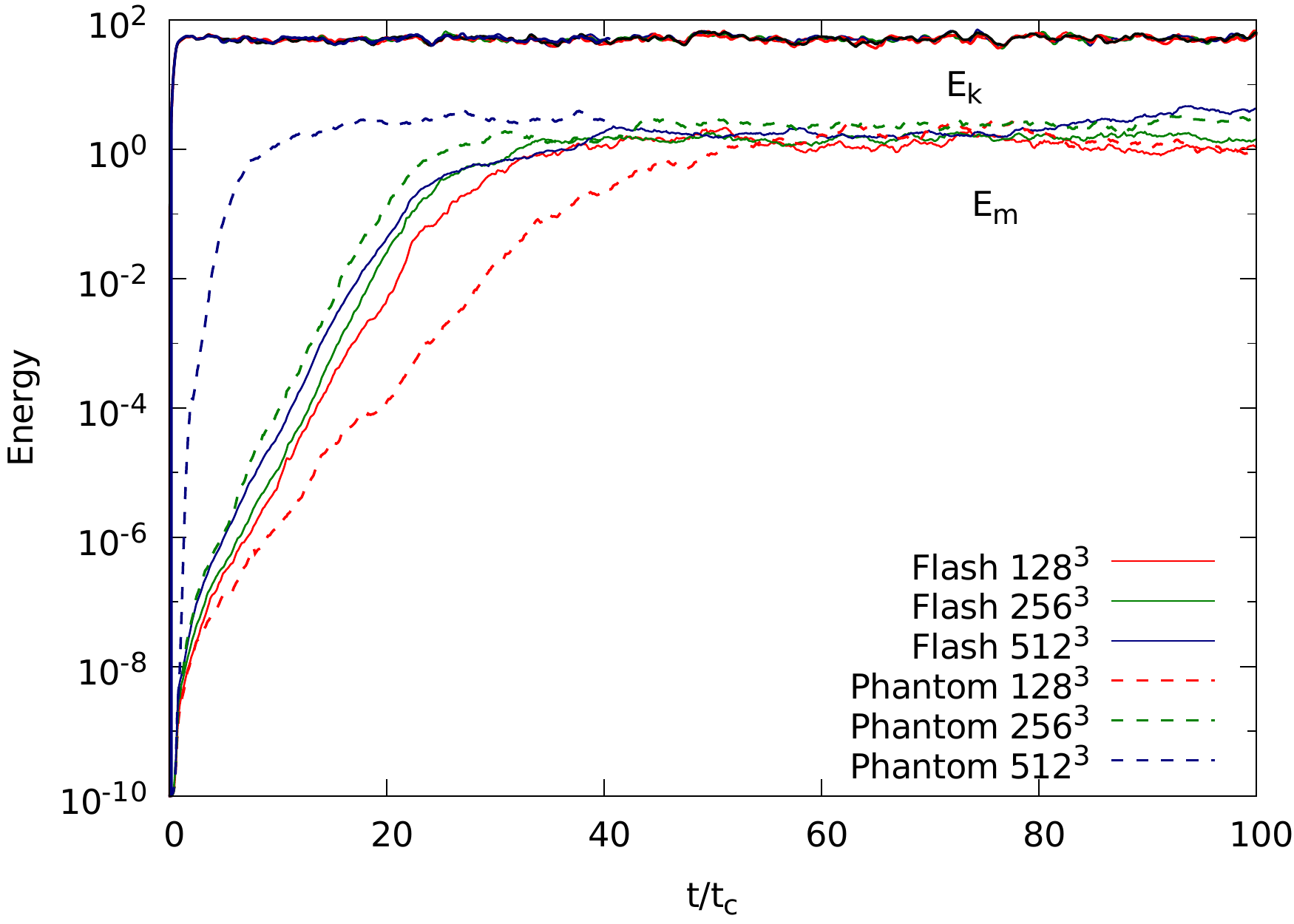}
\includegraphics[width=0.45\textwidth]{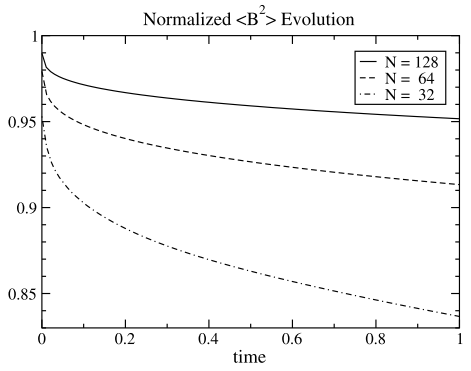}
\caption{Left: Growth of magnetic energy in supersonic driven turbulence simulations in SPMHD (dashed lines) and FV (solid lines) at different resolutions \citep{2016MNRAS.461.1260T}. Right: Evolution of magnetic energy in the advection of a magnetic field loop in 2 dimensions from \citet{2008JCoPh.227.4123G}.}
\label{fig:bgrowth}
\end{figure}

 These arguments extend also to magnetic fields, whose advection poses a challenging test for all Eulerian schemes. In figure \ref{fig:bgrowth}, right, we reproduce the time evolution of magnetic energy during the advection of a magnetic field loop in 2D with the {\small ATHENA} code at different resolutions \citep{2008JCoPh.227.4123G}. As the size of the field loop approaches the resolution scale, field energy is diffused more quickly.  Again, the diffusivity added by the scheme to keep local magnetic field divergence small varies with implementation and has to be determined by empirical tests. There are sizable differences even among CT schemes, which inherently conserve the divergence constraint to machine precision  \citep[see e.g.][]{2013JCoPh.243..269L}.
 
\subsubsection{Dynamos in Lagrangian Schemes} \label{sect.sph}

In adaptive Lagrangian cluster simulations, the resolution is a function of density and thus varies in space and time during the formation of a cluster or filament. Thus the dissipation scale and the Reynolds number are not well defined in Fourier space and turbulence can be strictly defined only on the coarsest resolution element in a given volume.  The effect of the adaptivity on the dynamo and especially the resulting field structure is not entirely clear. It seems reasonable to assume additional (magnetic) dissipation, if a magnetized gas parcel moves to a less-dense environment and is adiabatically expanded and divergence cleaned. While turbulent driving is correlated with over-densities in a cosmological context, the turbulent cascade is not. Thus density adaptivity, which is a very powerful approach in galaxy formation simulations, might { introduce a density bias to the magnetic field distribution in} strongly stratified media. { The growth rate of the turbulent dynamo depends on the eddy turnover time, which is smallest in highly resolved regions. Thus Lagrangian schemes might grow magnetic fields faster in high density regions (cluster cores) than in low density regions (cluster outskirts). However, it remains unclear how strongly current results are affected by this issue, simply because no Eulerian simulation with kpc resolution in the cluster outskirts is available.}\par

In cosmological simulations, \citep{1999A&A...348..351D,2002A&A...387..383D} reported sizeable cluster magnetic fields even with a traditional SPH algorithm and comparably low resolution. As we have shown, theory provides clear predictions for the evolution of a magnetic field in a turbulent dynamo, which have been successfully verified with Eulerian methods. For some Lagrangian methods \citep[e.g.][]{2011MNRAS.418.1392P}, it is reasonable to assume that at fixed resolution the result will be similar to the established dynamo theory, simply because their dissipation scale defaults to a finite volume method. For other new hybrid methods \citep{hop16} the situation is less clear. In general, the idealized magnetic dynamo in Lagrangian schemes is not well researched yet and we would encourage the community to close this gap.  \par
For traditional SPH algorithms, its ability to accurately model hydrodynamic turbulence was heavily debated \citep{2012MNRAS.423.2558B,2012MNRAS.420L..33P}. We note that computing a grid representation from an irregularly sampled vector field to obtain a power spectrum is a diffusive process and prone to aliasing \citep{2016MNRAS.455.2110B}. Modern SPH schemes have improved significantly, and it has been shown that sub-kernel re-meshing motions are required to maintain sampling accuracy \citep{2012JCoPh.231..759P}. The influence of these motions on the magnetic dynamo are not well understood, especially in the subsonic regime that is dominant in clusters. \par
In the supersonic regime, \citet{2016MNRAS.461.1260T} compared simulations with $M=10$ using the SPMHD code \texttt{PHANTOM} and the finite volume code \texttt{FLASH} with an HLL3R solver \citep{2011JCoPh.230.3331W} and both with Dedner cleaning. They found that the growth of magnetic energy in the SPMHD dynamo speeds up with increasing resolution. In contrast, the finite volume scheme converged (figure \ref{fig:bgrowth}, left). They found Prandtl numbers of $P_r=2$ and $P_r < 1$, respectively.  They argued that the growth in the SPMHD dynamo is due to the artificial viscosity and resistivity employed, which is negligibly small in the absence of shocks. \par
We note that these results cannot be simply transferred to galaxy cluster simulations. As mentioned before, cluster turbulence is largely sub-sonic, super-Alfv\'{e}nic and solenoidal, thus shocks do not play a role for the dissipation of turbulent energy. Cosmological codes usually do not include explicit dissipation terms, in contrast diffusivity is usually minimized. Driven subsonic turbulence simulations with SPMHD are  required to characterize the sub-sonic SPMHD dynamo in clusters and clarify the role particle noise could play even in early SPMHD cluster simulations. We note that some numerical amplification has been reported in SPMHD simulations of the galactic dynamo \citep{2015JCoPh.282..148S,2016MNRAS.461.4482D}.

\section{Magnetic Field Amplification at Shocks}
\label{sect.shocks}

Shocks amplify magnetic fields by a number of mechanisms, not all of which are well understood \citep{2012SSRv..166..187B}. Compression at the shock interface leads to  the amplification of the quasi-perpendicular part of the upstream magnetic field. Compressional amplification has the allure of explaining the large degrees of polarization in radio relics, but suffers from the limitation of small amplification factors. For amplification by pure compression, \citep{2012MNRAS.423.2781I} find for the ratio of magnetic fields : 
\begin{eqnarray}
	\frac{B_\mathrm{dw}}{B_\mathrm{uw}} &= \sqrt{\frac{2\sigma^2+1}{3}}, \label{eq.BampComp}
\end{eqnarray}
with the shock compression ratio $\sigma$. Thus, for typical shock strengths in cluster mergers, ($M \approx 2 - 3$), the amplification factor is limited to around 2.5, which results in inconsistencies of the minimum magnetic field strengths inferred in some radio relics with global magnetic field scalings \citep{2017MNRAS.471.4587D}. Similar expressions have been found for SNR \citep{1998ApJ...493..375R}.

\subsection{Shock-driven dynamo}

\begin{figure*}
  \includegraphics[width=0.49\textwidth]{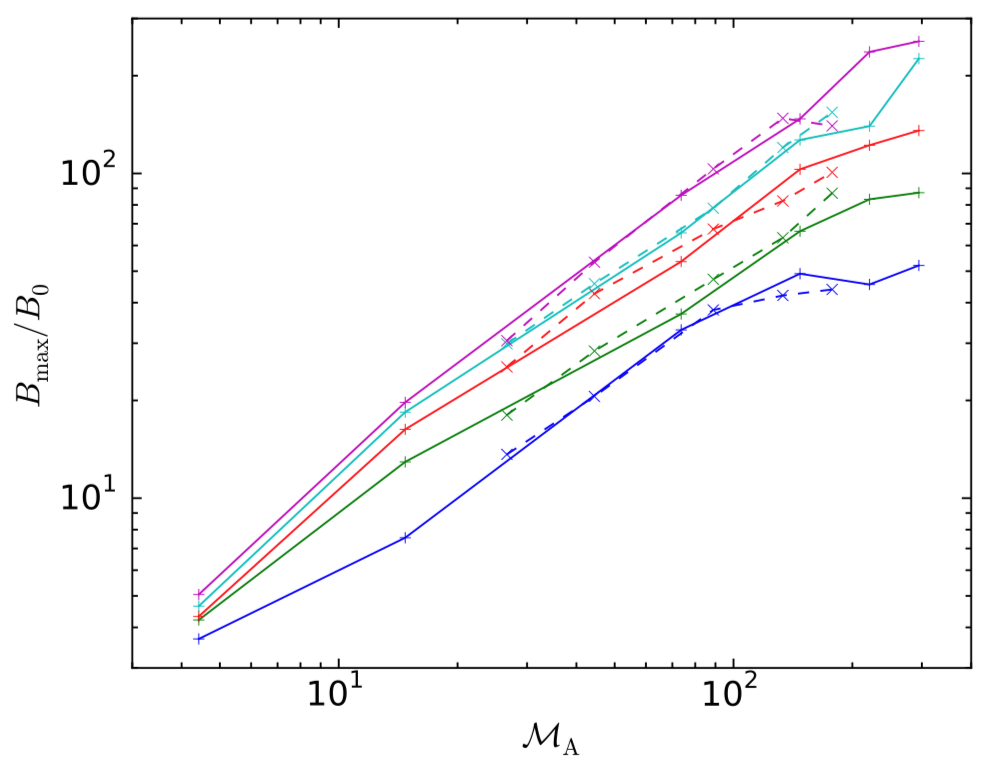}
  \includegraphics[width=0.49\textwidth]{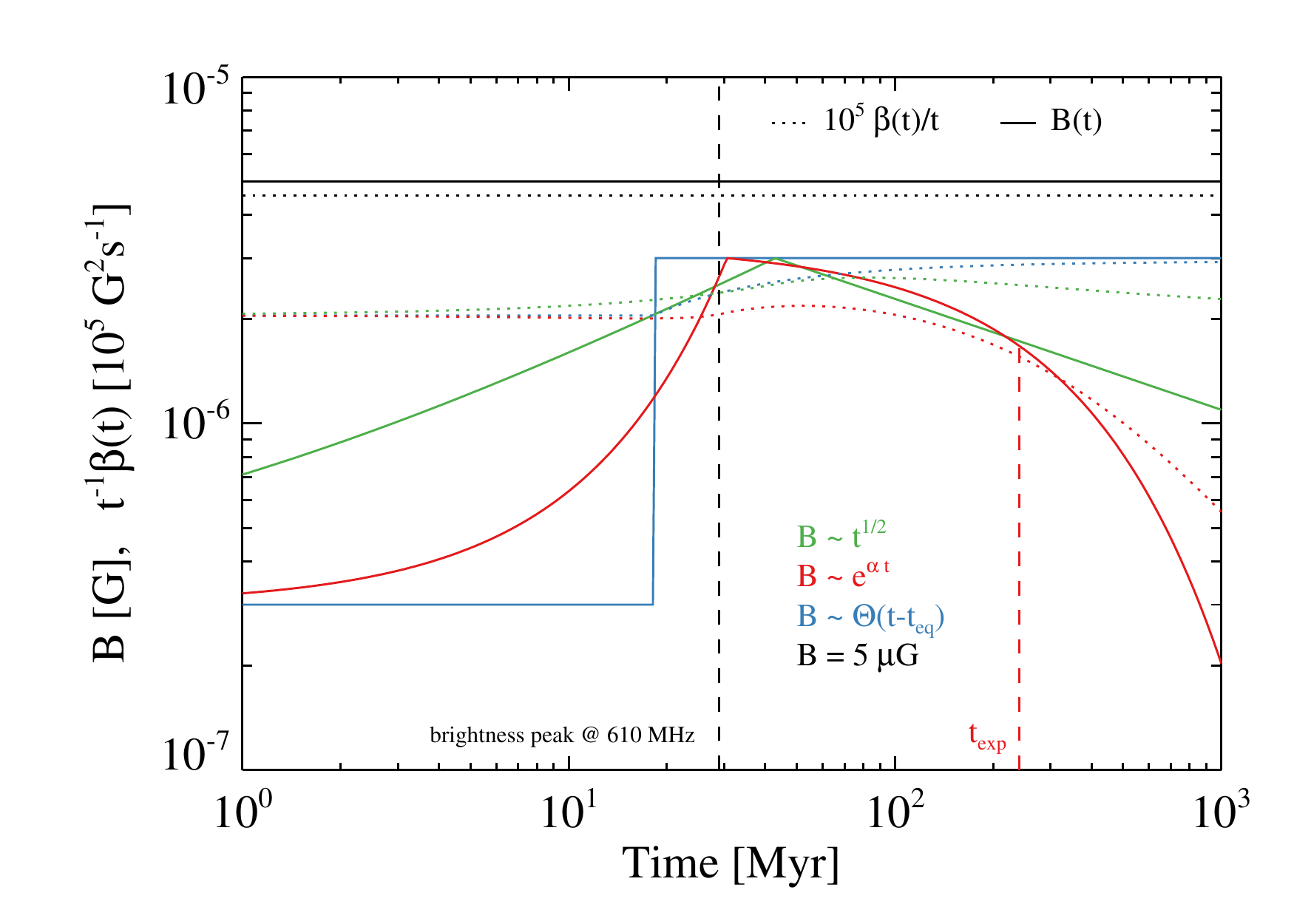}
\caption{Left: Magnetic field amplification over Alfv\'{e}nic Mach number in 2D MHD shock simulations from \citet{2016MNRAS.463.3989J}. Right: Models for magnetic field evolution inferred in the Sausage relic from \citet{2016MNRAS.462.2014D}}
\label{fig:shockampl}
\end{figure*}

Downstream of shocks, magnetic fields can be amplified by a small-scale dynamo that is driven by turbulence created at the shock front \citep{1974MNRAS.168...73B}. { This has been observed in supernova remnants (SNR) \citep{2006A&A...453..387P}.} This turbulence could be driven by the baroclinic vorticity that is generated for example by upstream inhomogeneities in gas density. For parameters relevant in SNR, \cite{2007ApJ...663L..41G} have demonstrated in MHD simulations that density inhomogeneities in the pre-shock fluid cause turbulence and magnetic field amplification in the post-shock fluid. Simulations by \citet{2009ApJ...695..825I} showed that the maximum amplification is set by the plasma beta parameter. \citet{2012ApJ...758..126S} argued that turbulence is injected by Richtmyer-Meshkow instabilities. \citet{2013ApJ...770...84F} derived an analytical approach for 2D SNR shocks. \citet{2012ApJ...747...98G} studied the interaction of a SNR shock propagating into a turbulent medium upstream. However, the relevant parameters in the shock and the upstream medium in SNR blast waves differ significantly from galaxy cluster shocks. In clusters, Mach numbers are lower ($<5$) and the plasma beta parameter is larger ($\beta_\mathrm{pl} \ge 100$). It is unclear if there results from SNR carry over to the ICM. \par
Literature on turbulent magnetic field amplification in ICM shocks remains scarce. \citet{2012MNRAS.423.2781I} studied the evolution of vorticity behind the shock. They argue that self-generated vorticity from the shock is not sufficient to drive a turbulent dynamo downstream, but that about $30\%$ of turbulent pressure is required upstream of the shock to explain observed magnetic field lower limits. \par 
\cite{2016MNRAS.463.3989J} studied magnetic field amplification in idealized MHD simulations of shocks. They found that amplification is independent of plasma beta for Mach numbers of a few, but is linearly dependent on the Alfv\'{e}nic Mach number in shocks. In figure \ref{fig:shockampl}, left, we show their results for 2D simulations at different resolutions, with the highest resolution in magenta. Below $M_A \approx 10$, compression dominates the amplification and results in magnetic field structures perpendicular to the shock normal. Above $M_A \approx 10$, turbulence injected by the shock amplifies magnetic fields to strengths significantly higher than expected by compression. In this limit, the field topology becomes mostly quasi-parallel, because velocity shear is largest in the direction of shock propagation. \par
Along these lines, in \cite{wi17b}, it has been found that the stretching motions dominate the evolution of turbulence in galaxy clusters. However, baroclinic motions are needed to generate turbulence. The enstrophy dissipation rate peaks when the enstrophy is maximal and this is the time when magnetic field amplification by a small-scale dynamo would be the strongest.

These results have important implications for radio relics. In most relics the lower limit for the downstream magnetic field is found to be around $1-3 \mu\mathrm{G}$ \citep[e.g.]{2010ApJ...715.1143F}. This is consistent with equipartition magnetic field strengths of $4-7\mu\mathrm{G}$ \citep[e.g.][]{2009PASJ...61..339N,2010ApJ...715.1143F,2014MNRAS.445.1213S}.\par
The ordered topology of magnetic fields expected by compressional amplification can explain the large degree of polarization found in some radio relics, thus dis-favouring turbulent amplification. However, given typical Mach numbers (2-3), the lower limits on magnetic field strengths in relics imply upstream fields of about  $1-2 \mu\mathrm{G}$ ahead of some shocks (eq. \ref{eq.BampComp}). As relics reside in the outskirts of clusters, this is difficult to explain with the common scaling of magnetic field strength with density/radius in the ICM \citep{bo10}. 

However, a recent model \citep{2016MNRAS.462.2014D} of the Sausage relic motivates Alfv\'{e}nic Mach numbers of around 100 in the shock and showed that exponential downstream field amplification (figure \ref{fig:shockampl}, right) can explain the steepening in the radio spectrum above 8 GHz found in the Sausage \citep{2013A&A...555A.110S}. More discussion will also be found in Van Weeren et al. (this volume).

\section{Magnetic field amplification from cold front motions} \label{sect.slosh}

Aside from turbulence and shocks, many galaxy clusters also possess subsonic bulk flows which can amplify magnetic fields in localized regions. The first evidence of these motions was provided shortly after the launch of the \textit{Chandra} X-ray Observatory. \textit{Chandra}'s sub-arcsecond spatial resolution revealed the presence of surface brightness edges in many clusters. Through spectroscopic analysis most of these edges, which superficially appear as shocks, were identified to be contact discontinuities, where the denser (brighter) side of the edge is colder than the lighter (dimmer) side. These features have been dubbed ``cold fronts'', and are believed to be the result of subsonic gas motions driven by cluster mergers and cosmic accretion \citep[for recent reviews see][]{MV07,zuh16}. Cold fronts have been described as forming via at least three processes: ``remnant-core'' fronts are formed by cool cores of sub-clusters or galaxies falling into or merging with larger, more diffuse structures, ``sloshing'' cold fronts which are formed in cool-core clusters by the displacement of the central low-entropy gas of the DM-dominated core, and ``stream'' cold fronts which are formed by collisions between coherent streams of gas \citep[][]{bir10,zuh16,zin18}. 

The relevance of such bulk motions for the amplification of the cluster magnetic field was first shown by \citet{lyu06}. They demonstrated that the subsonic motion of a dense gas cloud through the ICM would amplify and stretch magnetic fields, regardless of the initial geometry, along the contact discontinuity that forms, producing a thin ``magnetic draping layer''. The only condition is that the Alfv\'{e}nic mach number ${\cal M}_A > 1$, a condition readily satisfied in the ICM. The width of the layers is given roughly by $\Delta{r} \sim {L}/{\cal M}_A^2$, For typical conditions in the ICM and a mildly subsonic cloud with ${\cal M} \lesssim 1.0$, $\Delta{r} \sim 0.01L$. \citet{lyu06} also pointed out that such layers should be associated with a depletion of plasma. This is so because the total pressure should remain continuous within and around such a layer, given the subsonic motion of the gas, and thus an increase in magnetic pressure requires a decrease in thermal pressure. This indicates that such layers may be visible in X-ray observations of cold fronts, though in practice there will be large uncertainties given projection effects.

\begin{figure}
\begin{center}
\includegraphics[width=0.95\textwidth]{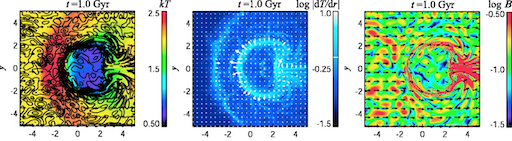}
\caption{Amplified magnetic fields produced by a remnant-core cold front in an MHD simulation detailed in \citet{asai07}. The panels show slices of temperature (left), temperature gradient (middle), and magnetic field strength (right). An amplified magnetic field appears in a draping layer around the cold front.\label{fig:asai}}
\end{center}
\end{figure}

The first numerical simulations used to examine this effect followed the evolution of a cold, dense core moving subsonically through a hot, magnetized ICM \citep{dursi2007,dursi2008,2004ApJ...606L.105A,asai05,asai07}, using a variety of field geometries. In all cases they confirmed the basic picture offered by \citet{lyu06} of magnetic field amplification in a thin layer ``draping'' the cold front surface which forms at the head of the cool core.  Figure \ref{fig:asai} shows an example MHD simulation of a remnant-core cold front producing an amplified magnetic field in a draping layer from \citet{asai07}.

\begin{figure}
\begin{center}
\includegraphics[width=0.95\textwidth]{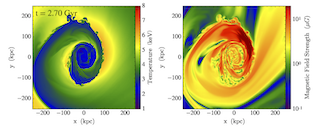}
\caption{Amplified magnetic fields produced by sloshing cold fronts in an MHD simulation. Left panel: Slice through the gas temperature in keV, showing the spiral-shaped cold fronts. Right panel: Slice through the magnetic field strength, showing amplified fields within the cold fronts. Figure reproduced from \citet{zuh16}.\label{fig:sloshing_mag}}
\end{center}
\end{figure}

A second type of cold front, the ``sloshing'' variety (figure \ref{fig:sloshing_mag}), occurs in more relaxed systems when the cold gas core is perturbed by infalling subclusters and is separated from the DM-dominated potential well. This gas then oscillates back and forth in the cluster center, producing a spiral-shaped pattern. Simulations have shown that sloshing cold fronts are also associated with amplified magnetic fields. The first simulations to demonstrate this were those of \citet{zuh11}, who simulated the evolution of initially tangled magnetic fields with a number of initial magnetic field strengths and correlation scales. The sloshing cold fronts are also associated with amplified magnetic layers, but unlike in the scenario envisaged by \citet{lyu06} the layers are on the {\it inside} of the front surface rather than outside, due to the fact that for the sloshing cold fronts the shear flow is predominantly inside. This results in increased magnetic fields within the volume bounded by the cold fronts (figure~\ref{fig:sloshing_mag}), an effect important for the generation of radio mini-halos (Section \ref{sec:minihalos}).

\subsection{Effects of cold front magnetic fields on the thermal plasma}\label{sec:bfield_effects}

\begin{figure}
\begin{center}
\includegraphics[width=0.3\textwidth]{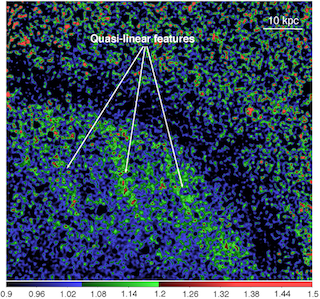}
\includegraphics[width=0.31\textwidth]{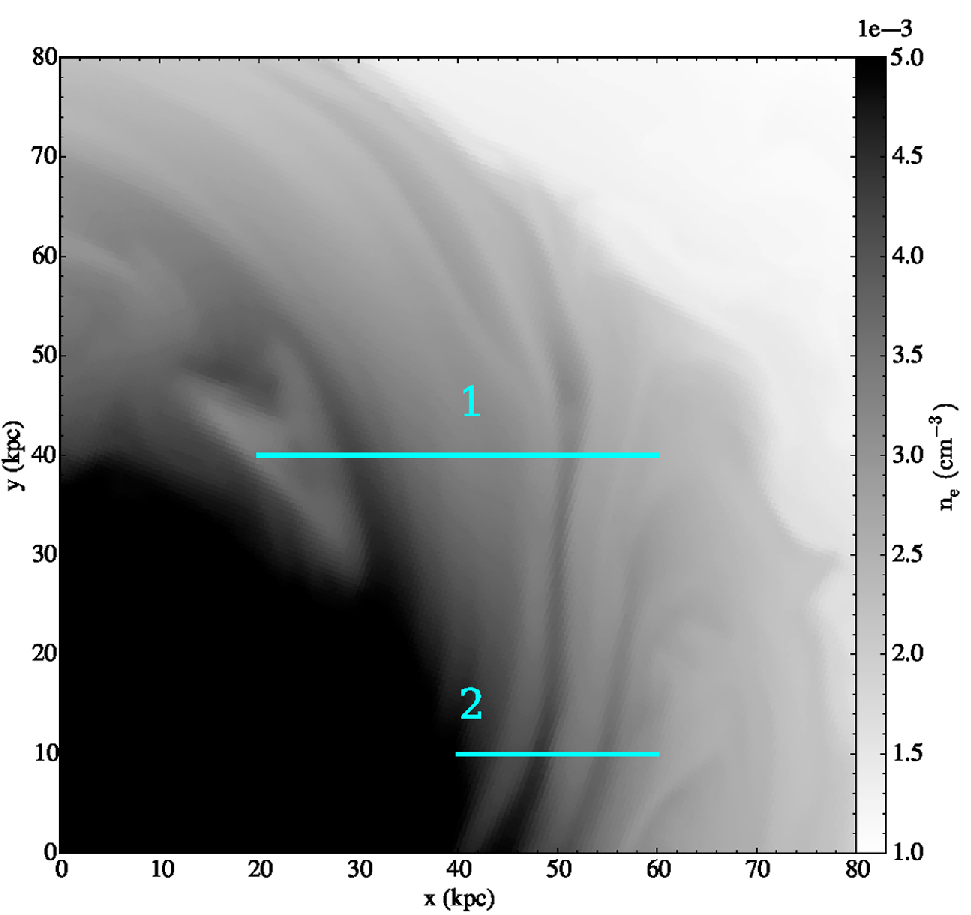}
\includegraphics[width=0.31\textwidth]{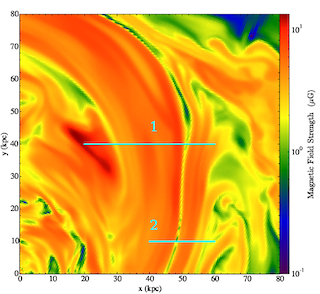}
\caption{Evidence for amplified magnetic fields in X-ray observations of the Virgo cluster from \citet{wer16}. Left panel: residual image of the X-ray surface brightness as seen by {\it Chandra} near the cold front. Three linear enhancements in surface brightness are apparent. Middle and right panels: Slices of gas density and magnetic field strength near the cold front surface in a MHD simulation of the Virgo cluster. Wide bands of strong magnetic field in between narrow channels of weak field produce linear features in density similar to those seen in Virgo.\label{fig:virgo}}
\end{center}
\end{figure}

The above considerations indicate that if a highly magnetized layer forms tangential to a cold front or otherwise because of shearing motions that it may produce a dip in X-ray surface brightness at this location. These dips were first noticed in MHD simulations of sloshing cold fronts by \citet{zuh11}. In these simulations, the layers reached magnetic field strengths with $\beta_\mathrm{pl} \sim 10$ and dips in density and temperature of roughly $\sim10-30\%$, which could produce dips in surface brightness of roughly $\sim5-10\%$, depending on the gas temperature. The evidence for such features in X-ray observations of clusters is so far inconclusive, but there are some tantalizing hints \citep[e.g.][]{wer16} (figure \ref{fig:virgo}).

The magnetic tension from a field stretched parallel to a front surface will suppress the growth of K-H instabilities if the field is strong enough \citep{chandra61}. The initial smooth appearance of many cold fronts as seen in \textit{Chandra} observations led readily to the proposal that such suppression was occurring. For example, the apparent smoothness of the merger-remnant cold front in A3667 led \citet{vik01,vik02} to estimate a magnetic field strength near the front surface between 6~$\mu$G $< B <$ 14~$\mu$G. More recently, \citet{hchen2017} estimated a magnetic field strength of $B \sim 20-30~\mu$G at the sloshing cold fronts in A2204 based on the lack of observed KHI. 

As deeper \textit{Chandra} observations of nearby clusters with longer exposures have been obtained over the years, some evidence for K-H instabilities has been been uncovered. \citet{rod15a,rod15b,kraft2017} presented evidence of gas stripping of caused by KHI in the elliptical galaxy M89 using deep X-ray observations and tailored simulations. \citet{ich2017} showed evidence for KHI in a longer combined exposure of A3667 than was available to \citet{vik01,vik02}, but did not make an updated estimate of the magnetic field strength. Finally, \citet{yysu17b} showed evidence for K-H instabilities at the interface of the cold front in NGC~1404, and used their presence to place an upper limit on the magnetic field strength at the front of 5~$\mu$G. 

However, the presence of \textit{some} degree of KHI in cold fronts is not inconsistent with the picture of magnetic draping layers per se---it is rather likely an indication of the strength of the fields in these layers. A recent series of papers has constrained the magnetic field strength in the Perseus cluster using MHD simulations and X-ray observations. \citet{swalker2017} showed convincing evidence of a giant KHI eddy at one of the cold front edges in the Perseus cluster. They compared the appearance of the cold fronts to the simulations from \citet{zuh11}, and suggested that a cluster with an initial $\beta \sim 200$ before the sloshing began could explain the presence of the KHI eddy-simulations with initially larger or smaller average magnetic field strengths produced results that were inconsistent in terms of having either too few or too many KHI eddies along the interface.

\subsection{Amplified magnetic fields and cosmic rays: radio mini-halos}\label{sec:minihalos}

Radio mini-halos are the smaller-scale siblings of the giant radio halos, hosted in cool-core clusters. Their emission is similarly diffuse and has a steep spectrum ($\alpha \sim 1-2$), but are nearly an order of magnitude smaller than radio halos and are confined to the core region. \citet{maz08} were the first to discover that the radio mini-halos in the clusters RX J1720.1+2638 and MS 1455.0+2232 were confined to the region on the sky bounded by sloshing cold fronts seen in the X-ray observations. Subsequent investigations of mini-halo emission from a number of cool-core clusters have confirmed the existence of sharp drops in radio emission at the position of the cold front surfaces in many cases \citep[][]{gia14a,gia14b,gia17}. 

Such radio emission requires a population of CRe with $\gamma \sim 10^3-10^4$, given the typical magnetic field strengths in clusters. Since CRe with such energies cool rapidly via synchrotron and Inverse-Compton losses, the existence of mini-halos requires a mechanism to replenish these electrons, either by reacceleration from a lower-energy population \citep[the ``reacceleration'' model][]{2007MNRAS.378..245B,bl11,bl11b} or as the byproducts of collisions of CRp with the ICM thermal proton population \citep[the ``hadronic'' model,][]{den80,pfr04,kl2010}, though this model is strongly constrained by the Fermi-LAT upper limits on gamma-ray emission in clusters \citep[][]{fermi14}, which are also produced by the same collisions. A review of these processes and their implications for non-thermal emission in clusters can be found in \citep{bj14}. 

\begin{figure}
\begin{center}
\raisebox{0.095\height}{\includegraphics[width=0.24\textwidth]{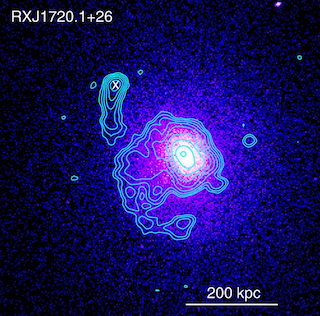}}
\includegraphics[width=0.66\textwidth]{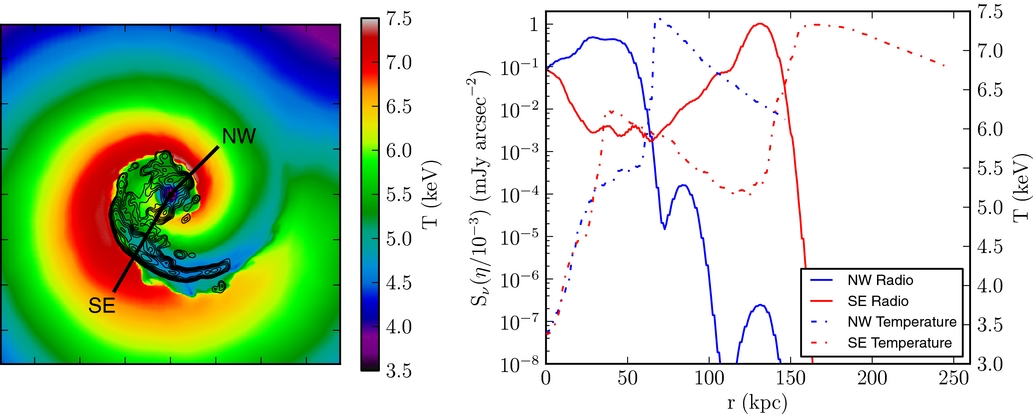}
\caption{Association of radio mini-halos with cold fronts from \citet{zuh13}. Left panel: {\it Chandra} observation of RXJ 1720.1+26, showing sloshing cold fronts, with 610~MHz radio contours overlaid. The radio emission is coincident with the sloshing cold fronts. Middle panel: Simulated 327~MHz radio contours overlaid on projected temperature from an MHD/CRe simulation of sloshing core gas which produces a mini-halo with emission bounded by the cold fronts. Right panel: Profiles of temperature and radio emission taken along the two directions shown in the middle panel, which show clearly that the radio emission drops steeply at the position of the cold fronts.\label{fig:minihalo}}
\end{center}
\end{figure}

As previously noted, sloshing cold fronts are very common in cool-core clusters, and these motions amplify magnetic fields. A stronger magnetic field within the core would lead to an enhancement of the mini-halo emission. Since this amplified magnetic field is largely confined to the volume bounded by the cold fronts, it may also explain the association of mini-halos with cold fronts and the steep drops in radio emission coincident with the front surfaces, as suggested by \citet{kl2010}.

The only simulations so far to directly test the reacceleration scenario for mini-halos were performed by \citet{zuh13}. They used a MHD simulation of gas sloshing in a cool-core cluster coupled with a simulation of the evolution of the CRe spectrum under reacceleration by turbulence and radiative and Coulomb losses along trajectories of passive tracer particles. It was found that reacceleration by turbulence coupled with the magnetic field amplification, both produced by the sloshing motions, could produce mini-halos which have the characteristic diffuse emission, steep spectrum, and spatial relationship to the cold fronts (figure \ref{fig:minihalo}). They noted that the mini-halo produced in their simulation had two further interesting characteristics: the emission was transient and brightest shortly after the beginning of the sloshing motions, and had a ``patchy'' appearance due to the intermittent and irregular distribution of turbulent gas motions in the core region. The latter prediction is perhaps supported by the recent JVLA 230-470~MHz observations of the Perseus mini-halo by \citet{mgm17}, which revealed a complex appearance of the radio emission.

The coincidence of mini-halos with cold fronts could also be explained by amplified magnetic fields in the hadronic model. \citet{kl2010} also suggested that these rapidly amplifying magnetic fields may be responsible for the steep spectrum of mini-halos. \citet{zuh15} tested this possibility using simulations that the fast amplification of magnetic fields by sloshing motions could produce diffuse, core-confined mini-halos with steep spectra by steepening the CRe spectrum. Though the amplification of the magnetic field strength within the sloshing cold fronts reproduced the observed spatial properties of mini-halo emission in this simulation, they found that only a small, observationally insignificant number of tracer particle trajectories experienced sufficiently rapid changes in magnetic field strength to steepen their radio spectra to $\alpha \sim 2$, and then only for brief periods of time. More complex mini-halo morphologies and spectral index properties may be produced in the context of hadronic models by taking into account other CRp physics such as diffusion, streaming, and advection by turbulence and/or bulk motions \citep[][]{ens11,pfr13,wie13,jac17a,jac17b,wie18}. 

\section{Concluding remarks}

We have presented the most important mechanisms that are expected to control and drive the amplification of magnetic fields observed in galaxy clusters at radio wavelengths. We gave a short introduction to turbulence, motivated MHD as a model for the intra-cluster-medium and introduced the basic principles of the small-scale MHD turbulent dynamo. We reviewed the outcome of (cosmological) numerical simulations of the growth of magnetic fields, under typical conditions in the intracluster medium. We provided a short discussion of numerical limitations of current approaches and established the demands of upcoming radio surveys of Faraday rotation measures and giant radio haloes. We also introduced magnetic field amplification at shocks and by cosmic-rays, which is evidenced by radio relics at cluster outskirts. Further we discussed magnetic field amplification by cold fronts and internal cluster motions, which are likely connected to radio mini halos. \par

The observed properties of magnetic fields in the intra-cluster medium require large amplification factors ($\geq 10^3$) even considering the effect of gas compression. Indeed, there is plenty of time and turbulent energy to boost the magnetic field energy up to observed values with a turbulent small-scale dynamo. The theoretical grounds of this small-scale dynamo model for the amplification of weak fields in random flows are robust and are covered in a significant amount of literature. A central outcome of these studies is the importance of the Alfv\'{e}n scale on the amplification and topology of magnetic fields in turbulent flows. \par
Given these expectations, the quantitative outcome of the small-scale dynamo in current cosmological simulations is likely not sufficient to robustly predict magnetic fields in the ICM for upcoming radio interferometers. Modern instruments require the robust prediction of magnetic field and shock structures down to a few kpc in the whole volume of a massive galaxy cluster. This means future cosmological simulations will need to resolve the small-scale dynamo down to at least similar resolutions. \par
A way forward may be higher order MHD methods suitable for cosmology, that resolve motions closer to the grid scale. Another possibility, especially for shocks, could be adaptive techniques to selectively refine the mesh, where the amplification is active. We note that ample computing power is available, as the largest computers approach $10^{18}$ floating point operations per second in the next years. \par
Given the large magnetic Reynolds number in the real intra-cluster medium, observed $\sim \mu G$ fields can be the result of either primordial $\sim \rm nG$ (co-moving) or $\sim \rm 10^{-6} nG$ fields, as well as from higher and more concentrated magnetic seeds  released by galactic winds or active galactic nuclei. Density adaptive techniques are well suited for these kinds of simulations. However, as opposed to galaxy formation, turbulence evolution and the SSD are only weakly correlated with density peaks in the large-scale structure. Thus density adaptive techniques might turn out to be inefficient for this problem, e.g. resolving shocks with a few kpc at the virial radius of a cluster would require exceedingly large particle numbers.\par
The inferred amplification factors in clusters shocks traced by radio relics and at the contact discontinuities generated by bulk motions in cluster cores are lower than from the SSD ($\sim 10-30$), yet amplification must again operate on small scales ($\leq 10^2$ kpc) and on short timescales ($\sim 10$ Myr). Fully reproducing these trends with simulations may still be a challenge for numerical simulations, again because of  resolution. Additionally, more complex interplay of magnetic fields and cosmic-rays are likely at work and plasma conditions across such discontinuities in the fluid may require a consideration of modified viscosity and thermal conduction.\par

\begin{table*}
\centering
\begin{tabular}{r|c|c|c|c|c|c|c}
 &  $B_{\rm obs}$ & $\beta_{\rm pl}$ &  $(\delta B/B_0)^2_{\rm tot}$ & $(\delta B/B_0)^2_{\rm dyn}$ & $t_{\rm dyn}$& $l_{\rm B}$ &  $\Gamma$ \\
 &  [$\mu G$] &  &  &     &   [$Gyr$] &  [$\rm kpc$] & [$\rm Gyr^{-1}$]\\  \hline \hline
{\bf SSD} & $\sim 0.1-5$ & $\sim 10^2$ & $\geq 10^8 $  & $\geq 10^3 $ & $\leq  10$ & $\leq 50-100$ & $\sim 1$ \\
{\bf shocks}& $\sim 1-3$ & $\sim 10$ & $\geq 100 $  & $\geq 30 $ & $\leq 0.01$  &  $\leq 10-10^2$& $\sim \rm 0.003$ \\
{\bf sloshing} & $\sim 10-20$ & $\sim 1$ & $\geq 10 $  &  $\geq 10 $ & $\leq  0.01$  & $\leq 1-10$ & $\sim  0.001$ 
\end{tabular}
\caption{Reference parameters for the amplification of magnetic fields in various locations of the intra-cluster-medium. 2nd column: typical observed value; 3rd column: plasma beta; 4th column: inferred (minimum) amplification factor of magnetic field energy (also including gas compression); 5th column: amplification factor only considering dynamo amplification; 6th column: minimum amplification time; 7th column: typical scale of field reversals; 8th column: estimated growth factor in magnetic energy.}
\label{tab:dynamos}
\end{table*}

In table \ref{tab:dynamos} we summarize these general trends in magnetic field amplification found from observations and guided by simulations, which we consider to be robust against numerical issues. We give the typically observed magnetic field, the estimated total magnetic energy growth (considering for each case the magnetic field before the process begins; the magnetic energy growth only due to small-scale dynamo (e.g. after removing for the $\propto n^{4/3}$ compression factor related to each process); the estimated (minimum) time for amplification; the typical energy containing scale related to each mechanism. The final column gives the estimated growth factor, $\gamma_{\rm growth}$ for each amplification mechanism, parameterized by $(\delta B/B_0)^2_{\rm dyn} \approx \exp (\Gamma  t_{\rm dyn})$  ). \par

 In this review, we were addressing the successes and limitations of numerical models for predictions of extragalactic magnetic fields. It seems obvious to foresee that the upcoming generation of radio observations (culminating in the SKA data) will pose ever more challenging questions to our theoretical and numerical models of rarefied space plasmas. Only in synergy will numerical techniques and radio observations exploit the next generation of radio telescopes to study plasma physics in the Universe. \par

\begin{acknowledgements}
The authors would like to thank the referee, D.C. Ellison, for valuable comments that improved the review. The authors thank K. Dolag for discussions on galactic dynamos. JD thanks T. Tricco, D. Price, D. Ryu and T.Jones for discussions on numerical MHD methods. JAZ thanks C. Pfrommer for useful comments. \par
F.V. acknowledges financial support from the ERC Starting Grant ''MAGCOW'', no.714196, and  the  usage of computational resources on the Piz-Daint supercluster at CSCS-ETHZ (Lugano, Switzerland) under project s701 and s805, and of the JURECA supercluster at JFZ (J\"{u}lich, Germany) under project hhh42 (HiMAG).  JD thanks H. Roettgering and Leiden observatory for the hospitality. This research has received funding from the People Programme (Marie Sklodowska Curie Actions) of the European Unions Eighth Framework Programme H2020 under REA grant agreement no 658912, ''CosmoPlasmas''. 

\end{acknowledgements}

\bibliographystyle{spbasic}      
\bibliography{franco2}   

\begin{thebibliography}{321}
\providecommand{\natexlab}[1]{#1}
\providecommand{\url}[1]{{#1}}
\providecommand{\urlprefix}{URL }
\expandafter\ifx\csname urlstyle\endcsname\relax
  \providecommand{\doi}[1]{DOI~\discretionary{}{}{}#1}\else
  \providecommand{\doi}{DOI~\discretionary{}{}{}\begingroup
  \urlstyle{rm}\Url}\fi
\providecommand{\eprint}[2][]{\url{#2}}

\bibitem[{{Ackermann} et~al(2014){Ackermann}, {Ajello}, {Albert}, {Allafort},
  {Atwood}, {Baldini}, {Ballet}, {Barbiellini}, {Bastieri}, {Bechtol},
  {Bellazzini}, {Bloom}, {Bonamente}, {Bottacini}, {Brandt}, {Bregeon},
  {Brigida}, {Bruel}, {Buehler}, {Buson}, {Caliandro}, {Cameron}, {Caraveo},
  {Cavazzuti}, {Chaves}, {Chiang}, {Chiaro}, {Ciprini}, {Claus},
  {Cohen-Tanugi}, {Conrad}, {D'Ammando}, {de Angelis}, {de Palma}, {Dermer},
  {Digel}, {Drell}, {Drlica-Wagner}, {Favuzzi}, {Franckowiak}, {Funk}, {Fusco},
  {Gargano}, {Gasparrini}, {Germani}, {Giglietto}, {Giordano}, {Giroletti},
  {Godfrey}, {Gomez-Vargas}, {Grenier}, {Guiriec}, {Gustafsson}, {Hadasch},
  {Hayashida}, {Hewitt}, {Hughes}, {Jeltema}, {J{\'o}hannesson}, {Johnson},
  {Kamae}, {Kataoka}, {Kn{\"o}dlseder}, {Kuss}, {Lande}, {Larsson},
  {Latronico}, {Llena Garde}, {Longo}, {Loparco}, {Lovellette}, {Lubrano},
  {Mayer}, {Mazziotta}, {McEnery}, {Michelson}, {Mitthumsiri}, {Mizuno},
  {Monzani}, {Morselli}, {Moskalenko}, {Murgia}, {Nemmen}, {Nuss}, {Ohsugi},
  {Orienti}, {Orlando}, {Ormes}, {Perkins}, {Pesce-Rollins}, {Piron}, {Pivato},
  {Rain{\`o}}, {Rando}, {Razzano}, {Razzaque}, {Reimer}, {Reimer}, {Ruan},
  {S{\'a}nchez-Conde}, {Schulz}, {Sgr{\`o}}, {Siskind}, {Spandre}, {Spinelli},
  {Storm}, {Strong}, {Suson}, {Takahashi}, {Thayer}, {Thayer}, {Thompson},
  {Tibaldo}, {Tinivella}, {Torres}, {Troja}, {Uchiyama}, {Usher},
  {Vandenbroucke}, {Vianello}, {Vitale}, {Winer}, {Wood}, {Zimmer}, {Fermi-LAT
  Collaboration}, {Pinzke}, and {Pfrommer}}]{fermi14}
{Ackermann} M, {Ajello} M, {Albert} A, {Allafort} A, {Atwood} WB, {Baldini} L,
  {Ballet} J, {Barbiellini} G, {Bastieri} D, {Bechtol} K, {Bellazzini} R,
  {Bloom} ED, {Bonamente} E, {Bottacini} E, {Brandt} TJ, {Bregeon} J, {Brigida}
  M, {Bruel} P, {Buehler} R, {Buson} S, {Caliandro} GA, {Cameron} RA, {Caraveo}
  PA, {Cavazzuti} E, {Chaves} RCG, {Chiang} J, {Chiaro} G, {Ciprini} S, {Claus}
  R, {Cohen-Tanugi} J, {Conrad} J, {D'Ammando} F, {de Angelis} A, {de Palma} F,
  {Dermer} CD, {Digel} SW, {Drell} PS, {Drlica-Wagner} A, {Favuzzi} C,
  {Franckowiak} A, {Funk} S, {Fusco} P, {Gargano} F, {Gasparrini} D, {Germani}
  S, {Giglietto} N, {Giordano} F, {Giroletti} M, {Godfrey} G, {Gomez-Vargas}
  GA, {Grenier} IA, {Guiriec} S, {Gustafsson} M, {Hadasch} D, {Hayashida} M,
  {Hewitt} J, {Hughes} RE, {Jeltema} TE, {J{\'o}hannesson} G, {Johnson} AS,
  {Kamae} T, {Kataoka} J, {Kn{\"o}dlseder} J, {Kuss} M, {Lande} J, {Larsson} S,
  {Latronico} L, {Llena Garde} M, {Longo} F, {Loparco} F, {Lovellette} MN,
  {Lubrano} P, {Mayer} M, {Mazziotta} MN, {McEnery} JE, {Michelson} PF,
  {Mitthumsiri} W, {Mizuno} T, {Monzani} ME, {Morselli} A, {Moskalenko} IV,
  {Murgia} S, {Nemmen} R, {Nuss} E, {Ohsugi} T, {Orienti} M, {Orlando} E,
  {Ormes} JF, {Perkins} JS, {Pesce-Rollins} M, {Piron} F, {Pivato} G,
  {Rain{\`o}} S, {Rando} R, {Razzano} M, {Razzaque} S, {Reimer} A, {Reimer} O,
  {Ruan} J, {S{\'a}nchez-Conde} M, {Schulz} A, {Sgr{\`o}} C, {Siskind} EJ,
  {Spandre} G, {Spinelli} P, {Storm} E, {Strong} AW, {Suson} DJ, {Takahashi} H,
  {Thayer} JG, {Thayer} JB, {Thompson} DJ, {Tibaldo} L, {Tinivella} M, {Torres}
  DF, {Troja} E, {Uchiyama} Y, {Usher} TL, {Vandenbroucke} J, {Vianello} G,
  {Vitale} V, {Winer} BL, {Wood} KS, {Zimmer} S, {Fermi-LAT Collaboration},
  {Pinzke} A, {Pfrommer} C (2014) {Search for Cosmic-Ray-induced Gamma-Ray
  Emission in Galaxy Clusters}. \apj 787:18, \doi{10.1088/0004-637X/787/1/18},
  \eprint{1308.5654}

\bibitem[{{Alfv{\'e}n}(1942)}]{1942Natur.150..405A}
{Alfv{\'e}n} H (1942) {Existence of Electromagnetic-Hydrodynamic Waves}. \nat
  150:405--406, \doi{10.1038/150405d0}

\bibitem[{{Arlen} et~al(2014){Arlen}, {Vassilev}, {Weisgarber}, {Wakely}, and
  {Yusef Shafi}}]{2014ApJ...796...18A}
{Arlen} TC, {Vassilev} VV, {Weisgarber} T, {Wakely} SP, {Yusef Shafi} S (2014)
  {Intergalactic Magnetic Fields and Gamma-Ray Observations of Extreme TeV
  Blazars}. \apj 796:18, \doi{10.1088/0004-637X/796/1/18}, \eprint{1210.2802}

\bibitem[{{Asai} et~al(2004){Asai}, {Fukuda}, and
  {Matsumoto}}]{2004ApJ...606L.105A}
{Asai} N, {Fukuda} N, {Matsumoto} R (2004) {Magnetohydrodynamic Simulations of
  the Formation of Cold Fronts in Clusters of Galaxies Including Heat
  Conduction}. \apjl 606:L105--L108, \doi{10.1086/421440},
  \eprint{astro-ph/0404160}

\bibitem[{{Asai} et~al(2005){Asai}, {Fukuda}, and {Matsumoto}}]{asai05}
{Asai} N, {Fukuda} N, {Matsumoto} R (2005) {Three-dimensional MHD simulations
  of X-ray emitting subcluster plasmas in cluster of galaxies}. Advances in
  Space Research 36:636--642, \doi{10.1016/j.asr.2005.04.041},
  \eprint{astro-ph/0504227}

\bibitem[{{Asai} et~al(2007){Asai}, {Fukuda}, and {Matsumoto}}]{asai07}
{Asai} N, {Fukuda} N, {Matsumoto} R (2007) {Three-dimensional
  Magnetohydrodynamic Simulations of Cold Fronts in Magnetically Turbulent
  ICM}. \apj 663:816--823, \doi{10.1086/518235}, \eprint{astro-ph/0703536}

\bibitem[{{Bai} et~al(2015){Bai}, {Caprioli}, {Sironi}, and
  {Spitkovsky}}]{2015ApJ...809...55B}
{Bai} XN, {Caprioli} D, {Sironi} L, {Spitkovsky} A (2015)
  {Magnetohydrodynamic-particle-in-cell Method for Coupling Cosmic Rays with a
  Thermal Plasma: Application to Non-relativistic Shocks}. \apj 809:55,
  \doi{10.1088/0004-637X/809/1/55}, \eprint{1412.1087}

\bibitem[{{Balsara}(2017)}]{2017LRCA....3....2B}
{Balsara} DS (2017) {Higher-order accurate space-time schemes for computational
  astrophysics - Part I: finite volume methods}. Living Reviews in
  Computational Astrophysics 3:2, \doi{10.1007/s41115-017-0002-8},
  \eprint{1703.01241}

\bibitem[{{Barrow} et~al(1997){Barrow}, {Ferreira}, and
  {Silk}}]{1997PhRvL..78.3610B}
{Barrow} JD, {Ferreira} PG, {Silk} J (1997) {Constraints on a Primordial
  Magnetic Field}. Physical Review Letters 78:3610--3613,
  \doi{10.1103/PhysRevLett.78.3610}, \eprint{astro-ph/9701063}

\bibitem[{{Batchelor}(1950)}]{1950RSPSA.201..405B}
{Batchelor} GK (1950) {On the Spontaneous Magnetic Field in a Conducting Liquid
  in Turbulent Motion}. Proceedings of the Royal Society of London Series A
  201:405--416, \doi{10.1098/rspa.1950.0069}

\bibitem[{{Battaglia} et~al(2012){Battaglia}, {Bond}, {Pfrommer}, and
  {Sievers}}]{2012ApJ...758...74B}
{Battaglia} N, {Bond} JR, {Pfrommer} C, {Sievers} JL (2012) {On the Cluster
  Physics of Sunyaev-Zel'dovich and X-Ray Surveys. I. The Influence of
  Feedback, Non-thermal Pressure, and Cluster Shapes on Y-M Scaling Relations}.
  \apj 758:74, \doi{10.1088/0004-637X/758/2/74}, \eprint{1109.3709}

\bibitem[{{Bauer} and {Springel}(2012)}]{2012MNRAS.423.2558B}
{Bauer} A, {Springel} V (2012) {Subsonic turbulence in smoothed particle
  hydrodynamics and moving-mesh simulations}. \mnras 423:2558--2578,
  \doi{10.1111/j.1365-2966.2012.21058.x}, \eprint{1109.4413}

\bibitem[{{Bauer} et~al(2016){Bauer}, {Schaal}, {Springel}, {Chandrashekar},
  {Pakmor}, and {Klingenberg}}]{2016arXiv160209079B}
{Bauer} A, {Schaal} K, {Springel} V, {Chandrashekar} P, {Pakmor} R,
  {Klingenberg} C (2016) {Simulating Turbulence Using the Astrophysical
  Discontinuous Galerkin Code TENET}. ArXiv e-prints \eprint{1602.09079}

\bibitem[{{Beck} et~al(2012){Beck}, {Lesch}, {Dolag}, {Kotarba}, {Geng}, and
  {Stasyszyn}}]{2012MNRAS.422.2152B}
{Beck} AM, {Lesch} H, {Dolag} K, {Kotarba} H, {Geng} A, {Stasyszyn} FA (2012)
  {Origin of strong magnetic fields in Milky Way-like galactic haloes}. \mnras
  422:2152--2163, \doi{10.1111/j.1365-2966.2012.20759.x}, \eprint{1202.3349}

\bibitem[{{Beck} et~al(2013{\natexlab{a}}){Beck}, {Dolag}, {Lesch}, and
  {Kronberg}}]{2013MNRAS.435.3575B}
{Beck} AM, {Dolag} K, {Lesch} H, {Kronberg} PP (2013{\natexlab{a}}) {Strong
  magnetic fields and large rotation measures in protogalaxies from supernova
  seeding}. \mnras 435:3575--3586, \doi{10.1093/mnras/stt1549},
  \eprint{1308.3440}

\bibitem[{{Beck} et~al(2013{\natexlab{b}}){Beck}, {Hanasz}, {Lesch}, {Remus},
  and {Stasyszyn}}]{beck13}
{Beck} AM, {Hanasz} M, {Lesch} H, {Remus} RS, {Stasyszyn} FA
  (2013{\natexlab{b}}) {On the magnetic fields in voids}. \mnras 429:L60--L64,
  \doi{10.1093/mnrasl/sls026}, \eprint{1210.8360}

\bibitem[{{Beck} et~al(2016){Beck}, {Murante}, {Arth}, {Remus}, {Teklu},
  {Donnert}, {Planelles}, {Beck}, {F{\"o}rster}, {Imgrund}, {Dolag}, and
  {Borgani}}]{2016MNRAS.455.2110B}
{Beck} AM, {Murante} G, {Arth} A, {Remus} RS, {Teklu} AF, {Donnert} JMF,
  {Planelles} S, {Beck} MC, {F{\"o}rster} P, {Imgrund} M, {Dolag} K, {Borgani}
  S (2016) {An improved SPH scheme for cosmological simulations}. \mnras
  455:2110--2130, \doi{10.1093/mnras/stv2443}, \eprint{1502.07358}

\bibitem[{{Beck} and {Krause}(2005)}]{2005AN....326..414B}
{Beck} R, {Krause} M (2005) {Revised equipartition and minimum energy formula
  for magnetic field strength estimates from radio synchrotron observations}.
  Astronomische Nachrichten 326:414--427, \doi{10.1002/asna.200510366},
  \eprint{arXiv:astro-ph/0507367}

\bibitem[{{Bell}(2004)}]{2004MNRAS.353..550B}
{Bell} AR (2004) {Turbulent amplification of magnetic field and diffusive shock
  acceleration of cosmic rays}. \mnras 353:550--558,
  \doi{10.1111/j.1365-2966.2004.08097.x}

\bibitem[{{Beresnyak} and {Lazarian}(2006)}]{2006ApJ...640L.175B}
{Beresnyak} A, {Lazarian} A (2006) {Polarization Intermittency and Its
  Influence on MHD Turbulence}. \apjl 640:L175--L178, \doi{10.1086/503708},
  \eprint{astro-ph/0512315}

\bibitem[{{Beresnyak} and {Miniati}(2016)}]{bm16}
{Beresnyak} A, {Miniati} F (2016) {Turbulent Amplification and Structure of the
  Intracluster Magnetic Field}. \apj 817:127,
  \doi{10.3847/0004-637X/817/2/127}, \eprint{1507.00342}

\bibitem[{{Bertone} et~al(2006){Bertone}, {Vogt}, and
  {En{\ss}lin}}]{2006MNRAS.370..319B}
{Bertone} S, {Vogt} C, {En{\ss}lin} T (2006) {Magnetic field seeding by
  galactic winds}. \mnras 370:319--330, \doi{10.1111/j.1365-2966.2006.10474.x},
  \eprint{arXiv:astro-ph/0604462}

\bibitem[{{Biermann} and {Schl{\"u}ter}(1951)}]{1951PhRv...82..863B}
{Biermann} L, {Schl{\"u}ter} A (1951) {Cosmic Radiation and Cosmic Magnetic
  Fields. II. Origin of Cosmic Magnetic Fields}. Physical Review 82:863--868,
  \doi{10.1103/PhysRev.82.863}

\bibitem[{{Binney}(1974)}]{1974MNRAS.168...73B}
{Binney} J (1974) {Galaxy formation without primordial turbulence: mechanisms
  for generating cosmic vorticity}. \mnras 168:73--92,
  \doi{10.1093/mnras/168.1.73}

\bibitem[{{Birnboim} et~al(2010){Birnboim}, {Keshet}, and {Hernquist}}]{bir10}
{Birnboim} Y, {Keshet} U, {Hernquist} L (2010) {Cold fronts by merging of
  shocks}. \mnras 408:199--212, \doi{10.1111/j.1365-2966.2010.17176.x},
  \eprint{1006.1892}

\bibitem[{{B{\"o}hringer} and {Werner}(2010)}]{2010A&ARv..18..127B}
{B{\"o}hringer} H, {Werner} N (2010) {X-ray spectroscopy of galaxy clusters:
  studying astrophysical processes in the largest celestial laboratories}.
  \aapr 18:127--196, \doi{10.1007/s00159-009-0023-3}

\bibitem[{{Bonafede} et~al(2010){Bonafede}, {Feretti}, {Murgia}, {Govoni},
  {Giovannini}, {Dallacasa}, {Dolag}, and {Taylor}}]{bo10}
{Bonafede} A, {Feretti} L, {Murgia} M, {Govoni} F, {Giovannini} G, {Dallacasa}
  D, {Dolag} K, {Taylor} GB (2010) {The Coma cluster magnetic field from
  Faraday rotation measures}. \aap 513:A30+, \doi{10.1051/0004-6361/200913696},
  \eprint{1002.0594}

\bibitem[{{Bonafede} et~al(2011){Bonafede}, {Dolag}, {Stasyszyn}, {Murante},
  and {Borgani}}]{2011MNRAS.418.2234B}
{Bonafede} A, {Dolag} K, {Stasyszyn} F, {Murante} G, {Borgani} S (2011) {A
  non-ideal magnetohydrodynamic GADGET: simulating massive galaxy clusters}.
  \mnras 418:2234--2250, \doi{10.1111/j.1365-2966.2011.19523.x},
  \eprint{1107.0968}

\bibitem[{{Bonafede} et~al(2013){Bonafede}, {Vazza}, {Br{\"u}ggen}, {Murgia},
  {Govoni}, {Feretti}, {Giovannini}, and {Ogrean}}]{bo13}
{Bonafede} A, {Vazza} F, {Br{\"u}ggen} M, {Murgia} M, {Govoni} F, {Feretti} L,
  {Giovannini} G, {Ogrean} G (2013) {Measurements and simulation of Faraday
  rotation across the Coma radio relic}. \mnras 433:3208--3226,
  \doi{10.1093/mnras/stt960}, \eprint{1305.7228}

\bibitem[{{Bonafede} et~al(2015){Bonafede}, {Vazza}, {Br{\"u}ggen}, {Akahori},
  {Carretti}, {Colafrancesco}, {Feretti}, {Ferrari}, {Giovannini}, {Govoni},
  {Johnston-Hollitt}, {Murgia}, {Scaife}, {Vacca}, {Govoni}, {Rudnick}, and
  {Scaife}}]{2015aska.confE..95B}
{Bonafede} A, {Vazza} F, {Br{\"u}ggen} M, {Akahori} T, {Carretti} E,
  {Colafrancesco} S, {Feretti} L, {Ferrari} C, {Giovannini} G, {Govoni} F,
  {Johnston-Hollitt} M, {Murgia} M, {Scaife} A, {Vacca} V, {Govoni} F,
  {Rudnick} L, {Scaife} A (2015) {Unravelling the origin of large-scale
  magnetic fields in galaxy clusters and beyond through Faraday Rotation
  Measures with the SKA}. Advancing Astrophysics with the Square Kilometre
  Array (AASKA14) 95, \eprint{1501.00321}

\bibitem[{{Borgani} et~al(2008){Borgani}, {Diaferio}, {Dolag}, and
  {Schindler}}]{borgani08}
{Borgani} S, {Diaferio} A, {Dolag} K, {Schindler} S (2008) {Thermodynamical
  Properties of the ICM from Hydrodynamical Simulations}. \ssr 134:269--293,
  \doi{10.1007/s11214-008-9317-4}, \eprint{0801.1032}

\bibitem[{{Brandenburg}(2011)}]{2011ApJ...741...92B}
{Brandenburg} A (2011) {Nonlinear Small-scale Dynamos at Low Magnetic Prandtl
  Numbers}. \apj 741:92, \doi{10.1088/0004-637X/741/2/92}, \eprint{1106.5777}

\bibitem[{{Brandenburg} and {Subramanian}(2005)}]{2005PhR...417....1B}
{Brandenburg} A, {Subramanian} K (2005) {Astrophysical magnetic fields and
  nonlinear dynamo theory}. \physrep 417:1--209,
  \doi{10.1016/j.physrep.2005.06.005}, \eprint{astro-ph/0405052}

\bibitem[{{Broderick} et~al(2012){Broderick}, {Chang}, and
  {Pfrommer}}]{2012ApJ...752...22B}
{Broderick} AE, {Chang} P, {Pfrommer} C (2012) {The Cosmological Impact of
  Luminous TeV Blazars. I. Implications of Plasma Instabilities for the
  Intergalactic Magnetic Field and Extragalactic Gamma-Ray Background}. \apj
  752:22, \doi{10.1088/0004-637X/752/1/22}, \eprint{1106.5494}

\bibitem[{{Br{\"u}ggen}(2013)}]{2013MNRAS.436..294B}
{Br{\"u}ggen} M (2013) {Magnetic field amplification by cosmic ray-driven
  turbulence - I. Isotropic CR diffusion}. \mnras 436:294--303,
  \doi{10.1093/mnras/stt1566}, \eprint{1308.5230}

\bibitem[{{Br{\"u}ggen} et~al(2005{\natexlab{a}}){Br{\"u}ggen}, {Hoeft}, and
  {Ruszkowski}}]{2005ApJ...628..153B}
{Br{\"u}ggen} M, {Hoeft} M, {Ruszkowski} M (2005{\natexlab{a}}) {X-Ray Line
  Tomography of AGN-induced Motion in Clusters of Galaxies}. \apj 628:153--159,
  \doi{10.1086/430732}, \eprint{arXiv:astro-ph/0503656}

\bibitem[{{Br{\"u}ggen} et~al(2005{\natexlab{b}}){Br{\"u}ggen}, {Ruszkowski},
  {Simionescu}, {Hoeft}, and {Dalla Vecchia}}]{br05}
{Br{\"u}ggen} M, {Ruszkowski} M, {Simionescu} A, {Hoeft} M, {Dalla Vecchia} C
  (2005{\natexlab{b}}) {Simulations of Magnetic Fields in Filaments}. \apjl
  631:L21--L24, \doi{10.1086/497004}, \eprint{astro-ph/0508231}

\bibitem[{{Br{\"u}ggen} et~al(2012){Br{\"u}ggen}, {Bykov}, {Ryu}, and
  {R{\"o}ttgering}}]{2012SSRv..166..187B}
{Br{\"u}ggen} M, {Bykov} A, {Ryu} D, {R{\"o}ttgering} H (2012) {Magnetic
  Fields, Relativistic Particles, and Shock Waves in Cluster Outskirts}. \ssr
  166:187--213, \doi{10.1007/s11214-011-9785-9}, \eprint{1107.5223}

\bibitem[{{Brunetti} and {Jones}(2014)}]{bj14}
{Brunetti} G, {Jones} TW (2014) {Cosmic Rays in Galaxy Clusters and Their
  Nonthermal Emission}. International Journal of Modern Physics D
  23:1430007-98, \doi{10.1142/S0218271814300079}, \eprint{1401.7519}

\bibitem[{{Brunetti} and {Lazarian}(2007)}]{2007MNRAS.378..245B}
{Brunetti} G, {Lazarian} A (2007) {Compressible turbulence in galaxy clusters:
  physics and stochastic particle re-acceleration}. \mnras 378:245--275,
  \doi{10.1111/j.1365-2966.2007.11771.x}, \eprint{astro-ph/0703591}

\bibitem[{{Brunetti} and {Lazarian}(2011{\natexlab{a}})}]{bl11b}
{Brunetti} G, {Lazarian} A (2011{\natexlab{a}}) {Acceleration of primary and
  secondary particles in galaxy clusters by compressible MHD turbulence: from
  radio haloes to gamma-rays}. \mnras 410:127--142,
  \doi{10.1111/j.1365-2966.2010.17457.x}, \eprint{1008.0184}

\bibitem[{{Brunetti} and {Lazarian}(2011{\natexlab{b}})}]{bl11}
{Brunetti} G, {Lazarian} A (2011{\natexlab{b}}) {Particle reacceleration by
  compressible turbulence in galaxy clusters: effects of a reduced mean free
  path}. \mnras 412:817--824, \doi{10.1111/j.1365-2966.2010.17937.x},
  \eprint{1011.1198}

\bibitem[{{Brunetti} and {Lazarian}(2011{\natexlab{c}})}]{2011MNRAS.412..817B}
{Brunetti} G, {Lazarian} A (2011{\natexlab{c}}) {Particle reacceleration by
  compressible turbulence in galaxy clusters: effects of a reduced mean free
  path}. \mnras 412:817--824, \doi{10.1111/j.1365-2966.2010.17937.x},
  \eprint{1011.1198}

\bibitem[{{Brunetti} et~al(2004){Brunetti}, {Blasi}, {Cassano}, and
  {Gabici}}]{2004MNRAS.350.1174B}
{Brunetti} G, {Blasi} P, {Cassano} R, {Gabici} S (2004) {Alfv{\' e}nic
  reacceleration of relativistic particles in galaxy clusters: MHD waves,
  leptons and hadrons}. \mnras 350:1174--1194

\bibitem[{{Burbidge}(1958)}]{1958ApJ...128....1B}
{Burbidge} GR (1958) {Possible Sources of Radio Emission in Clusters of
  Galaxies.} \apj 128:1, \doi{10.1086/146509}

\bibitem[{Burgers(1939)}]{Burgers1939}
Burgers JM (1939) {Some considerations on the fields of stress connected with
  dislocations in a regular crystal lattice I}. Proceedings of the Koninklijke
  Nederlandse Akademie van Wetenschappen 42:293--378

\bibitem[{{Burns} et~al(2010){Burns}, {Skillman}, and
  {O'Shea}}]{2010ApJ...721.1105B}
{Burns} JO, {Skillman} SW, {O'Shea} BW (2010) {Galaxy Clusters at the Edge:
  Temperature, Entropy, and Gas Dynamics Near the Virial Radius}. \apj
  721:1105--1112, \doi{10.1088/0004-637X/721/2/1105}, \eprint{1004.3553}

\bibitem[{{Butsky} et~al(2017){Butsky}, {Zrake}, {Kim}, {Yang}, and
  {Abel}}]{2017ApJ...843..113B}
{Butsky} I, {Zrake} J, {Kim} Jh, {Yang} HI, {Abel} T (2017) {Ab Initio
  Simulations of a Supernova-driven Galactic Dynamo in an Isolated Disk
  Galaxy}. \apj 843:113, \doi{10.3847/1538-4357/aa799f}, \eprint{1610.08528}

\bibitem[{{Bykov} et~al(2009){Bykov}, {Osipov}, and
  {Toptygin}}]{2009AstL...35..555B}
{Bykov} AM, {Osipov} SM, {Toptygin} IN (2009) {Long-wavelength MHD instability
  in the prefront of collisionless shocks with accelerated particles}.
  Astronomy Letters 35:555--563, \doi{10.1134/S1063773709080052}

\bibitem[{{Bykov} et~al(2011){Bykov}, {Osipov}, and
  {Ellison}}]{2011MNRAS.410...39B}
{Bykov} AM, {Osipov} SM, {Ellison} DC (2011) {Cosmic ray current driven
  turbulence in shocks with efficient particle acceleration: the oblique,
  long-wavelength mode instability}. \mnras 410:39--52,
  \doi{10.1111/j.1365-2966.2010.17421.x}, \eprint{1010.0408}

\bibitem[{{Bykov} et~al(2014){Bykov}, {Ellison}, {Osipov}, and
  {Vladimirov}}]{2014ApJ...789..137B}
{Bykov} AM, {Ellison} DC, {Osipov} SM, {Vladimirov} AE (2014) {Magnetic Field
  Amplification in Nonlinear Diffusive Shock Acceleration Including Resonant
  and Non-resonant Cosmic-Ray Driven Instabilities}. \apj 789:137,
  \doi{10.1088/0004-637X/789/2/137}, \eprint{1406.0084}

\bibitem[{{Campanelli}(2009)}]{2009IJMPD..18.1395C}
{Campanelli} L (2009) {Helical Magnetic Fields from Inflation}. International
  Journal of Modern Physics D 18:1395--1411, \doi{10.1142/S0218271809015175},
  \eprint{0805.0575}

\bibitem[{{Caprini} and {Gabici}(2015)}]{2015PhRvD..91l3514C}
{Caprini} C, {Gabici} S (2015) {Gamma-ray observations of blazars and the
  intergalactic magnetic field spectrum}. \prd 91(12):123514,
  \doi{10.1103/PhysRevD.91.123514}, \eprint{1504.00383}

\bibitem[{{Caprioli}(2012)}]{2012JCAP...07..038C}
{Caprioli} D (2012) {Cosmic-ray acceleration in supernova remnants: non-linear
  theory revised}. \jcap 7:038, \doi{10.1088/1475-7516/2012/07/038},
  \eprint{1206.1360}

\bibitem[{{Caprioli} and
  {Spitkovsky}(2014{\natexlab{a}})}]{2014ApJ...783...91C}
{Caprioli} D, {Spitkovsky} A (2014{\natexlab{a}}) {Simulations of Ion
  Acceleration at Non-relativistic Shocks. I. Acceleration Efficiency}. \apj
  783:91, \doi{10.1088/0004-637X/783/2/91}, \eprint{1310.2943}

\bibitem[{{Caprioli} and
  {Spitkovsky}(2014{\natexlab{b}})}]{2014ApJ...794...46C}
{Caprioli} D, {Spitkovsky} A (2014{\natexlab{b}}) {Simulations of Ion
  Acceleration at Non-relativistic Shocks. II. Magnetic Field Amplification}.
  \apj 794:46, \doi{10.1088/0004-637X/794/1/46}, \eprint{1401.7679}

\bibitem[{{Chandrasekhar}(1961)}]{chandra61}
{Chandrasekhar} S (1961) {Hydrodynamic and hydromagnetic stability}

\bibitem[{{Chen} et~al(2017){Chen}, {Jones}, {Andrade-Santos}, {ZuHone}, and
  {Li}}]{hchen2017}
{Chen} H, {Jones} C, {Andrade-Santos} F, {ZuHone} JA, {Li} Z (2017) {Gas
  Sloshing in Abell 2204: Constraining the Properties of the Magnetized
  Intracluster Medium}. \apj 838:38, \doi{10.3847/1538-4357/aa64de},
  \eprint{1703.01895}

\bibitem[{{Chen} et~al(2015){Chen}, {Buckley}, and
  {Ferrer}}]{2015PhRvL.115u1103C}
{Chen} W, {Buckley} JH, {Ferrer} F (2015) {Search for GeV {$\gamma$} -Ray Pair
  Halos Around Low Redshift Blazars}. Physical Review Letters 115(21):211103,
  \doi{10.1103/PhysRevLett.115.211103}, \eprint{1410.7717}

\bibitem[{{Chernin}(1967)}]{1967SvA....10..634C}
{Chernin} AD (1967) {A. Cosmological Model with a Disordered Magnetic Field}.
  \sovast 10:634

\bibitem[{{Cho} et~al(2009){Cho}, {Vishniac}, {Beresnyak}, {Lazarian}, and
  {Ryu}}]{2009ApJ...693.1449C}
{Cho} J, {Vishniac} ET, {Beresnyak} A, {Lazarian} A, {Ryu} D (2009) {Growth of
  Magnetic Fields Induced by Turbulent Motions}. \apj 693:1449--1461,
  \doi{10.1088/0004-637X/693/2/1449}, \eprint{0812.0817}

\bibitem[{{Churazov} et~al(2004){Churazov}, {Forman}, {Jones}, {Sunyaev}, and
  {B{\"o}hringer}}]{chu04}
{Churazov} E, {Forman} W, {Jones} C, {Sunyaev} R, {B{\"o}hringer} H (2004)
  {XMM-Newton observations of the Perseus cluster - II. Evidence for gas
  motions in the core}. \mnras 347:29--35,
  \doi{10.1111/j.1365-2966.2004.07201.x}, \eprint{arXiv:astro-ph/0309427}

\bibitem[{{De Young}(1992)}]{1992ApJ...386..464D}
{De Young} DS (1992) {Galaxy-driven turbulence and the growth of intracluster
  magnetic fields}. \apj 386:464--472, \doi{10.1086/171032}

\bibitem[{{Dedner} et~al(2002){Dedner}, {Kemm}, {Kr{\"o}ner}, {Munz},
  {Schnitzer}, and {Wesenberg}}]{ded02}
{Dedner} A, {Kemm} F, {Kr{\"o}ner} D, {Munz} CD, {Schnitzer} T, {Wesenberg} M
  (2002) {Hyperbolic Divergence Cleaning for the MHD Equations}. Journal of
  Computational Physics 175:645--673, \doi{10.1006/jcph.2001.6961}

\bibitem[{{Dennison}(1980)}]{den80}
{Dennison} B (1980) {Formation of radio halos in clusters of galaxies from
  cosmic-ray protons}. \apjl 239:L93--L96, \doi{10.1086/183300}

\bibitem[{{Dobbs} et~al(2016){Dobbs}, {Price}, {Pettitt}, {Bate}, and
  {Tricco}}]{2016MNRAS.461.4482D}
{Dobbs} CL, {Price} DJ, {Pettitt} AR, {Bate} MR, {Tricco} TS (2016) {Magnetic
  field evolution and reversals in spiral galaxies}. \mnras 461:4482--4495,
  \doi{10.1093/mnras/stw1625}, \eprint{1607.05532}

\bibitem[{{Dobler} et~al(2003){Dobler}, {Haugen}, {Yousef}, and
  {Brandenburg}}]{2003PhRvE..68b6304D}
{Dobler} W, {Haugen} NE, {Yousef} TA, {Brandenburg} A (2003) {Bottleneck effect
  in three-dimensional turbulence simulations}. \pre 68(2):026304,
  \doi{10.1103/PhysRevE.68.026304}, \eprint{astro-ph/0303324}

\bibitem[{{Dolag} and {En{\ss}lin}(2000)}]{2000A&A...362..151D}
{Dolag} K, {En{\ss}lin} TA (2000) {Radio halos of galaxy clusters from hadronic
  secondary electron injection in realistic magnetic field configurations}.
  \aap 362:151--157

\bibitem[{{Dolag} and {Schindler}(2000)}]{2000A&A...364..491D}
{Dolag} K, {Schindler} S (2000) {The effect of magnetic fields on the mass
  determination of clusters of galaxies}. \aap 364:491--496,
  \eprint{astro-ph/0010296}

\bibitem[{{Dolag} and {Stasyszyn}(2009)}]{2009MNRAS.398.1678D}
{Dolag} K, {Stasyszyn} F (2009) {An MHD GADGET for cosmological simulations}.
  \mnras 398:1678--1697, \doi{10.1111/j.1365-2966.2009.15181.x},
  \eprint{0807.3553}

\bibitem[{{Dolag} et~al(1999){Dolag}, {Bartelmann}, and
  {Lesch}}]{1999A&A...348..351D}
{Dolag} K, {Bartelmann} M, {Lesch} H (1999) {SPH simulations of magnetic fields
  in galaxy clusters}. \aap 348:351--363, \eprint{astro-ph/0202272}

\bibitem[{{Dolag} et~al(2001){Dolag}, {Evrard}, and
  {Bartelmann}}]{2001A&A...369...36D}
{Dolag} K, {Evrard} A, {Bartelmann} M (2001) {The temperature-mass relation in
  magnetized galaxy clusters}. \aap 369:36--41,
  \doi{10.1051/0004-6361:20010094}, \eprint{astro-ph/0101302}

\bibitem[{{Dolag} et~al(2002){Dolag}, {Bartelmann}, and
  {Lesch}}]{2002A&A...387..383D}
{Dolag} K, {Bartelmann} M, {Lesch} H (2002) {Evolution and structure of
  magnetic fields in simulated galaxy clusters}. \aap 387:383--395,
  \doi{10.1051/0004-6361:20020241}

\bibitem[{{Dolag} et~al(2005{\natexlab{a}}){Dolag}, {Grasso}, {Springel}, and
  {Tkachev}}]{2005JCAP...01..009D}
{Dolag} K, {Grasso} D, {Springel} V, {Tkachev} I (2005{\natexlab{a}})
  {Constrained simulations of the magnetic field in the local Universe and the
  propagation of ultrahigh energy cosmic rays}. \jcap 1:009,
  \doi{10.1088/1475-7516/2005/01/009}, \eprint{astro-ph/0410419}

\bibitem[{{Dolag} et~al(2005{\natexlab{b}}){Dolag}, {Vazza}, {Brunetti}, and
  {Tormen}}]{2005MNRAS.364..753D}
{Dolag} K, {Vazza} F, {Brunetti} G, {Tormen} G (2005{\natexlab{b}}) {Turbulent
  gas motions in galaxy cluster simulations: the role of smoothed particle
  hydrodynamics viscosity}. \mnras 364:753--772,
  \doi{10.1111/j.1365-2966.2005.09630.x}, \eprint{arXiv:astro-ph/0507480}

\bibitem[{{Dolag} et~al(2008){Dolag}, {Bykov}, and
  {Diaferio}}]{2008SSRv..134..311D}
{Dolag} K, {Bykov} AM, {Diaferio} A (2008) {Non-Thermal Processes in
  Cosmological Simulations}. \ssr 134:311--335,
  \doi{10.1007/s11214-008-9319-2}, \eprint{0801.1048}

\bibitem[{{Dolag} et~al(2009){Dolag}, {Kachelrie{\ss}}, {Ostapchenko}, and
  {Tom{\`a}s}}]{2009ApJ...703.1078D}
{Dolag} K, {Kachelrie{\ss}} M, {Ostapchenko} S, {Tom{\`a}s} R (2009) {Blazar
  Halos as Probe for Extragalactic Magnetic Fields and Maximal Acceleration
  Energy}. \apj 703:1078--1085, \doi{10.1088/0004-637X/703/1/1078},
  \eprint{0903.2842}

\bibitem[{{Dolag} et~al(2011){Dolag}, {Kachelriess}, {Ostapchenko}, and
  {Tom{\`a}s}}]{2011ApJ...727L...4D}
{Dolag} K, {Kachelriess} M, {Ostapchenko} S, {Tom{\`a}s} R (2011) {Lower Limit
  on the Strength and Filling Factor of Extragalactic Magnetic Fields}. \apjl
  727:L4, \doi{10.1088/2041-8205/727/1/L4}, \eprint{1009.1782}

\bibitem[{{Dolag} et~al(2015){Dolag}, {Gaensler}, {Beck}, and
  {Beck}}]{2015MNRAS.451.4277D}
{Dolag} K, {Gaensler} BM, {Beck} AM, {Beck} MC (2015) {Constraints on the
  distribution and energetics of fast radio bursts using cosmological
  hydrodynamic simulations}. \mnras 451:4277--4289,
  \doi{10.1093/mnras/stv1190}, \eprint{1412.4829}

\bibitem[{{Donnert} and {Brunetti}(2014)}]{2014MNRAS.443.3564D}
{Donnert} J, {Brunetti} G (2014) {An efficient Fokker-Planck solver and its
  application to stochastic particle acceleration in galaxy clusters}. \mnras
  443:3564--3577, \doi{10.1093/mnras/stu1417}, \eprint{1407.2735}

\bibitem[{{Donnert} et~al(2009){Donnert}, {Dolag}, {Lesch}, and
  {M{\"u}ller}}]{donn09}
{Donnert} J, {Dolag} K, {Lesch} H, {M{\"u}ller} E (2009) {Cluster magnetic
  fields from galactic outflows}. \mnras 392:1008--1021,
  \doi{10.1111/j.1365-2966.2008.14132.x}, \eprint{0808.0919}

\bibitem[{{Donnert} et~al(2010){Donnert}, {Dolag}, {Brunetti}, {Cassano}, and
  {Bonafede}}]{2010MNRAS.401...47D}
{Donnert} J, {Dolag} K, {Brunetti} G, {Cassano} R, {Bonafede} A (2010) {Radio
  haloes from simulations and hadronic models - I. The Coma cluster}. \mnras
  401:47--54, \doi{10.1111/j.1365-2966.2009.15655.x}, \eprint{0905.2418}

\bibitem[{{Donnert} et~al(2016){Donnert}, {Stroe}, {Brunetti}, {Hoang}, and
  {Roettgering}}]{2016MNRAS.462.2014D}
{Donnert} JMF, {Stroe} A, {Brunetti} G, {Hoang} D, {Roettgering} H (2016)
  {Magnetic field evolution in giant radio relics using the example of CIZA
  J2242.8+5301}. \mnras 462:2014--2032, \doi{10.1093/mnras/stw1792},
  \eprint{1603.06570}

\bibitem[{{Donnert} et~al(2017){Donnert}, {Beck}, {Dolag}, and
  {R{\"o}ttgering}}]{2017MNRAS.471.4587D}
{Donnert} JMF, {Beck} AM, {Dolag} K, {R{\"o}ttgering} HJA (2017) {Simulations
  of the galaxy cluster CIZA J2242.8+5301 - I. Thermal model and shock
  properties}. \mnras 471:4587--4605, \doi{10.1093/mnras/stx1819},
  \eprint{1703.05682}

\bibitem[{{Drury} and {Downes}(2012)}]{2012MNRAS.427.2308D}
{Drury} LO, {Downes} TP (2012) {Turbulent magnetic field amplification driven
  by cosmic ray pressure gradients}. \mnras 427:2308--2313,
  \doi{10.1111/j.1365-2966.2012.22106.x}, \eprint{1205.6823}

\bibitem[{{Dubois} and {Teyssier}(2008)}]{2008A&A...482L..13D}
{Dubois} Y, {Teyssier} R (2008) {Cosmological MHD simulation of a cooling flow
  cluster}. \aap 482:L13--L16, \doi{10.1051/0004-6361:200809513},
  \eprint{0802.0490}

\bibitem[{{Dubois} et~al(2009){Dubois}, {Devriendt}, {Slyz}, and
  {Silk}}]{2009MNRAS.399L..49D}
{Dubois} Y, {Devriendt} J, {Slyz} A, {Silk} J (2009) {Influence of AGN jets on
  the magnetized ICM}. \mnras 399:L49--L53,
  \doi{10.1111/j.1745-3933.2009.00721.x}, \eprint{0905.3345}

\bibitem[{{Durrer} and {Neronov}(2013)}]{2013A&ARv..21...62D}
{Durrer} R, {Neronov} A (2013) {Cosmological magnetic fields: their generation,
  evolution and observation}. \aapr 21:62, \doi{10.1007/s00159-013-0062-7},
  \eprint{1303.7121}

\bibitem[{{Dursi}(2007)}]{dursi2007}
{Dursi} LJ (2007) {Bubble Wrap for Bullets: The Stability Imparted by a Thin
  Magnetic Layer}. \apj 670:221--230, \doi{10.1086/521997}, \eprint{0706.3216}

\bibitem[{{Dursi} and {Pfrommer}(2008)}]{dursi2008}
{Dursi} LJ, {Pfrommer} C (2008) {Draping of Cluster Magnetic Fields over
  Bullets and Bubbles---Morphology and Dynamic Effects}. \apj 677:993-1018,
  \doi{10.1086/529371}, \eprint{0711.0213}

\bibitem[{{En{\ss}lin} et~al(2011){En{\ss}lin}, {Pfrommer}, {Miniati}, and
  {Subramanian}}]{ens11}
{En{\ss}lin} T, {Pfrommer} C, {Miniati} F, {Subramanian} K (2011) {Cosmic ray
  transport in galaxy clusters: implications for radio halos, gamma-ray
  signatures, and cool core heating}. \aap 527:A99,
  \doi{10.1051/0004-6361/201015652}, \eprint{1008.4717}

\bibitem[{{En{\ss}lin} and {Vogt}(2006)}]{2006A&A...453..447E}
{En{\ss}lin} TA, {Vogt} C (2006) {Magnetic turbulence in cool cores of galaxy
  clusters}. \aap 453:447--458, \doi{10.1051/0004-6361:20053518},
  \eprint{astro-ph/0505517}

\bibitem[{{Ettori} et~al(2013){Ettori}, {Pratt}, {de Plaa}, {Eckert},
  {Nevalainen}, {Battistelli}, {Borgani}, {Croston}, {Finoguenov}, {Kaastra},
  {Gaspari}, {Gastaldello}, {Gitti}, {Molendi}, {Pointecouteau}, {Ponman},
  {Reiprich}, {Roncarelli}, {Rossetti}, {Sanders}, {Sun}, {Trinchieri},
  {Vazza}, {Arnaud}, {B{\"o}ringher}, {Brighenti}, {Dahle}, {De Grandi},
  {Mohr}, {Moretti}, and {Schindler}}]{2013arXiv1306.2322E}
{Ettori} S, {Pratt} GW, {de Plaa} J, {Eckert} D, {Nevalainen} J, {Battistelli}
  ES, {Borgani} S, {Croston} JH, {Finoguenov} A, {Kaastra} J, {Gaspari} M,
  {Gastaldello} F, {Gitti} M, {Molendi} S, {Pointecouteau} E, {Ponman} TJ,
  {Reiprich} TH, {Roncarelli} M, {Rossetti} M, {Sanders} JS, {Sun} M,
  {Trinchieri} G, {Vazza} F, {Arnaud} M, {B{\"o}ringher} H, {Brighenti} F,
  {Dahle} H, {De Grandi} S, {Mohr} JJ, {Moretti} A, {Schindler} S (2013) {The
  Hot and Energetic Universe: The astrophysics of galaxy groups and clusters}.
  ArXiv e-prints \eprint{1306.2322}

\bibitem[{{Evrard}(1990)}]{1990ApJ...363..349E}
{Evrard} AE (1990) {Formation and evolution of X-ray clusters - A hydrodynamic
  simulation of the intracluster medium}. \apj 363:349--366,
  \doi{10.1086/169350}

\bibitem[{{Federrath}(2016)}]{2016JPlPh..82f5301F}
{Federrath} C (2016) {Magnetic field amplification in turbulent astrophysical
  plasmas}. Journal of Plasma Physics 82(6):535820601,
  \doi{10.1017/S0022377816001069}, \eprint{1610.08132}

\bibitem[{{Federrath} et~al(2011{\natexlab{a}}){Federrath}, {Chabrier},
  {S2015aska.confE..95Bber}, {Banerjee}, {Klessen}, and
  {Schleicher}}]{2011PhRvL.107k4504F}
{Federrath} C, {Chabrier} G, {S2015askaconfE95Bber} J, {Banerjee} R, {Klessen}
  RS, {Schleicher} DRG (2011{\natexlab{a}}) {Mach Number Dependence of
  Turbulent Magnetic Field Amplification: Solenoidal versus Compressive Flows}.
  Physical Review Letters 107(11):114504, \doi{10.1103/PhysRevLett.107.114504},
  \eprint{1109.1760}

\bibitem[{{Federrath} et~al(2011{\natexlab{b}}){Federrath}, {Sur},
  {Schleicher}, {Banerjee}, and {Klessen}}]{2011ApJ...731...62F}
{Federrath} C, {Sur} S, {Schleicher} DRG, {Banerjee} R, {Klessen} RS
  (2011{\natexlab{b}}) {A New Jeans Resolution Criterion for (M)HD Simulations
  of Self-gravitating Gas: Application to Magnetic Field Amplification by
  Gravity-driven Turbulence}. \apj 731:62, \doi{10.1088/0004-637X/731/1/62},
  \eprint{1102.0266}

\bibitem[{{Federrath} et~al(2014){Federrath}, {Schober}, {Bovino}, and
  {Schleicher}}]{fed14}
{Federrath} C, {Schober} J, {Bovino} S, {Schleicher} DRG (2014) {The Turbulent
  Dynamo in Highly Compressible Supersonic Plasmas}. \apjl 797:L19,
  \doi{10.1088/2041-8205/797/2/L19}, \eprint{1411.4707}

\bibitem[{{Ferrari} et~al(2008){Ferrari}, {Govoni}, {Schindler}, {Bykov}, and
  {Rephaeli}}]{2008SSRv..134...93F}
{Ferrari} C, {Govoni} F, {Schindler} S, {Bykov} AM, {Rephaeli} Y (2008)
  {Observations of Extended Radio Emission in Clusters}. \ssr 134:93--118,
  \doi{10.1007/s11214-008-9311-x}, \eprint{0801.0985}

\bibitem[{{Finoguenov} et~al(2010){Finoguenov}, {Sarazin}, {Nakazawa}, {Wik},
  and {Clarke}}]{2010ApJ...715.1143F}
{Finoguenov} A, {Sarazin} CL, {Nakazawa} K, {Wik} DR, {Clarke} TE (2010)
  {XMM-Newton Observation of the Northwest Radio Relic Region in A3667}. \apj
  715:1143--1151, \doi{10.1088/0004-637X/715/2/1143}, \eprint{1004.2331}

\bibitem[{{Fraschetti}(2013)}]{2013ApJ...770...84F}
{Fraschetti} F (2013) {Turbulent Amplification of a Magnetic Field Driven by
  the Dynamo Effect at Rippled Shocks}. \apj 770:84,
  \doi{10.1088/0004-637X/770/2/84}, \eprint{1304.4956}

\bibitem[{{Furlanetto} and {Loeb}(2001)}]{2001ApJ...556..619F}
{Furlanetto} SR, {Loeb} A (2001) {Intergalactic Magnetic Fields from Quasar
  Outflows}. \apj 556:619--634, \doi{10.1086/321630}, \eprint{astro-ph/0102076}

\bibitem[{{Gaensler} et~al(2010){Gaensler}, {Landecker}, {Taylor}, and {POSSUM
  Collaboration}}]{2010AAS...21547013G}
{Gaensler} BM, {Landecker} TL, {Taylor} AR, {POSSUM Collaboration} (2010)
  {Survey Science with ASKAP: Polarization Sky Survey of the Universe's
  Magnetism (POSSUM)}. In: American Astronomical Society Meeting Abstracts
  \#215, Bulletin of the American Astronomical Society, vol~42, p 515

\bibitem[{{Gardiner} and {Stone}(2008)}]{2008JCoPh.227.4123G}
{Gardiner} TA, {Stone} JM (2008) {An unsplit Godunov method for ideal MHD via
  constrained transport in three dimensions}. Journal of Computational Physics
  227:4123--4141, \doi{10.1016/j.jcp.2007.12.017}, \eprint{0712.2634}

\bibitem[{{Gaspari} et~al(2011){Gaspari}, {Melioli}, {Brighenti}, and
  {D'Ercole}}]{gaspari11a}
{Gaspari} M, {Melioli} C, {Brighenti} F, {D'Ercole} A (2011) {The dance of
  heating and cooling in galaxy clusters: three-dimensional simulations of
  self-regulated active galactic nuclei outflows}. \mnras 411:349--372,
  \doi{10.1111/j.1365-2966.2010.17688.x}, \eprint{1007.0674}

\bibitem[{{Gendron-Marsolais} et~al(2017){Gendron-Marsolais},
  {Hlavacek-Larrondo}, {van Weeren}, {Clarke}, {Fabian}, {Intema}, {Taylor},
  {Blundell}, and {Sanders}}]{mgm17}
{Gendron-Marsolais} M, {Hlavacek-Larrondo} J, {van Weeren} RJ, {Clarke} T,
  {Fabian} AC, {Intema} HT, {Taylor} GB, {Blundell} KM, {Sanders} JS (2017)
  {Deep 230-470 MHz VLA observations of the mini-halo in the Perseus cluster}.
  \mnras 469:3872--3880, \doi{10.1093/mnras/stx1042}, \eprint{1701.03791}

\bibitem[{{Gheller} et~al(2016){Gheller}, {Vazza}, {Br{\"u}ggen}, {Alpaslan},
  {Holwerda}, {Hopkins}, and {Liske}}]{2016MNRAS.462..448G}
{Gheller} C, {Vazza} F, {Br{\"u}ggen} M, {Alpaslan} M, {Holwerda} BW, {Hopkins}
  AM, {Liske} J (2016) {Evolution of cosmic filaments and of their galaxy
  population from MHD cosmological simulations}. \mnras 462:448--463,
  \doi{10.1093/mnras/stw1595}, \eprint{1607.01406}

\bibitem[{{Giacalone} and {Jokipii}(2007)}]{2007ApJ...663L..41G}
{Giacalone} J, {Jokipii} JR (2007) {Magnetic Field Amplification by Shocks in
  Turbulent Fluids}. \apjl 663:L41--L44, \doi{10.1086/519994}

\bibitem[{{Giacintucci} et~al(2014{\natexlab{a}}){Giacintucci}, {Markevitch},
  {Brunetti}, {ZuHone}, {Venturi}, {Mazzotta}, and {Bourdin}}]{gia14b}
{Giacintucci} S, {Markevitch} M, {Brunetti} G, {ZuHone} JA, {Venturi} T,
  {Mazzotta} P, {Bourdin} H (2014{\natexlab{a}}) {Mapping the Particle
  Acceleration in the Cool Core of the Galaxy Cluster RX J1720.1+2638}. \apj
  795:73, \doi{10.1088/0004-637X/795/1/73}, \eprint{1403.2820}

\bibitem[{{Giacintucci} et~al(2014{\natexlab{b}}){Giacintucci}, {Markevitch},
  {Venturi}, {Clarke}, {Cassano}, and {Mazzotta}}]{gia14a}
{Giacintucci} S, {Markevitch} M, {Venturi} T, {Clarke} TE, {Cassano} R,
  {Mazzotta} P (2014{\natexlab{b}}) {New Detections of Radio Minihalos in Cool
  Cores of Galaxy Clusters}. \apj 781:9, \doi{10.1088/0004-637X/781/1/9},
  \eprint{1311.5248}

\bibitem[{{Giacintucci} et~al(2017){Giacintucci}, {Markevitch}, {Cassano},
  {Venturi}, {Clarke}, and {Brunetti}}]{gia17}
{Giacintucci} S, {Markevitch} M, {Cassano} R, {Venturi} T, {Clarke} TE,
  {Brunetti} G (2017) {Occurrence of Radio Minihalos in a Mass-limited Sample
  of Galaxy Clusters}. \apj 841:71, \doi{10.3847/1538-4357/aa7069},
  \eprint{1701.01364}

\bibitem[{{Goldreich} and {Sridhar}(1997)}]{1997ApJ...485..680G}
{Goldreich} P, {Sridhar} S (1997) {Magnetohydrodynamic Turbulence Revisited}.
  \apj 485:680--688, \doi{10.1086/304442}, \eprint{astro-ph/9612243}

\bibitem[{{Goldshmidt} and {Rephaeli}(1993)}]{1993ApJ...411..518G}
{Goldshmidt} O, {Rephaeli} Y (1993) {Turbulent generation of intracluster
  magnetic fields and Faraday rotation measurements}. \apj 411:518--528,
  \doi{10.1086/172853}

\bibitem[{{Govoni} et~al(2010){Govoni}, {Dolag}, {Murgia}, {Feretti},
  {Schindler}, {Giovannini}, {Boschin}, {Vacca}, and
  {Bonafede}}]{2010A&A...522A.105G}
{Govoni} F, {Dolag} K, {Murgia} M, {Feretti} L, {Schindler} S, {Giovannini} G,
  {Boschin} W, {Vacca} V, {Bonafede} A (2010) {Rotation measures of radio
  sources in hot galaxy clusters}. \aap 522:A105,
  \doi{10.1051/0004-6361/200913665}, \eprint{1007.5207}

\bibitem[{{Govoni} et~al(2014){Govoni}, {Johnston-Hollitt}, {Agudo}, {Akahori},
  {Beck}, {Bonafede}, {Carozzi}, {Colafrancesco}, {Feretti}, {Ferriere},
  {Gaensler}, {Harvey-Smith}, {Haverkorn}, {Heald}, {Mao}, {Rudnick},
  {Schnitzeler}, {Scaife}, {Stil}, {Takahashi}, {Taylor}, and
  {Wucknitz}}]{2014skao.rept.....G}
{Govoni} F, {Johnston-Hollitt} M, {Agudo} I, {Akahori} T, {Beck} R, {Bonafede}
  A, {Carozzi} TD, {Colafrancesco} S, {Feretti} L, {Ferriere} K, {Gaensler} BM,
  {Harvey-Smith} L, {Haverkorn} M, {Heald} GH, {Mao} SA, {Rudnick} L,
  {Schnitzeler} D, {Scaife} A, {Stil} JM, {Takahashi} K, {Taylor} AR,
  {Wucknitz} O (2014) {Cosmic Magnetism Science in the SKA1 Era}. Tech. rep.

\bibitem[{{Govoni} et~al(2017){Govoni}, {Murgia}, {Vacca}, {Loi}, {Girardi},
  {Gastaldello}, {Giovannini}, {Feretti}, {Paladino}, {Carretti}, {Concu},
  {Melis}, {Poppi}, {Valente}, {Bernardi}, {Bonafede}, {Boschin}, {Brienza},
  {Clarke}, {Colafrancesco}, {de Gasperin}, {Eckert}, {En{\ss}lin}, {Ferrari},
  {Gregorini}, {Johnston-Hollitt}, {Junklewitz}, {Orr{\`u}}, {Parma}, {Perley},
  {Rossetti}, {B Taylor}, and {Vazza}}]{2017A&A...603A.122G}
{Govoni} F, {Murgia} M, {Vacca} V, {Loi} F, {Girardi} M, {Gastaldello} F,
  {Giovannini} G, {Feretti} L, {Paladino} R, {Carretti} E, {Concu} R, {Melis}
  A, {Poppi} S, {Valente} G, {Bernardi} G, {Bonafede} A, {Boschin} W, {Brienza}
  M, {Clarke} TE, {Colafrancesco} S, {de Gasperin} F, {Eckert} D, {En{\ss}lin}
  TA, {Ferrari} C, {Gregorini} L, {Johnston-Hollitt} M, {Junklewitz} H,
  {Orr{\`u}} E, {Parma} P, {Perley} R, {Rossetti} M, {B Taylor} G, {Vazza} F
  (2017) {Sardinia Radio Telescope observations of Abell 194. The intra-cluster
  magnetic field power spectrum}. \aap 603:A122,
  \doi{10.1051/0004-6361/201630349}, \eprint{1703.08688}

\bibitem[{{Guo} et~al(2012){Guo}, {Li}, {Li}, {Giacalone}, {Jokipii}, and
  {Li}}]{2012ApJ...747...98G}
{Guo} F, {Li} S, {Li} H, {Giacalone} J, {Jokipii} JR, {Li} D (2012) {On the
  Amplification of Magnetic Field by a Supernova Blast Shock Wave in a
  Turbulent Medium}. \apj 747:98, \doi{10.1088/0004-637X/747/2/98},
  \eprint{1112.6373}

\bibitem[{{Guo} et~al(2016){Guo}, {Li}, {Li}, {Daughton}, {Zhang},
  {Lloyd-Ronning}, {Liu}, {Zhang}, and {Deng}}]{2016ApJ...818L...9G}
{Guo} F, {Li} X, {Li} H, {Daughton} W, {Zhang} B, {Lloyd-Ronning} N, {Liu} YH,
  {Zhang} H, {Deng} W (2016) {Efficient Production of High-energy Nonthermal
  Particles during Magnetic Reconnection in a Magnetically Dominated
  Ion-Electron Plasma}. \apjl 818:L9, \doi{10.3847/2041-8205/818/1/L9},
  \eprint{1511.01434}

\bibitem[{{Harrison}(1973)}]{1973MNRAS.165..185H}
{Harrison} ER (1973) {Magnetic fields in the early Universe}. \mnras 165:185,
  \doi{10.1093/mnras/165.2.185}

\bibitem[{{Haugen} and {Brandenburg}(2004)}]{2004PhRvE..70c6408H}
{Haugen} NEL, {Brandenburg} A (2004) {Suppression of small scale dynamo action
  by an imposed magnetic field}. \pre 70(3):036408,
  \doi{10.1103/PhysRevE.70.036408}, \eprint{astro-ph/0402281}

\bibitem[{{Haugen} et~al(2004){Haugen}, {Brandenburg}, and
  {Mee}}]{2004MNRAS.353..947H}
{Haugen} NEL, {Brandenburg} A, {Mee} AJ (2004) {Mach number dependence of the
  onset of dynamo action}. \mnras 353:947--952,
  \doi{10.1111/j.1365-2966.2004.08127.x}, \eprint{astro-ph/0405453}

\bibitem[{{Hitomi Collaboration} et~al(2016){Hitomi Collaboration},
  {Aharonian}, {Akamatsu}, {Akimoto}, {Allen}, {Anabuki}, {Angelini}, {Arnaud},
  {Audard}, {Awaki}, {Axelsson}, {Bamba}, {Bautz}, {Blandford}, {Brenneman},
  {Brown}, {Bulbul}, {Cackett}, {Chernyakova}, {Chiao}, {Coppi}, {Costantini},
  {de Plaa}, {den Herder}, {Done}, {Dotani}, {Ebisawa}, {Eckart}, {Enoto},
  {Ezoe}, {Fabian}, {Ferrigno}, {Foster}, {Fujimoto}, {Fukazawa}, {Furuzawa},
  {Galeazzi}, {Gallo}, {Gandhi}, {Giustini}, {Goldwurm}, {Gu}, {Guainazzi},
  {Haba}, {Hagino}, {Hamaguchi}, {Harrus}, {Hatsukade}, {Hayashi}, {Hayashi},
  {Hayashida}, {Hiraga}, {Hornschemeier}, {Hoshino}, {Hughes}, {Iizuka},
  {Inoue}, {Inoue}, {Ishibashi}, {Ishida}, {Ishikawa}, {Ishisaki}, {Itoh},
  {Iyomoto}, {Kaastra}, {Kallman}, {Kamae}, {Kara}, {Kataoka}, {Katsuda},
  {Katsuta}, {Kawaharada}, {Kawai}, {Kelley}, {Khangulyan}, {Kilbourne},
  {King}, {Kitaguchi}, {Kitamoto}, {Kitayama}, {Kohmura}, {Kokubun}, {Koyama},
  {Koyama}, {Kretschmar}, {Krimm}, {Kubota}, {Kunieda}, {Laurent}, {Lebrun},
  {Lee}, {Leutenegger}, {Limousin}, {Loewenstein}, {Long}, {Lumb}, {Madejski},
  {Maeda}, {Maier}, {Makishima}, {Markevitch}, {Matsumoto}, {Matsushita},
  {McCammon}, {McNamara}, {Mehdipour}, {Miller}, {Miller}, {Mineshige},
  {Mitsuda}, {Mitsuishi}, {Miyazawa}, {Mizuno}, {Mori}, {Mori}, {Moseley},
  {Mukai}, {Murakami}, {Murakami}, {Mushotzky}, {Nagino}, {Nakagawa},
  {Nakajima}, {Nakamori}, {Nakano}, {Nakashima}, {Nakazawa}, {Nobukawa},
  {Noda}, {Nomachi}, {O'Dell}, {Odaka}, {Ohashi}, {Ohno}, {Okajima}, {Ota},
  {Ozaki}, {Paerels}, {Paltani}, {Parmar}, {Petre}, {Pinto}, {Pohl}, {Porter},
  {Pottschmidt}, {Ramsey}, {Reynolds}, {Russell}, {Safi-Harb}, {Saito},
  {Sakai}, {Sameshima}, {Sato}, {Sato}, {Sato}, {Sawada}, {Schartel},
  {Serlemitsos}, {Seta}, {Shidatsu}, {Simionescu}, {Smith}, {Soong}, {Stawarz},
  {Sugawara}, {Sugita}, {Szymkowiak}, {Tajima}, {Takahashi}, {Takahashi},
  {Takeda}, {Takei}, {Tamagawa}, {Tamura}, {Tamura}, {Tanaka}, {Tanaka},
  {Tanaka}, {Tashiro}, {Tawara}, {Terada}, {Terashima}, {Tombesi}, {Tomida},
  {Tsuboi}, {Tsujimoto}, {Tsunemi}, {Tsuru}, {Uchida}, {Uchiyama}, {Uchiyama},
  {Ueda}, {Ueda}, {Ueno}, {Uno}, {Urry}, {Ursino}, {de Vries}, {Watanabe},
  {Werner}, {Wik}, {Wilkins}, {Williams}, {Yamada}, {Yamaguchi}, {Yamaoka},
  {Yamasaki}, {Yamauchi}, {Yamauchi}, {Yaqoob}, {Yatsu}, {Yonetoku}, {Yoshida},
  {Yuasa}, {Zhuravleva}, and {Zoghbi}}]{2016Natur.535..117H}
{Hitomi Collaboration}, {Aharonian} F, {Akamatsu} H, {Akimoto} F, {Allen} SW,
  {Anabuki} N, {Angelini} L, {Arnaud} K, {Audard} M, {Awaki} H, {Axelsson} M,
  {Bamba} A, {Bautz} M, {Blandford} R, {Brenneman} L, {Brown} GV, {Bulbul} E,
  {Cackett} E, {Chernyakova} M, {Chiao} M, {Coppi} P, {Costantini} E, {de Plaa}
  J, {den Herder} JW, {Done} C, {Dotani} T, {Ebisawa} K, {Eckart} M, {Enoto} T,
  {Ezoe} Y, {Fabian} AC, {Ferrigno} C, {Foster} A, {Fujimoto} R, {Fukazawa} Y,
  {Furuzawa} A, {Galeazzi} M, {Gallo} L, {Gandhi} P, {Giustini} M, {Goldwurm}
  A, {Gu} L, {Guainazzi} M, {Haba} Y, {Hagino} K, {Hamaguchi} K, {Harrus} I,
  {Hatsukade} I, {Hayashi} K, {Hayashi} T, {Hayashida} K, {Hiraga} J,
  {Hornschemeier} A, {Hoshino} A, {Hughes} J, {Iizuka} R, {Inoue} H, {Inoue} Y,
  {Ishibashi} K, {Ishida} M, {Ishikawa} K, {Ishisaki} Y, {Itoh} M, {Iyomoto} N,
  {Kaastra} J, {Kallman} T, {Kamae} T, {Kara} E, {Kataoka} J, {Katsuda} S,
  {Katsuta} J, {Kawaharada} M, {Kawai} N, {Kelley} R, {Khangulyan} D,
  {Kilbourne} C, {King} A, {Kitaguchi} T, {Kitamoto} S, {Kitayama} T, {Kohmura}
  T, {Kokubun} M, {Koyama} S, {Koyama} K, {Kretschmar} P, {Krimm} H, {Kubota}
  A, {Kunieda} H, {Laurent} P, {Lebrun} F, {Lee} SH, {Leutenegger} M,
  {Limousin} O, {Loewenstein} M, {Long} KS, {Lumb} D, {Madejski} G, {Maeda} Y,
  {Maier} D, {Makishima} K, {Markevitch} M, {Matsumoto} H, {Matsushita} K,
  {McCammon} D, {McNamara} B, {Mehdipour} M, {Miller} E, {Miller} J,
  {Mineshige} S, {Mitsuda} K, {Mitsuishi} I, {Miyazawa} T, {Mizuno} T, {Mori}
  H, {Mori} K, {Moseley} H, {Mukai} K, {Murakami} H, {Murakami} T, {Mushotzky}
  R, {Nagino} R, {Nakagawa} T, {Nakajima} H, {Nakamori} T, {Nakano} T,
  {Nakashima} S, {Nakazawa} K, {Nobukawa} M, {Noda} H, {Nomachi} M, {O'Dell} S,
  {Odaka} H, {Ohashi} T, {Ohno} M, {Okajima} T, {Ota} N, {Ozaki} M, {Paerels}
  F, {Paltani} S, {Parmar} A, {Petre} R, {Pinto} C, {Pohl} M, {Porter} FS,
  {Pottschmidt} K, {Ramsey} B, {Reynolds} C, {Russell} H, {Safi-Harb} S,
  {Saito} S, {Sakai} K, {Sameshima} H, {Sato} G, {Sato} K, {Sato} R, {Sawada}
  M, {Schartel} N, {Serlemitsos} P, {Seta} H, {Shidatsu} M, {Simionescu} A,
  {Smith} R, {Soong} Y, {Stawarz} L, {Sugawara} Y, {Sugita} S, {Szymkowiak} A,
  {Tajima} H, {Takahashi} H, {Takahashi} T, {Takeda} S, {Takei} Y, {Tamagawa}
  T, {Tamura} K, {Tamura} T, {Tanaka} T, {Tanaka} Y, {Tanaka} Y, {Tashiro} M,
  {Tawara} Y, {Terada} Y, {Terashima} Y, {Tombesi} F, {Tomida} H, {Tsuboi} Y,
  {Tsujimoto} M, {Tsunemi} H, {Tsuru} T, {Uchida} H, {Uchiyama} H, {Uchiyama}
  Y, {Ueda} S, {Ueda} Y, {Ueno} S, {Uno} S, {Urry} M, {Ursino} E, {de Vries} C,
  {Watanabe} S, {Werner} N, {Wik} D, {Wilkins} D, {Williams} B, {Yamada} S,
  {Yamaguchi} H, {Yamaoka} K, {Yamasaki} NY, {Yamauchi} M, {Yamauchi} S,
  {Yaqoob} T, {Yatsu} Y, {Yonetoku} D, {Yoshida} A, {Yuasa} T, {Zhuravleva} I,
  {Zoghbi} A (2016) {The quiescent intracluster medium in the core of the
  Perseus cluster}. \nat 535:117--121, \doi{10.1038/nature18627},
  \eprint{1607.04487}

\bibitem[{{Hoang} et~al(2017){Hoang}, {Shimwell}, {Stroe}, {Akamatsu},
  {Brunetti}, {Donnert}, {Intema}, {Mulcahy}, {R{\"o}ttgering}, {van Weeren},
  {Bonafede}, {Br{\"u}ggen}, {Cassano}, {Chyzy}, {En{\ss}lin}, {Ferrari}, {de
  Gasperin}, {Gu}, {Hoeft}, {Miley}, {Orr{\'u}}, {Pizzo}, and
  {White}}]{2017MNRAS.471.1107H}
{Hoang} DN, {Shimwell} TW, {Stroe} A, {Akamatsu} H, {Brunetti} G, {Donnert}
  JMF, {Intema} HT, {Mulcahy} DD, {R{\"o}ttgering} HJA, {van Weeren} RJ,
  {Bonafede} A, {Br{\"u}ggen} M, {Cassano} R, {Chyzy} KT, {En{\ss}lin} T,
  {Ferrari} C, {de Gasperin} F, {Gu} L, {Hoeft} M, {Miley} GK, {Orr{\'u}} E,
  {Pizzo} R, {White} GJ (2017) {Deep LOFAR observations of the merging galaxy
  cluster CIZA J2242.8+5301}. \mnras 471:1107--1125,
  \doi{10.1093/mnras/stx1645}, \eprint{1706.09903}

\bibitem[{{Hopkins}(2015)}]{2015MNRAS.450...53H}
{Hopkins} PF (2015) {A new class of accurate, mesh-free hydrodynamic simulation
  methods}. \mnras 450:53--110, \doi{10.1093/mnras/stv195}, \eprint{1409.7395}

\bibitem[{{Hopkins} and {Raives}(2016)}]{hop16}
{Hopkins} PF, {Raives} MJ (2016) {Accurate, meshless methods for
  magnetohydrodynamics}. \mnras 455:51--88, \doi{10.1093/mnras/stv2180},
  \eprint{1505.02783}

\bibitem[{{Hutschenreuter} et~al(2018){Hutschenreuter}, {Dorn}, {Jasche},
  {Vazza}, {Paoletti}, {Lavaux}, and {En{\ss}lin}}]{2018arXiv180302629H}
{Hutschenreuter} S, {Dorn} S, {Jasche} J, {Vazza} F, {Paoletti} D, {Lavaux} G,
  {En{\ss}lin} TA (2018) {The primordial magnetic field in our cosmic
  backyard}. ArXiv e-prints \eprint{1803.02629}

\bibitem[{{Iapichino} and {Br{\"u}ggen}(2012)}]{2012MNRAS.423.2781I}
{Iapichino} L, {Br{\"u}ggen} M (2012) {Magnetic field amplification by shocks
  in galaxy clusters: application to radio relics}. \mnras 423:2781--2788,
  \doi{10.1111/j.1365-2966.2012.21084.x}, \eprint{1204.2455}

\bibitem[{{Iapichino} et~al(2011){Iapichino}, {Schmidt}, {Niemeyer}, and
  {Merklein}}]{iapichino11}
{Iapichino} L, {Schmidt} W, {Niemeyer} JC, {Merklein} J (2011) {Turbulence
  production and turbulent pressure support in the intergalactic medium}.
  \mnras 414:2297--2308, \doi{10.1111/j.1365-2966.2011.18550.x},
  \eprint{1102.3352}

\bibitem[{{Iapichino} et~al(2017){Iapichino}, {Federrath}, and
  {Klessen}}]{2017MNRAS.469.3641I}
{Iapichino} L, {Federrath} C, {Klessen} RS (2017) {Adaptive mesh refinement
  simulations of a galaxy cluster merger - I. Resolving and modelling the
  turbulent flow in the cluster outskirts}. \mnras 469:3641--3655,
  \doi{10.1093/mnras/stx882}, \eprint{1704.02922}

\bibitem[{{Ichinohe} et~al(2017){Ichinohe}, {Simionescu}, {Werner}, and
  {Takahashi}}]{ich2017}
{Ichinohe} Y, {Simionescu} A, {Werner} N, {Takahashi} T (2017) {An azimuthally
  resolved study of the cold front in Abell 3667}. \mnras 467:3662--3676,
  \doi{10.1093/mnras/stx280}, \eprint{1702.01026}

\bibitem[{{Inoue} et~al(2009){Inoue}, {Yamazaki}, and
  {Inutsuka}}]{2009ApJ...695..825I}
{Inoue} T, {Yamazaki} R, {Inutsuka} Si (2009) {Turbulence and Magnetic Field
  Amplification in Supernova Remnants: Interactions Between a Strong Shock Wave
  and Multiphase Interstellar Medium}. \apj 695:825--833,
  \doi{10.1088/0004-637X/695/2/825}, \eprint{0901.0486}

\bibitem[{{Jacob} and {Pfrommer}(2017{\natexlab{a}})}]{jac17a}
{Jacob} S, {Pfrommer} C (2017{\natexlab{a}}) {Cosmic ray heating in cool core
  clusters - I. Diversity of steady state solutions}. \mnras 467:1449--1477,
  \doi{10.1093/mnras/stx131}, \eprint{1609.06321}

\bibitem[{{Jacob} and {Pfrommer}(2017{\natexlab{b}})}]{jac17b}
{Jacob} S, {Pfrommer} C (2017{\natexlab{b}}) {Cosmic ray heating in cool core
  clusters - II. Self-regulation cycle and non-thermal emission}. \mnras
  467:1478--1495, \doi{10.1093/mnras/stx132}, \eprint{1609.06322}

\bibitem[{{Jaffe}(1980)}]{1980ApJ...241..925J}
{Jaffe} W (1980) {On the morphology of the magnetic field in galaxy clusters}.
  \apj 241:925--927, \doi{10.1086/158407}

\bibitem[{{Ji} et~al(2016){Ji}, {Oh}, {Ruszkowski}, and
  {Markevitch}}]{2016MNRAS.463.3989J}
{Ji} S, {Oh} SP, {Ruszkowski} M, {Markevitch} M (2016) {The efficiency of
  magnetic field amplification at shocks by turbulence}. \mnras 463:3989--4003,
  \doi{10.1093/mnras/stw2320}, \eprint{1603.08518}

\bibitem[{{Jones} et~al(2011){Jones}, {Porter}, {Ryu}, and {Cho}}]{jones11}
{Jones} TW, {Porter} DH, {Ryu} D, {Cho} J (2011) {Cluster turbulence:
  simulation insights}. \memsai 82:588, \eprint{1101.4050}

\bibitem[{{Kahniashvili} et~al(2010){Kahniashvili}, {Tevzadze}, {Sethi},
  {Pandey}, and {Ratra}}]{2010PhRvD..82h3005K}
{Kahniashvili} T, {Tevzadze} AG, {Sethi} SK, {Pandey} K, {Ratra} B (2010)
  {Primordial magnetic field limits from cosmological data}. \prd 82(8):083005,
  \doi{10.1103/PhysRevD.82.083005}, \eprint{1009.2094}

\bibitem[{{Kahniashvili} et~al(2011){Kahniashvili}, {Tevzadze}, and
  {Ratra}}]{2011ApJ...726...78K}
{Kahniashvili} T, {Tevzadze} AG, {Ratra} B (2011) {Phase Transition Generated
  Cosmological Magnetic Field at Large Scales}. \apj 726:78,
  \doi{10.1088/0004-637X/726/2/78}, \eprint{0907.0197}

\bibitem[{{Kahniashvili} et~al(2016){Kahniashvili}, {Brandenburg}, and
  {Tevzadze}}]{2016PhyS...91j4008K}
{Kahniashvili} T, {Brandenburg} A, {Tevzadze} AG (2016) {The evolution of
  primordial magnetic fields since their generation}. \physscr 91(10):104008,
  \doi{10.1088/0031-8949/91/10/104008}, \eprint{1507.00510}

\bibitem[{{Kang} et~al(2007){Kang}, {Ryu}, {Cen}, and {Ostriker}}]{ka07}
{Kang} H, {Ryu} D, {Cen} R, {Ostriker} JP (2007) \apj 669:729--740,
  \doi{10.1086/521717}, \eprint{arXiv:0704.1521}

\bibitem[{{Kazantsev}(1968)}]{1968JETP...26.1031K}
{Kazantsev} AP (1968) {Enhancement of a Magnetic Field by a Conducting Fluid}.
  Soviet Journal of Experimental and Theoretical Physics 26:1031

\bibitem[{{Keshet} and {Loeb}(2010)}]{kl2010}
{Keshet} U, {Loeb} A (2010) {Using Radio Halos and Minihalos to Measure the
  Distributions of Magnetic Fields and Cosmic Rays in Galaxy Clusters}. \apj
  722:737--749, \doi{10.1088/0004-637X/722/1/737}, \eprint{1003.1133}

\bibitem[{{Khatri} and {Gaspari}(2016)}]{2016MNRAS.463..655K}
{Khatri} R, {Gaspari} M (2016) {Thermal SZ fluctuations in the ICM: probing
  turbulence and thermodynamics in Coma cluster with Planck}. \mnras
  463:655--669, \doi{10.1093/mnras/stw2027}, \eprint{1604.03106}

\bibitem[{{Klein} and {Fletcher}(2015)}]{2015gimf.book.....K}
{Klein} U, {Fletcher} A (2015) {Galactic and Intergalactic Magnetic Fields}

\bibitem[{{Kolmogorov}(1941)}]{1941DoSSR..30..301K}
{Kolmogorov} A (1941) {The Local Structure of Turbulence in Incompressible
  Viscous Fluid for Very Large Reynolds' Numbers}. Akademiia Nauk SSSR Doklady
  30:301--305

\bibitem[{{Kolmogorov}(1991)}]{1991RSPSA.434....9K}
{Kolmogorov} AN (1991) {The local structure of turbulence in incompressible
  viscous fluid for very large Reynolds numbers}. Proceedings of the Royal
  Society of London Series A 434:9--13, \doi{10.1098/rspa.1991.0075}

\bibitem[{{Kraft} et~al(2017){Kraft}, {Roediger}, {Machacek}, {Forman},
  {Nulsen}, {Jones}, {Churazov}, {Randall}, {Su}, and {Sheardown}}]{kraft2017}
{Kraft} RP, {Roediger} E, {Machacek} M, {Forman} WR, {Nulsen} PEJ, {Jones} C,
  {Churazov} E, {Randall} S, {Su} Y, {Sheardown} A (2017) {Stripped Elliptical
  Galaxies as Probes of ICM Physics. III. Deep Chandra Observations of NGC
  4552: Measuring the Viscosity of the Intracluster Medium}. \apj 848:27,
  \doi{10.3847/1538-4357/aa8a6e}, \eprint{1709.02784}

\bibitem[{{Kraichnan} and {Nagarajan}(1967)}]{1967PhFl...10..859K}
{Kraichnan} RH, {Nagarajan} S (1967) {Growth of Turbulent Magnetic Fields}.
  Physics of Fluids 10:859--870, \doi{10.1063/1.1762201}

\bibitem[{{Kritsuk} et~al(2011){Kritsuk}, {Nordlund}, {Collins}, {Padoan},
  {Norman}, {Abel}, {Banerjee}, {Federrath}, {Flock}, {Lee}, {Li},
  {M{\"u}ller}, {Teyssier}, {Ustyugov}, {Vogel}, and
  {Xu}}]{2011ApJ...737...13K}
{Kritsuk} AG, {Nordlund} {\AA}, {Collins} D, {Padoan} P, {Norman} ML, {Abel} T,
  {Banerjee} R, {Federrath} C, {Flock} M, {Lee} D, {Li} PS, {M{\"u}ller} WC,
  {Teyssier} R, {Ustyugov} SD, {Vogel} C, {Xu} H (2011) {Comparing Numerical
  Methods for Isothermal Magnetized Supersonic Turbulence}. \apj 737:13,
  \doi{10.1088/0004-637X/737/1/13}, \eprint{1103.5525}

\bibitem[{{Kronberg} et~al(1999){Kronberg}, {Lesch}, and
  {Hopp}}]{Kronberg..1999ApJ}
{Kronberg} PP, {Lesch} H, {Hopp} U (1999) {Magnetization of the Intergalactic
  Medium by Primeval Galaxies}. \apj 511:56--64

\bibitem[{{Kuchar} and {En{\ss}lin}(2011)}]{2011A&A...529A..13K}
{Kuchar} P, {En{\ss}lin} TA (2011) {Magnetic power spectra from Faraday
  rotation maps. REALMAF and its use on Hydra A}. \aap 529:A13,
  \doi{10.1051/0004-6361/200913918}, \eprint{0912.3930}

\bibitem[{{Kulsrud} and {Anderson}(1992)}]{1992ApJ...396..606K}
{Kulsrud} RM, {Anderson} SW (1992) {The spectrum of random magnetic fields in
  the mean field dynamo theory of the Galactic magnetic field}. \apj
  396:606--630, \doi{10.1086/171743}

\bibitem[{{Kulsrud} and {Ostriker}(2006)}]{2006PhT....59a..58K}
{Kulsrud} RM, {Ostriker} EC (2006) {Plasma Physics for Astrophysics}. Physics
  Today 59(1):58, \doi{10.1063/1.2180179}

\bibitem[{{Kulsrud} et~al(1997){Kulsrud}, {Cen}, {Ostriker}, and
  {Ryu}}]{1997ApJ...480..481K}
{Kulsrud} RM, {Cen} R, {Ostriker} JP, {Ryu} D (1997) {The Protogalactic Origin
  for Cosmic Magnetic Fields}. \apj 480:481--491, \doi{10.1086/303987},
  \eprint{astro-ph/9607141}

\bibitem[{{Kunz}(2011)}]{2011MNRAS.417..602K}
{Kunz} MW (2011) {Dynamical stability of a thermally stratified intracluster
  medium with anisotropic momentum and heat transport}. \mnras 417:602--616,
  \doi{10.1111/j.1365-2966.2011.19303.x}, \eprint{1104.3595}

\bibitem[{{Kunz} et~al(2014){Kunz}, {Schekochihin}, and
  {Stone}}]{2014PhRvL.112t5003K}
{Kunz} MW, {Schekochihin} AA, {Stone} JM (2014) {Firehose and Mirror
  Instabilities in a Collisionless Shearing Plasma}. Physical Review Letters
  112(20):205003, \doi{10.1103/PhysRevLett.112.205003}, \eprint{1402.0010}

\bibitem[{{Laing} et~al(2008){Laing}, {Bridle}, {Parma}, and
  {Murgia}}]{2008MNRAS.391..521L}
{Laing} RA, {Bridle} AH, {Parma} P, {Murgia} M (2008) {Structures of the
  magnetoionic media around the Fanaroff-Riley Class I radio galaxies 3C31 and
  Hydra A}. \mnras 391:521--549, \doi{10.1111/j.1365-2966.2008.13895.x},
  \eprint{0809.2411}

\bibitem[{{Landau} and {Lifshitz}(1966)}]{1966hydr.book.....L}
{Landau} LD, {Lifshitz} EM (1966) {Hydrodynamik}

\bibitem[{{Landau} et~al(1961){Landau}, {Lifshitz}, and
  {King}}]{1961AmJPh..29..647L}
{Landau} LD, {Lifshitz} EM, {King} AL (1961) {Electrodynamics of Continuous
  Media}. American Journal of Physics 29:647--648, \doi{10.1119/1.1937882}

\bibitem[{{Langer} et~al(2005){Langer}, {Aghanim}, and
  {Puget}}]{2005A&A...443..367L}
{Langer} M, {Aghanim} N, {Puget} JL (2005) {Magnetic fields from reionisation}.
  \aap 443:367--372, \doi{10.1051/0004-6361:20053372},
  \eprint{astro-ph/0508173}

\bibitem[{{Latif} et~al(2013){Latif}, {Schleicher}, {Schmidt}, and
  {Niemeyer}}]{2013MNRAS.432..668L}
{Latif} MA, {Schleicher} DRG, {Schmidt} W, {Niemeyer} J (2013) {The small-scale
  dynamo and the amplification of magnetic fields in massive primordial
  haloes}. \mnras 432:668--678, \doi{10.1093/mnras/stt503}, \eprint{1212.1619}

\bibitem[{{Lau} et~al(2009){Lau}, {Kravtsov}, and {Nagai}}]{lau09}
{Lau} ET, {Kravtsov} AV, {Nagai} D (2009) {Residual Gas Motions in the
  Intracluster Medium and Bias in Hydrostatic Measurements of Mass Profiles of
  Clusters}. \apj 705:1129--1138, \doi{10.1088/0004-637X/705/2/1129},
  \eprint{0903.4895}

\bibitem[{{Lazarian} et~al(2009){Lazarian}, {Beresnyak}, {Yan}, {Opher}, and
  {Liu}}]{2009SSRv..143..387L}
{Lazarian} A, {Beresnyak} A, {Yan} H, {Opher} M, {Liu} Y (2009) {Properties and
  Selected Implications of Magnetic Turbulence for Interstellar Medium, Local
  Bubble and Solar Wind}. \ssr 143:387--413, \doi{10.1007/s11214-008-9452-y},
  \eprint{0811.0826}

\bibitem[{{Lazarian} et~al(2012){Lazarian}, {Vlahos}, {Kowal}, {Yan},
  {Beresnyak}, and {de Gouveia Dal Pino}}]{2012SSRv..173..557L}
{Lazarian} A, {Vlahos} L, {Kowal} G, {Yan} H, {Beresnyak} A, {de Gouveia Dal
  Pino} EM (2012) {Turbulence, Magnetic Reconnection in Turbulent Fluids and
  Energetic Particle Acceleration}. \ssr 173:557--622,
  \doi{10.1007/s11214-012-9936-7}, \eprint{1211.0008}

\bibitem[{{Lee}(2013)}]{2013JCoPh.243..269L}
{Lee} D (2013) {A solution accurate, efficient and stable unsplit staggered
  mesh scheme for three dimensional magnetohydrodynamics}. Journal of
  Computational Physics 243:269--292, \doi{10.1016/j.jcp.2013.02.049},
  \eprint{1303.6988}

\bibitem[{{Li} et~al(2008){Li}, {Li}, and {Cen}}]{2008ApJS..174....1L}
{Li} S, {Li} H, {Cen} R (2008) {CosmoMHD: A Cosmological Magnetohydrodynamics
  Code}. \apjs 174:1-12, \doi{10.1086/521302}, \eprint{astro-ph/0611863}

\bibitem[{{Longair}(2011)}]{2011hea..book.....L}
{Longair} MS (2011) {High Energy Astrophysics}

\bibitem[{{Lucek} and {Bell}(2000)}]{2000MNRAS.314...65L}
{Lucek} SG, {Bell} AR (2000) {Non-linear amplification of a magnetic field
  driven by cosmic ray streaming}. \mnras 314:65--74,
  \doi{10.1046/j.1365-8711.2000.03363.x}

\bibitem[{{Lyutikov}(2006)}]{lyu06}
{Lyutikov} M (2006) {Magnetic draping of merging cores and radio bubbles in
  clusters of galaxies}. \mnras 373:73--78,
  \doi{10.1111/j.1365-2966.2006.10835.x}, \eprint{astro-ph/0604178}

\bibitem[{{Maier} et~al(2009){Maier}, {Iapichino}, {Schmidt}, and
  {Niemeyer}}]{2009ApJ...707...40M}
{Maier} A, {Iapichino} L, {Schmidt} W, {Niemeyer} JC (2009) {Adaptively Refined
  Large Eddy Simulations of a Galaxy Cluster: Turbulence Modeling and the
  Physics of the Intracluster Medium}. \apj 707:40--54,
  \doi{10.1088/0004-637X/707/1/40}, \eprint{0909.1800}

\bibitem[{{Malkov} and {O'C Drury}(2001)}]{md01}
{Malkov} MA, {O'C Drury} L (2001) {Nonlinear theory of diffusive acceleration
  of particles by shock waves}. Reports on Progress in Physics 64:429--481,
  \doi{10.1088/0034-4885/64/4/201}

\bibitem[{{Marcowith} et~al(2016){Marcowith}, {Bret}, {Bykov}, {Dieckman}, {O'C
  Drury}, {Lemb{\`e}ge}, {Lemoine}, {Morlino}, {Murphy}, {Pelletier},
  {Plotnikov}, {Reville}, {Riquelme}, {Sironi}, and {Stockem
  Novo}}]{2016RPPh...79d6901M}
{Marcowith} A, {Bret} A, {Bykov} A, {Dieckman} ME, {O'C Drury} L, {Lemb{\`e}ge}
  B, {Lemoine} M, {Morlino} G, {Murphy} G, {Pelletier} G, {Plotnikov} I,
  {Reville} B, {Riquelme} M, {Sironi} L, {Stockem Novo} A (2016) {The
  microphysics of collisionless shock waves}. Reports on Progress in Physics
  79(4):046901, \doi{10.1088/0034-4885/79/4/046901}, \eprint{1604.00318}

\bibitem[{{Marinacci} et~al(2015){Marinacci}, {Vogelsberger}, {Mocz}, and
  {Pakmor}}]{2015MNRAS.453.3999M}
{Marinacci} F, {Vogelsberger} M, {Mocz} P, {Pakmor} R (2015) {The large-scale
  properties of simulated cosmological magnetic fields}. \mnras 453:3999--4019,
  \doi{10.1093/mnras/stv1692}, \eprint{1506.00005}

\bibitem[{{Marinacci} et~al(2018{\natexlab{a}}){Marinacci}, {Vogelsberger},
  {Kannan}, {Mocz}, {Pakmor}, and {Springel}}]{2018MNRAS.476.2476M}
{Marinacci} F, {Vogelsberger} M, {Kannan} R, {Mocz} P, {Pakmor} R, {Springel} V
  (2018{\natexlab{a}}) {Non-ideal magnetohydrodynamics on a moving mesh}.
  \mnras 476:2476--2492, \doi{10.1093/mnras/sty397}, \eprint{1710.10265}

\bibitem[{{Marinacci} et~al(2018{\natexlab{b}}){Marinacci}, {Vogelsberger},
  {Pakmor}, {Torrey}, {Springel}, {Hernquist}, {Nelson}, {Weinberger},
  {Pillepich}, {Naiman}, and {Genel}}]{ma17}
{Marinacci} F, {Vogelsberger} M, {Pakmor} R, {Torrey} P, {Springel} V,
  {Hernquist} L, {Nelson} D, {Weinberger} R, {Pillepich} A, {Naiman} J, {Genel}
  S (2018{\natexlab{b}}) {First results from the IllustrisTNG simulations:
  radio haloes and magnetic fields}. \mnras 480:5113--5139,
  \doi{10.1093/mnras/sty2206}, \eprint{1707.03396}

\bibitem[{{Markevitch} and {Vikhlinin}(2007)}]{MV07}
{Markevitch} M, {Vikhlinin} A (2007) \physrep 443:1--53,
  \doi{10.1016/j.physrep.2007.01.001}, \eprint{arXiv:astro-ph/0701821}

\bibitem[{{Martin-Alvarez} et~al(2018){Martin-Alvarez}, {Devriendt}, {Slyz},
  and {Teyssier}}]{2018MNRAS.tmp.1551M}
{Martin-Alvarez} S, {Devriendt} J, {Slyz} A, {Teyssier} R (2018) {A three-phase
  amplification of the cosmic magnetic field in galaxies}. \mnras
  \doi{10.1093/mnras/sty1623}, \eprint{1806.06866}

\bibitem[{{Mazzotta} and {Giacintucci}(2008)}]{maz08}
{Mazzotta} P, {Giacintucci} S (2008) {Do Radio Core-Halos and Cold Fronts in
  Non-Major-Merging Clusters Originate from the Same Gas Sloshing?} \apjl
  675:L9, \doi{10.1086/529433}, \eprint{0801.1905}

\bibitem[{{Meinecke} et~al(2015){Meinecke}, {Tzeferacos}, {Bell}, {Bingham},
  {Clarke}, {Churazov}, {Crowston}, {Doyle}, {Drake}, {Heathcote}, {Koenig},
  {Kuramitsu}, {Kuranz}, {Lee}, {MacDonald}, {Murphy}, {Notley}, {Park},
  {Pelka}, {Ravasio}, {Reville}, {Sakawa}, {Wan}, {Woolsey}, {Yurchak},
  {Miniati}, {Schekochihin}, {Lamb}, and {Gregori}}]{2015PNAS..112.8211M}
{Meinecke} J, {Tzeferacos} P, {Bell} A, {Bingham} R, {Clarke} R, {Churazov} E,
  {Crowston} R, {Doyle} H, {Drake} RP, {Heathcote} R, {Koenig} M, {Kuramitsu}
  Y, {Kuranz} C, {Lee} D, {MacDonald} M, {Murphy} C, {Notley} M, {Park} HS,
  {Pelka} A, {Ravasio} A, {Reville} B, {Sakawa} Y, {Wan} W, {Woolsey} N,
  {Yurchak} R, {Miniati} F, {Schekochihin} A, {Lamb} D, {Gregori} G (2015)
  {Developed turbulence and nonlinear amplification of magnetic fields in
  laboratory and astrophysical plasmas}. Proceedings of the National Academy of
  Science 112:8211--8215, \doi{10.1073/pnas.1502079112}

\bibitem[{{Melville} et~al(2016){Melville}, {Schekochihin}, and
  {Kunz}}]{2016MNRAS.459.2701M}
{Melville} S, {Schekochihin} AA, {Kunz} MW (2016) {Pressure-anisotropy-driven
  microturbulence and magnetic-field evolution in shearing, collisionless
  plasma}. \mnras 459:2701--2720, \doi{10.1093/mnras/stw793},
  \eprint{1512.08131}

\bibitem[{{Miniati}(2014)}]{2014ApJ...782...21M}
{Miniati} F (2014) {The Matryoshka Run: A Eulerian Refinement Strategy to Study
  the Statistics of Turbulence in Virialized Cosmic Structures}. \apj 782:21,
  \doi{10.1088/0004-637X/782/1/21}, \eprint{1310.2951}

\bibitem[{{Miniati}(2015)}]{2015ApJ...800...60M}
{Miniati} F (2015) {The Matryoshka Run. II. Time-dependent Turbulence
  Statistics, Stochastic Particle Acceleration, and Microphysics Impact in a
  Massive Galaxy Cluster}. \apj 800:60, \doi{10.1088/0004-637X/800/1/60},
  \eprint{1409.3576}

\bibitem[{{Miniati} and {Bell}(2011)}]{mb11}
{Miniati} F, {Bell} AR (2011) {Resistive Magnetic Field Generation at Cosmic
  Dawn}. \apj 729:73, \doi{10.1088/0004-637X/729/1/73}, \eprint{1001.2011}

\bibitem[{{Miniati} and {Beresnyak}(2015)}]{2015Natur.523...59M}
{Miniati} F, {Beresnyak} A (2015) {Self-similar energetics in large clusters of
  galaxies}. \nat 523:59--62, \doi{10.1038/nature14552}, \eprint{1507.01940}

\bibitem[{{Miniati} et~al(2000){Miniati}, {Ryu}, {Kang}, {Jones}, {Cen}, and
  {Ostriker}}]{2000ApJ...542..608M}
{Miniati} F, {Ryu} D, {Kang} H, {Jones} TW, {Cen} R, {Ostriker} JP (2000)
  {Properties of Cosmic Shock Waves in Large-Scale Structure Formation}. \apj
  542:608--621, \doi{10.1086/317027}, \eprint{astro-ph/0005444}

\bibitem[{{Mitchell} et~al(2009){Mitchell}, {McCarthy}, {Bower}, {Theuns}, and
  {Crain}}]{2009MNRAS.395..180M}
{Mitchell} NL, {McCarthy} IG, {Bower} RG, {Theuns} T, {Crain} RA (2009) {On the
  origin of cores in simulated galaxy clusters}. \mnras 395:180--196,
  \doi{10.1111/j.1365-2966.2009.14550.x}, \eprint{0812.1750}

\bibitem[{{Mo} et~al(2010){Mo}, {van den Bosch}, and
  {White}}]{2010gfe..book.....M}
{Mo} H, {van den Bosch} FC, {White} S (2010) {Galaxy Formation and Evolution}

\bibitem[{{Mogavero} and {Schekochihin}(2014)}]{2014MNRAS.440.3226M}
{Mogavero} F, {Schekochihin} AA (2014) {Models of magnetic field evolution and
  effective viscosity in weakly collisional extragalactic plasmas}. \mnras
  440:3226--3242, \doi{10.1093/mnras/stu433}, \eprint{1312.3672}

\bibitem[{{Naab} and {Ostriker}(2017)}]{2017ARA&A..55...59N}
{Naab} T, {Ostriker} JP (2017) {Theoretical Challenges in Galaxy Formation}.
  \araa 55:59--109, \doi{10.1146/annurev-astro-081913-040019},
  \eprint{1612.06891}

\bibitem[{{Nagai} et~al(2007){Nagai}, {Kravtsov}, and
  {Vikhlinin}}]{2007ApJ...668....1N}
{Nagai} D, {Kravtsov} AV, {Vikhlinin} A (2007) {Effects of Galaxy Formation on
  Thermodynamics of the Intracluster Medium}. \apj 668:1--14,
  \doi{10.1086/521328}, \eprint{arXiv:astro-ph/0703661}

\bibitem[{{Nakazawa} et~al(2009){Nakazawa}, {Sarazin}, {Kawaharada},
  {Kitaguchi}, {Okuyama}, {Makishima}, {Kawano}, {Fukazawa}, {Inoue},
  {Takizawa}, {Wik}, {Finoguenov}, and {Clarke}}]{2009PASJ...61..339N}
{Nakazawa} K, {Sarazin} CL, {Kawaharada} M, {Kitaguchi} T, {Okuyama} S,
  {Makishima} K, {Kawano} N, {Fukazawa} Y, {Inoue} S, {Takizawa} M, {Wik} DR,
  {Finoguenov} A, {Clarke} TE (2009) {Hard X-Ray Properties of the Merging
  Cluster Abell 3667 as Observed with Suzaku}. \pasj 61:339--355,
  \doi{10.1093/pasj/61.2.339}, \eprint{0812.1438}

\bibitem[{{Nelson} et~al(2014){Nelson}, {Lau}, and
  {Nagai}}]{2014ApJ...792...25N}
{Nelson} K, {Lau} ET, {Nagai} D (2014) {Hydrodynamic Simulation of Non-thermal
  Pressure Profiles of Galaxy Clusters}. \apj 792:25,
  \doi{10.1088/0004-637X/792/1/25}, \eprint{1404.4636}

\bibitem[{{Neronov} and {Vovk}(2010)}]{2010Sci...328...73N}
{Neronov} A, {Vovk} I (2010) {Evidence for Strong Extragalactic Magnetic Fields
  from Fermi Observations of TeV Blazars}. Science 328:73--,
  \doi{10.1126/science.1184192}, \eprint{1006.3504}

\bibitem[{{Owen} et~al(2014){Owen}, {Rudnick}, {Eilek}, {Rau}, {Bhatnagar}, and
  {Kogan}}]{2014ApJ...794...24O}
{Owen} FN, {Rudnick} L, {Eilek} J, {Rau} U, {Bhatnagar} S, {Kogan} L (2014)
  {Wideband Very Large Array Observations of A2256. I. Continuum, Rotation
  Measure, and Spectral Imaging}. \apj 794:24,
  \doi{10.1088/0004-637X/794/1/24}, \eprint{1408.5931}

\bibitem[{{Pakmor} et~al(2011){Pakmor}, {Bauer}, and
  {Springel}}]{2011MNRAS.418.1392P}
{Pakmor} R, {Bauer} A, {Springel} V (2011) {Magnetohydrodynamics on an
  unstructured moving grid}. \mnras 418:1392--1401,
  \doi{10.1111/j.1365-2966.2011.19591.x}, \eprint{1108.1792}

\bibitem[{{Pakmor} et~al(2017){Pakmor}, {G{\'o}mez}, {Grand}, {Marinacci},
  {Simpson}, {Springel}, {Campbell}, {Frenk}, {Guillet}, {Pfrommer}, and
  {White}}]{2017MNRAS.469.3185P}
{Pakmor} R, {G{\'o}mez} FA, {Grand} RJJ, {Marinacci} F, {Simpson} CM,
  {Springel} V, {Campbell} DJR, {Frenk} CS, {Guillet} T, {Pfrommer} C, {White}
  SDM (2017) {Magnetic field formation in the Milky Way like disc galaxies of
  the Auriga project}. \mnras 469:3185--3199, \doi{10.1093/mnras/stx1074},
  \eprint{1701.07028}

\bibitem[{{Parizot} et~al(2006){Parizot}, {Marcowith}, {Ballet}, and
  {Gallant}}]{2006A&A...453..387P}
{Parizot} E, {Marcowith} A, {Ballet} J, {Gallant} YA (2006) {Observational
  constraints on energetic particle diffusion in young supernovae remnants:
  amplified magnetic field and maximum energy}. \aap 453:387--395,
  \doi{10.1051/0004-6361:20064985}, \eprint{astro-ph/0603723}

\bibitem[{{Paul} et~al(2011){Paul}, {Iapichino}, {Miniati}, {Bagchi}, and
  {Mannheim}}]{2011ApJ...726...17P}
{Paul} S, {Iapichino} L, {Miniati} F, {Bagchi} J, {Mannheim} K (2011)
  {Evolution of Shocks and Turbulence in Major Cluster Mergers}. \apj 726:17,
  \doi{10.1088/0004-637X/726/1/17}, \eprint{1001.1170}

\bibitem[{{Pfrommer}(2013)}]{pfr13}
{Pfrommer} C (2013) {Toward a Comprehensive Model for Feedback by Active
  Galactic Nuclei: New Insights from M87 Observations by LOFAR, Fermi, and
  H.E.S.S.} \apj 779:10, \doi{10.1088/0004-637X/779/1/10}, \eprint{1303.5443}

\bibitem[{{Pfrommer} and {En{\ss}lin}(2004)}]{pfr04}
{Pfrommer} C, {En{\ss}lin} TA (2004) {Constraining the population of cosmic ray
  protons in cooling flow clusters with {$\gamma$}-ray and radio observations:
  Are radio mini-halos of hadronic origin?} \aap 413:17--36,
  \doi{10.1051/0004-6361:20031464}

\bibitem[{{Pfrommer} et~al(2006){Pfrommer}, {Springel}, {En{\ss}lin}, and
  {Jubelgas}}]{pf06}
{Pfrommer} C, {Springel} V, {En{\ss}lin} TA, {Jubelgas} M (2006) \mnras
  367:113--131, \doi{10.1111/j.1365-2966.2005.09953.x},
  \eprint{arXiv:astro-ph/0603483}

\bibitem[{{Planck Collaboration} et~al(2016){Planck Collaboration}, {Ade},
  {Aghanim}, {Arnaud}, {Arroja}, {Ashdown}, {Aumont}, {Baccigalupi},
  {Ballardini}, {Banday}, and et~al.}]{PLANCK2015}
{Planck Collaboration}, {Ade} PAR, {Aghanim} N, {Arnaud} M, {Arroja} F,
  {Ashdown} M, {Aumont} J, {Baccigalupi} C, {Ballardini} M, {Banday} AJ, et~al
  (2016) {Planck 2015 results. XIX. Constraints on primordial magnetic fields}.
  \aap 594:A19, \doi{10.1051/0004-6361/201525821}, \eprint{1502.01594}

\bibitem[{{Planelles} et~al(2015){Planelles}, {Schleicher}, and
  {Bykov}}]{2015SSRv..188...93P}
{Planelles} S, {Schleicher} DRG, {Bykov} AM (2015) {Large-Scale Structure
  Formation: From the First Non-linear Objects to Massive Galaxy Clusters}.
  \ssr 188:93--139, \doi{10.1007/s11214-014-0045-7}, \eprint{1404.3956}

\bibitem[{{Planelles} et~al(2016){Planelles}, {Schleicher}, and
  {Bykov}}]{2016mssf.book...93P}
{Planelles} S, {Schleicher} DRG, {Bykov} AM (2016) {Large-Scale Structure
  Formation: From the First Non-linear Objects to Massive Galaxy Clusters}, pp
  93--139. \doi{10.1007/978-1-4939-3547-5_4}

\bibitem[{{Porter} et~al(2015){Porter}, {Jones}, and {Ryu}}]{po15}
{Porter} DH, {Jones} TW, {Ryu} D (2015) {Vorticity, Shocks, and Magnetic Fields
  in Subsonic, ICM-like Turbulence}. \apj 810:93,
  \doi{10.1088/0004-637X/810/2/93}, \eprint{1507.08737}

\bibitem[{{Price}(2012{\natexlab{a}})}]{2012MNRAS.420L..33P}
{Price} DJ (2012{\natexlab{a}}) {Resolving high Reynolds numbers in smoothed
  particle hydrodynamics simulations of subsonic turbulence}. \mnras
  420:L33--L37, \doi{10.1111/j.1745-3933.2011.01187.x}, \eprint{1111.1255}

\bibitem[{{Price}(2012{\natexlab{b}})}]{2012JCoPh.231..759P}
{Price} DJ (2012{\natexlab{b}}) {Smoothed particle hydrodynamics and
  magnetohydrodynamics}. Journal of Computational Physics 231:759--794,
  \doi{10.1016/j.jcp.2010.12.011}, \eprint{1012.1885}

\bibitem[{{Rajpurohit} et~al(2018){Rajpurohit}, {Hoeft}, {van Weeren},
  {Rudnick}, {R{\"o}ttgering}, {Forman}, {Br{\"u}ggen}, {Croston},
  {Andrade-Santos}, {Dawson}, {Intema}, {Kraft}, {Jones}, and
  {Jee}}]{2018ApJ...852...65R}
{Rajpurohit} K, {Hoeft} M, {van Weeren} RJ, {Rudnick} L, {R{\"o}ttgering} HJA,
  {Forman} WR, {Br{\"u}ggen} M, {Croston} JH, {Andrade-Santos} F, {Dawson} WA,
  {Intema} HT, {Kraft} RP, {Jones} C, {Jee} MJ (2018) {Deep VLA Observations of
  the Cluster 1RXS J0603.3+4214 in the Frequency Range of 1-2 GHz}. \apj
  852:65, \doi{10.3847/1538-4357/aa9f13}, \eprint{1712.01327}

\bibitem[{{Reville} and {Bell}(2013)}]{2013MNRAS.430.2873R}
{Reville} B, {Bell} AR (2013) {Universal behaviour of shock precursors in the
  presence of efficient cosmic ray acceleration}. \mnras 430:2873--2884,
  \doi{10.1093/mnras/stt100}, \eprint{1301.3173}

\bibitem[{{Reville} et~al(2008){Reville}, {O'Sullivan}, {Duffy}, and
  {Kirk}}]{2008MNRAS.386..509R}
{Reville} B, {O'Sullivan} S, {Duffy} P, {Kirk} JG (2008) {The transport of
  cosmic rays in self-excited magnetic turbulence}. \mnras 386:509--515,
  \doi{10.1111/j.1365-2966.2008.13059.x}, \eprint{0802.0109}

\bibitem[{{Reynolds}(1998)}]{1998ApJ...493..375R}
{Reynolds} SP (1998) {Models of Synchrotron X-Rays from Shell Supernova
  Remnants}. \apj 493:375--396, \doi{10.1086/305103}

\bibitem[{{Ricker} and {Sarazin}(2001)}]{2001ApJ...561..621R}
{Ricker} PM, {Sarazin} CL (2001) {Off-Axis Cluster Mergers: Effects of a
  Strongly Peaked Dark Matter Profile}. \apj 561:621--644

\bibitem[{{Rieder} and {Teyssier}(2016)}]{2016MNRAS.457.1722R}
{Rieder} M, {Teyssier} R (2016) {A small-scale dynamo in feedback-dominated
  galaxies as the origin of cosmic magnetic fields - I. The kinematic phase}.
  \mnras 457:1722--1738, \doi{10.1093/mnras/stv2985}, \eprint{1506.00849}

\bibitem[{{Rincon} et~al(2016){Rincon}, {Califano}, {Schekochihin}, and
  {Valentini}}]{2016PNAS..113.3950R}
{Rincon} F, {Califano} F, {Schekochihin} AA, {Valentini} F (2016) {Turbulent
  dynamo in a collisionless plasma}. Proceedings of the National Academy of
  Science 113:3950--3953, \doi{10.1073/pnas.1525194113}, \eprint{1512.06455}

\bibitem[{{Riquelme} and {Spitkovsky}(2010)}]{2010ApJ...717.1054R}
{Riquelme} MA, {Spitkovsky} A (2010) {Magnetic Amplification by Magnetized
  Cosmic Rays in Supernova Remnant Shocks}. \apj 717:1054--1066,
  \doi{10.1088/0004-637X/717/2/1054}, \eprint{0912.4990}

\bibitem[{{Riquelme} and {Spitkovsky}(2011)}]{2011ApJ...733...63R}
{Riquelme} MA, {Spitkovsky} A (2011) {Electron Injection by Whistler Waves in
  Non-relativistic Shocks}. \apj 733:63, \doi{10.1088/0004-637X/733/1/63},
  \eprint{1009.3319}

\bibitem[{{Robertson} et~al(2010){Robertson}, {Kravtsov}, {Gnedin}, {Abel}, and
  {Rudd}}]{2010MNRAS.401.2463R}
{Robertson} BE, {Kravtsov} AV, {Gnedin} NY, {Abel} T, {Rudd} DH (2010)
  {Computational Eulerian hydrodynamics and Galilean invariance}. \mnras
  401:2463--2476, \doi{10.1111/j.1365-2966.2009.15823.x}, \eprint{0909.0513}

\bibitem[{{Roediger} et~al(2015{\natexlab{a}}){Roediger}, {Kraft}, {Nulsen},
  {Forman}, {Machacek}, {Randall}, {Jones}, {Churazov}, and
  {Kokotanekova}}]{rod15a}
{Roediger} E, {Kraft} RP, {Nulsen} PEJ, {Forman} WR, {Machacek} M, {Randall} S,
  {Jones} C, {Churazov} E, {Kokotanekova} R (2015{\natexlab{a}}) {Stripped
  Elliptical Galaxies as Probes of ICM Physics: I. Tails, Wakes, and Flow
  Patterns in and Around Stripped Ellipticals}. \apj 806:103,
  \doi{10.1088/0004-637X/806/1/103}, \eprint{1409.6300}

\bibitem[{{Roediger} et~al(2015{\natexlab{b}}){Roediger}, {Kraft}, {Nulsen},
  {Forman}, {Machacek}, {Randall}, {Jones}, {Churazov}, and
  {Kokotanekova}}]{rod15b}
{Roediger} E, {Kraft} RP, {Nulsen} PEJ, {Forman} WR, {Machacek} M, {Randall} S,
  {Jones} C, {Churazov} E, {Kokotanekova} R (2015{\natexlab{b}}) {Stripped
  Elliptical Galaxies as Probes of ICM Physics: II. Stirred, but Mixed? Viscous
  and Inviscid Gas Stripping of the Virgo Elliptical M89}. \apj 806:104,
  \doi{10.1088/0004-637X/806/1/104}, \eprint{1409.6312}

\bibitem[{{Roettiger} et~al(1993){Roettiger}, {Burns}, and
  {Loken}}]{1993ApJ...407L..53R}
{Roettiger} K, {Burns} J, {Loken} C (1993) {When clusters collide - A numerical
  Hydro/N-body simulation of merging galaxy clusters}. \apjl 407:L53--L56,
  \doi{10.1086/186804}

\bibitem[{{Roettiger} et~al(1997){Roettiger}, {Loken}, and
  {Burns}}]{1997ApJS..109..307R}
{Roettiger} K, {Loken} C, {Burns} JO (1997) {Numerical Simulations of Merging
  Clusters of Galaxies}. \apjs 109:307--+

\bibitem[{{Roettiger} et~al(1999){Roettiger}, {Stone}, and
  {Burns}}]{1999ApJ...518..594R}
{Roettiger} K, {Stone} JM, {Burns} JO (1999) {Magnetic Field Evolution in
  Merging Clusters of Galaxies}. \apj 518:594--602

\bibitem[{{Roland}(1981)}]{1981A&A....93..407R}
{Roland} J (1981) {On the origin of the intergalactic magnetic field and of the
  radio halo associated with the Coma cluster of galaxies}. \aap 93:407--410

\bibitem[{{Roncarelli} et~al(2018){Roncarelli}, {Gaspari}, {Ettori}, {Biffi},
  {Brighenti}, {Bulbul}, {Clerc}, {Cucchetti}, {Pointecouteau}, and
  {Rasia}}]{2018arXiv180502577R}
{Roncarelli} M, {Gaspari} M, {Ettori} S, {Biffi} V, {Brighenti} F, {Bulbul} E,
  {Clerc} N, {Cucchetti} E, {Pointecouteau} E, {Rasia} E (2018) {Measuring
  turbulence and gas motions in galaxy clusters via synthetic
  $\{$$\backslash$it Athena$\}$ X-IFU observations}. ArXiv e-prints
  \eprint{1805.02577}

\bibitem[{{Ruszkowski} and {Oh}(2011)}]{2011MNRAS.414.1493R}
{Ruszkowski} M, {Oh} SP (2011) {Galaxy motions, turbulence and conduction in
  clusters of galaxies}. \mnras 414:1493--1507,
  \doi{10.1111/j.1365-2966.2011.18482.x}, \eprint{1008.5016}

\bibitem[{{Ruszkowski} et~al(2011){Ruszkowski}, {Lee}, {Br{\"u}ggen},
  {Parrish}, and {Oh}}]{ruszkowski11}
{Ruszkowski} M, {Lee} D, {Br{\"u}ggen} M, {Parrish} I, {Oh} SP (2011)
  {Cosmological Magnetohydrodynamic Simulations of Cluster Formation with
  Anisotropic Thermal Conduction}. \apj 740:81,
  \doi{10.1088/0004-637X/740/2/81}, \eprint{1010.2277}

\bibitem[{{Ruzmaikin} et~al(1989){Ruzmaikin}, {Sokolov}, and
  {Shukurov}}]{1989MNRAS.241....1R}
{Ruzmaikin} A, {Sokolov} D, {Shukurov} A (1989) {The dynamo origin of magnetic
  fields in galaxy clusters}. \mnras 241:1--14, \doi{10.1093/mnras/241.1.1}

\bibitem[{{Ryu} et~al(2003){Ryu}, {Kang}, {Hallman}, and {Jones}}]{ry03}
{Ryu} D, {Kang} H, {Hallman} E, {Jones} TW (2003) \apj 593:599--610,
  \doi{10.1086/376723}, \eprint{arXiv:astro-ph/0305164}

\bibitem[{{Ryu} et~al(2008){Ryu}, {Kang}, {Cho}, and {Das}}]{ry08}
{Ryu} D, {Kang} H, {Cho} J, {Das} S (2008) {Turbulence and Magnetic Fields in
  the Large-Scale Structure of the Universe}. Science 320:909--,
  \doi{10.1126/science.1154923}, \eprint{0805.2466}

\bibitem[{{Ryu} et~al(2012){Ryu}, {Schleicher}, {Treumann}, {Tsagas}, and
  {Widrow}}]{2012SSRv..166....1R}
{Ryu} D, {Schleicher} DRG, {Treumann} RA, {Tsagas} CG, {Widrow} LM (2012)
  {Magnetic Fields in the Large-Scale Structure of the Universe}. \ssr
  166:1--35, \doi{10.1007/s11214-011-9839-z}, \eprint{1109.4055}

\bibitem[{{Samui} et~al(2017){Samui}, {Subramanian}, and {Srianand}}]{sam17}
{Samui} S, {Subramanian} K, {Srianand} R (2017) {Efficient cold outflows driven
  by cosmic rays in high redshift galaxies and their global effects on the
  IGM}. ArXiv e-prints \eprint{1706.01890}

\bibitem[{{S{\'a}nchez-Salcedo} et~al(1998){S{\'a}nchez-Salcedo},
  {Brandenburg}, and {Shukurov}}]{1998Ap&SS.263...87S}
{S{\'a}nchez-Salcedo} FJ, {Brandenburg} A, {Shukurov} A (1998) {Turbulence and
  Magnetic Fields in Clusters of Galaxies}. \apss 263:87--90,
  \doi{10.1023/A:1002144413095}

\bibitem[{{Sano} et~al(2012){Sano}, {Nishihara}, {Matsuoka}, and
  {Inoue}}]{2012ApJ...758..126S}
{Sano} T, {Nishihara} K, {Matsuoka} C, {Inoue} T (2012) {Magnetic Field
  Amplification Associated with the Richtmyer-Meshkov Instability}. \apj
  758:126, \doi{10.1088/0004-637X/758/2/126}, \eprint{1209.0961}

\bibitem[{{Santos-Lima} et~al(2014){Santos-Lima}, {de Gouveia Dal Pino},
  {Kowal}, {Falceta-Gon{\c c}alves}, {Lazarian}, and
  {Nakwacki}}]{2014ApJ...781...84S}
{Santos-Lima} R, {de Gouveia Dal Pino} EM, {Kowal} G, {Falceta-Gon{\c c}alves}
  D, {Lazarian} A, {Nakwacki} MS (2014) {Magnetic Field Amplification and
  Evolution in Turbulent Collisionless Magnetohydrodynamics: An Application to
  the Intracluster Medium}. \apj 781:84, \doi{10.1088/0004-637X/781/2/84},
  \eprint{1305.5654}

\bibitem[{{Santos-Lima} et~al(2017){Santos-Lima}, {de Gouveia Dal Pino},
  {Falceta-Gon{\c c}alves}, {Nakwacki}, and {Kowal}}]{2017MNRAS.465.4866S}
{Santos-Lima} R, {de Gouveia Dal Pino} EM, {Falceta-Gon{\c c}alves} DA,
  {Nakwacki} MS, {Kowal} G (2017) {Features of collisionless turbulence in the
  intracluster medium from simulated Faraday rotation maps - II. The effects of
  instabilities feedback}. \mnras 465:4866--4871, \doi{10.1093/mnras/stw3050},
  \eprint{1611.10183}

\bibitem[{Sarazin(1986)}]{SA86.1}
Sarazin CL (1986) X-ray emission from clusters of galaxies. RvMP 58:1

\bibitem[{Sarazin(1988)}]{SA88.1}
Sarazin CL (1988) X-ray emission from clusters of galaxies. Cambridge
  University Press, Cambridge

\bibitem[{{Sarazin}(2002)}]{2002mpgc.book....1S}
{Sarazin} CL (2002) {The Physics of Cluster Mergers}, ASSL Vol.~272: Merging
  Processes in Galaxy Clusters, pp 1--38

\bibitem[{{Schekochihin} et~al(2002{\natexlab{a}}){Schekochihin}, {Cowley},
  {Maron}, and {Malyshkin}}]{2002PhRvE..65a6305S}
{Schekochihin} A, {Cowley} S, {Maron} J, {Malyshkin} L (2002{\natexlab{a}})
  {Structure of small-scale magnetic fields in the kinematic dynamo theory}.
  \pre 65(1):016305, \doi{10.1103/PhysRevE.65.016305},
  \eprint{astro-ph/0105322}

\bibitem[{{Schekochihin} et~al(2005{\natexlab{a}}){Schekochihin}, {Cowley},
  {Kulsrud}, {Hammett}, and {Sharma}}]{2005mpge.conf...86S}
{Schekochihin} A, {Cowley} S, {Kulsrud} R, {Hammett} G, {Sharma} P
  (2005{\natexlab{a}}) {Magnetised plasma turbulence in clusters of galaxies}.
  In: {Chyzy} KT, {Otmianowska-Mazur} K, {Soida} M, {Dettmar} RJ (eds) The
  Magnetized Plasma in Galaxy Evolution, pp 86--92, \eprint{astro-ph/0411781}

\bibitem[{{Schekochihin} and {Cowley}(2007)}]{2007mhet.book...85S}
{Schekochihin} AA, {Cowley} SC (2007) {Turbulence and Magnetic Fields in
  Astrophysical Plasmas}, Springer, p~85

\bibitem[{{Schekochihin} et~al(2002{\natexlab{b}}){Schekochihin}, {Maron},
  {Cowley}, and {McWilliams}}]{2002ApJ...576..806S}
{Schekochihin} AA, {Maron} JL, {Cowley} SC, {McWilliams} JC
  (2002{\natexlab{b}}) {The Small-Scale Structure of Magnetohydrodynamic
  Turbulence with Large Magnetic Prandtl Numbers}. \apj 576:806--813,
  \doi{10.1086/341814}, \eprint{astro-ph/0203219}

\bibitem[{{Schekochihin} et~al(2004){Schekochihin}, {Cowley}, {Taylor},
  {Maron}, and {McWilliams}}]{2004ApJ...612..276S}
{Schekochihin} AA, {Cowley} SC, {Taylor} SF, {Maron} JL, {McWilliams} JC (2004)
  {Simulations of the Small-Scale Turbulent Dynamo}. \apj 612:276--307,
  \doi{10.1086/422547}, \eprint{astro-ph/0312046}

\bibitem[{{Schekochihin} et~al(2005{\natexlab{b}}){Schekochihin}, {Cowley},
  {Kulsrud}, {Hammett}, and {Sharma}}]{sch05}
{Schekochihin} AA, {Cowley} SC, {Kulsrud} RM, {Hammett} GW, {Sharma} P
  (2005{\natexlab{b}}) {Plasma Instabilities and Magnetic Field Growth in
  Clusters of Galaxies}. \apj 629:139--142, \doi{10.1086/431202},
  \eprint{arXiv:astro-ph/0501362}

\bibitem[{{Schekochihin} et~al(2008){Schekochihin}, {Cowley}, {Kulsrud},
  {Rosin}, and {Heinemann}}]{2008PhRvL.100h1301S}
{Schekochihin} AA, {Cowley} SC, {Kulsrud} RM, {Rosin} MS, {Heinemann} T (2008)
  {Nonlinear Growth of Firehose and Mirror Fluctuations in Astrophysical
  Plasmas}. Physical Review Letters 100(8):081301,
  \doi{10.1103/PhysRevLett.100.081301}, \eprint{0709.3828}

\bibitem[{{Schekochihin} et~al(2009){Schekochihin}, {Cowley}, {Dorland},
  {Hammett}, {Howes}, {Quataert}, and {Tatsuno}}]{2009ApJS..182..310S}
{Schekochihin} AA, {Cowley} SC, {Dorland} W, {Hammett} GW, {Howes} GG,
  {Quataert} E, {Tatsuno} T (2009) {Astrophysical Gyrokinetics: Kinetic and
  Fluid Turbulent Cascades in Magnetized Weakly Collisional Plasmas}. \apjs
  182:310--377, \doi{10.1088/0067-0049/182/1/310}, \eprint{0704.0044}

\bibitem[{{Schindler} and {Mueller}(1993)}]{1993A&A...272..137S}
{Schindler} S, {Mueller} E (1993) {Simulations of the evolution of galaxy
  clusters. 11. Dynamics of the intra-cluster gas}. \aap 272:137

\bibitem[{{Schleicher} et~al(2010){Schleicher}, {Banerjee}, {Sur}, {Arshakian},
  {Klessen}, {Beck}, and {Spaans}}]{2010A&A...522A.115S}
{Schleicher} DRG, {Banerjee} R, {Sur} S, {Arshakian} TG, {Klessen} RS, {Beck}
  R, {Spaans} M (2010) {Small-scale dynamo action during the formation of the
  first stars and galaxies. I. The ideal MHD limit}. \aap 522:A115,
  \doi{10.1051/0004-6361/201015184}, \eprint{1003.1135}

\bibitem[{{Schleicher} et~al(2013){Schleicher}, {Schober}, {Federrath},
  {Bovino}, and {Schmidt}}]{2013NJPh...15b3017S}
{Schleicher} DRG, {Schober} J, {Federrath} C, {Bovino} S, {Schmidt} W (2013)
  {The small-scale dynamo: breaking universality at high Mach numbers}. New
  Journal of Physics 15(2):023017, \doi{10.1088/1367-2630/15/2/023017},
  \eprint{1301.4371}

\bibitem[{{Schlickeiser}(2002)}]{2002cra..book.....S}
{Schlickeiser} R (2002) {Cosmic Ray Astrophysics}

\bibitem[{{Schlickeiser} et~al(2012){Schlickeiser}, {Ibscher}, and
  {Supsar}}]{2012ApJ...758..102S}
{Schlickeiser} R, {Ibscher} D, {Supsar} M (2012) {Plasma Effects on Fast Pair
  Beams in Cosmic Voids}. \apj 758:102, \doi{10.1088/0004-637X/758/2/102}

\bibitem[{{Schmidt} et~al(2017){Schmidt}, {Byrohl}, {Engels}, {Behrens}, and
  {Niemeyer}}]{2017MNRAS.470..142S}
{Schmidt} W, {Byrohl} C, {Engels} JF, {Behrens} C, {Niemeyer} JC (2017)
  {Viscosity, pressure and support of the gas in simulations of merging
  cool-core clusters}. \mnras 470:142--156, \doi{10.1093/mnras/stx1274},
  \eprint{1705.06933}

\bibitem[{{Schober} et~al(2013){Schober}, {Schleicher}, and
  {Klessen}}]{schober13}
{Schober} J, {Schleicher} DRG, {Klessen} RS (2013) {Magnetic field
  amplification in young galaxies}. \aap 560:A87,
  \doi{10.1051/0004-6361/201322185}, \eprint{1310.0853}

\bibitem[{{Schuecker} et~al(2004{\natexlab{a}}){Schuecker}, {Finoguenov},
  {Miniati}, {B{\"o}hringer}, and {Briel}}]{sc04}
{Schuecker} P, {Finoguenov} A, {Miniati} F, {B{\"o}hringer} H, {Briel} UG
  (2004{\natexlab{a}}) {Probing turbulence in the Coma galaxy cluster}. \aap
  426:387--397, \doi{10.1051/0004-6361:20041039},
  \eprint{arXiv:astro-ph/0404132}

\bibitem[{{Schuecker} et~al(2004{\natexlab{b}}){Schuecker}, {Finoguenov},
  {Miniati}, {B{\"o}hringer}, and {Briel}}]{2004A&A...426..387S}
{Schuecker} P, {Finoguenov} A, {Miniati} F, {B{\"o}hringer} H, {Briel} UG
  (2004{\natexlab{b}}) {Probing turbulence in the Coma galaxy cluster}. \aap
  426:387--397, \doi{10.1051/0004-6361:20041039}, \eprint{astro-ph/0404132}

\bibitem[{{Semikoz} and {Sokoloff}(2005)}]{2005A&A...433L..53S}
{Semikoz} VB, {Sokoloff} D (2005) {Magnetic helicity and cosmological magnetic
  field}. \aap 433:L53--L56, \doi{10.1051/0004-6361:200500094},
  \eprint{astro-ph/0411496}

\bibitem[{{Shaw} et~al(2010){Shaw}, {Nagai}, {Bhattacharya}, and
  {Lau}}]{2010ApJ...725.1452S}
{Shaw} LD, {Nagai} D, {Bhattacharya} S, {Lau} ET (2010) {Impact of Cluster
  Physics on the Sunyaev-Zel'dovich Power Spectrum}. \apj 725:1452--1465,
  \doi{10.1088/0004-637X/725/2/1452}, \eprint{1006.1945}

\bibitem[{{Shi} and {Komatsu}(2014)}]{2014MNRAS.442..521S}
{Shi} X, {Komatsu} E (2014) {Analytical model for non-thermal pressure in
  galaxy clusters}. \mnras 442:521--532, \doi{10.1093/mnras/stu858},
  \eprint{1401.7657}

\bibitem[{{Shi} et~al(2015){Shi}, {Komatsu}, {Nelson}, and
  {Nagai}}]{2015MNRAS.448.1020S}
{Shi} X, {Komatsu} E, {Nelson} K, {Nagai} D (2015) {Analytical model for
  non-thermal pressure in galaxy clusters - II. Comparison with cosmological
  hydrodynamics simulation}. \mnras 448:1020--1029, \doi{10.1093/mnras/stv036},
  \eprint{1408.3832}

\bibitem[{{Shi} et~al(2016){Shi}, {Komatsu}, {Nagai}, and
  {Lau}}]{2016MNRAS.455.2936S}
{Shi} X, {Komatsu} E, {Nagai} D, {Lau} ET (2016) {Analytical model for
  non-thermal pressure in galaxy clusters - III. Removing the hydrostatic mass
  bias}. \mnras 455:2936--2944, \doi{10.1093/mnras/stv2504},
  \eprint{1507.04338}

\bibitem[{{Shimwell} et~al(2017){Shimwell}, {R{\"o}ttgering}, {Best},
  {Williams}, {Dijkema}, {de Gasperin}, {Hardcastle}, {Heald}, {Hoang},
  {Horneffer}, {Intema}, {Mahony}, {Mandal}, {Mechev}, {Morabito}, {Oonk},
  {Rafferty}, {Retana-Montenegro}, {Sabater}, {Tasse}, {van Weeren},
  {Br{\"u}ggen}, {Brunetti}, {Chyzy}, {Conway}, {Haverkorn}, {Jackson},
  {Jarvis}, {McKean}, {Miley}, {Morganti}, {White}, {Wise}, {van Bemmel},
  {Beck}, {Brienza}, {Bonafede}, {Calistro Rivera}, {Cassano}, {Clarke},
  {Cseh}, {Deller}, {Drabent}, {van Driel}, {Engels}, {Falcke}, {Ferrari},
  {Fr{\"o}hlich}, {Garrett}, {Harwood}, {Heesen}, {Hoeft}, {Horellou},
  {Israel}, {Kapi{\'n}ska}, {Kunert-Bajraszewska}, {McKay}, {Mohan},
  {Orr{\'u}}, {Pizzo}, {Prandoni}, {Schwarz}, {Shulevski}, {Sipior}, {Smith},
  {Sridhar}, {Steinmetz}, {Stroe}, {Varenius}, {van der Werf}, {Zensus}, and
  {Zwart}}]{2017A&A...598A.104S}
{Shimwell} TW, {R{\"o}ttgering} HJA, {Best} PN, {Williams} WL, {Dijkema} TJ,
  {de Gasperin} F, {Hardcastle} MJ, {Heald} GH, {Hoang} DN, {Horneffer} A,
  {Intema} H, {Mahony} EK, {Mandal} S, {Mechev} AP, {Morabito} L, {Oonk} JBR,
  {Rafferty} D, {Retana-Montenegro} E, {Sabater} J, {Tasse} C, {van Weeren} RJ,
  {Br{\"u}ggen} M, {Brunetti} G, {Chyzy} KT, {Conway} JE, {Haverkorn} M,
  {Jackson} N, {Jarvis} MJ, {McKean} JP, {Miley} GK, {Morganti} R, {White} GJ,
  {Wise} MW, {van Bemmel} IM, {Beck} R, {Brienza} M, {Bonafede} A, {Calistro
  Rivera} G, {Cassano} R, {Clarke} AO, {Cseh} D, {Deller} A, {Drabent} A, {van
  Driel} W, {Engels} D, {Falcke} H, {Ferrari} C, {Fr{\"o}hlich} S, {Garrett}
  MA, {Harwood} JJ, {Heesen} V, {Hoeft} M, {Horellou} C, {Israel} FP,
  {Kapi{\'n}ska} AD, {Kunert-Bajraszewska} M, {McKay} DJ, {Mohan} NR,
  {Orr{\'u}} E, {Pizzo} RF, {Prandoni} I, {Schwarz} DJ, {Shulevski} A, {Sipior}
  M, {Smith} DJB, {Sridhar} SS, {Steinmetz} M, {Stroe} A, {Varenius} E, {van
  der Werf} PP, {Zensus} JA, {Zwart} JTL (2017) {The LOFAR Two-metre Sky
  Survey. I. Survey description and preliminary data release}. \aap 598:A104,
  \doi{10.1051/0004-6361/201629313}, \eprint{1611.02700}

\bibitem[{{Skillman} et~al(2013){Skillman}, {Xu}, {Hallman}, {O'Shea}, {Burns},
  {Li}, {Collins}, and {Norman}}]{sk13}
{Skillman} SW, {Xu} H, {Hallman} EJ, {O'Shea} BW, {Burns} JO, {Li} H, {Collins}
  DC, {Norman} ML (2013) {Cosmological Magnetohydrodynamic Simulations of
  Galaxy Cluster Radio Relics: Insights and Warnings for Observations}. \apj
  765:21, \doi{10.1088/0004-637X/765/1/21}, \eprint{1211.3122}

\bibitem[{{Slane} et~al(2014){Slane}, {Lee}, {Ellison}, {Patnaude}, {Hughes},
  {Eriksen}, {Castro}, and {Nagataki}}]{2014ApJ...783...33S}
{Slane} P, {Lee} SH, {Ellison} DC, {Patnaude} DJ, {Hughes} JP, {Eriksen} KA,
  {Castro} D, {Nagataki} S (2014) {A CR-hydro-NEI Model of the Structure and
  Broadband Emission from Tycho's Supernova Remnant}. \apj 783:33,
  \doi{10.1088/0004-637X/783/1/33}

\bibitem[{{Spitzer}(1956)}]{1956pfig.book.....S}
{Spitzer} L (1956) {Physics of Fully Ionized Gases}

\bibitem[{{Springel}(2010)}]{2010MNRAS.401..791S}
{Springel} V (2010) {E pur si muove: Galilean-invariant cosmological
  hydrodynamical simulations on a moving mesh}. \mnras 401:791--851,
  \doi{10.1111/j.1365-2966.2009.15715.x}, \eprint{0901.4107}

\bibitem[{{Stasyszyn} and {Elstner}(2015)}]{2015JCoPh.282..148S}
{Stasyszyn} FA, {Elstner} D (2015) {A vector potential implementation for
  smoothed particle magnetohydrodynamics}. Journal of Computational Physics
  282:148--156, \doi{10.1016/j.jcp.2014.11.011}, \eprint{1411.3290}

\bibitem[{{Steinwandel} et~al(2018){Steinwandel}, {Beck}, {Arth}, {Dolag},
  {Moster}, and {Nielaba}}]{2018arXiv180809975S}
{Steinwandel} UP, {Beck} MC, {Arth} A, {Dolag} K, {Moster} BP, {Nielaba} P
  (2018) {Magnetic buoyancy in simulated galactic discs with a realistic circum
  galactic medium}. ArXiv e-prints \eprint{1808.09975}

\bibitem[{{Stroe} et~al(2013){Stroe}, {van Weeren}, {Intema}, {R{\"o}ttgering},
  {Br{\"u}ggen}, and {Hoeft}}]{2013A&A...555A.110S}
{Stroe} A, {van Weeren} RJ, {Intema} HT, {R{\"o}ttgering} HJA, {Br{\"u}ggen} M,
  {Hoeft} M (2013) {Discovery of spectral curvature in the shock downstream
  region: CIZA J2242.8+5301}. \aap 555:A110, \doi{10.1051/0004-6361/201321267},
  \eprint{1305.0005}

\bibitem[{{Stroe} et~al(2014){Stroe}, {Harwood}, {Hardcastle}, and
  {R{\"o}ttgering}}]{2014MNRAS.445.1213S}
{Stroe} A, {Harwood} JJ, {Hardcastle} MJ, {R{\"o}ttgering} HJA (2014) {Spectral
  age modelling of the `Sausage' cluster radio relic}. \mnras 445:1213--1222,
  \doi{10.1093/mnras/stu1839}, \eprint{1409.1579}

\bibitem[{{Su} et~al(2017){Su}, {Kraft}, {Roediger}, {Nulsen}, {Forman},
  {Churazov}, {Randall}, {Jones}, and {Machacek}}]{yysu17b}
{Su} Y, {Kraft} RP, {Roediger} E, {Nulsen} P, {Forman} WR, {Churazov} E,
  {Randall} SW, {Jones} C, {Machacek} ME (2017) {Deep Chandra Observations of
  NGC 1404: Cluster Plasma Physics Revealed by an Infalling Early-type Galaxy}.
  \apj 834:74, \doi{10.3847/1538-4357/834/1/74}, \eprint{1612.00535}

\bibitem[{{Subramanian}(2016)}]{sub16}
{Subramanian} K (2016) {The origin, evolution and signatures of primordial
  magnetic fields}. Reports on Progress in Physics 79(7):076901,
  \doi{10.1088/0034-4885/79/7/076901}, \eprint{1504.02311}

\bibitem[{{Subramanian} et~al(2006){Subramanian}, {Shukurov}, and
  {Haugen}}]{su06}
{Subramanian} K, {Shukurov} A, {Haugen} NEL (2006) {Evolving turbulence and
  magnetic fields in galaxy clusters}. \mnras 366:1437--1454,
  \doi{10.1111/j.1365-2966.2006.09918.x}, \eprint{arXiv:astro-ph/0505144}

\bibitem[{{Sur} et~al(2012){Sur}, {Federrath}, {Schleicher}, {Banerjee}, and
  {Klessen}}]{2012MNRAS.423.3148S}
{Sur} S, {Federrath} C, {Schleicher} DRG, {Banerjee} R, {Klessen} RS (2012)
  {Magnetic field amplification during gravitational collapse - influence of
  turbulence, rotation and gravitational compression}. \mnras 423:3148--3162,
  \doi{10.1111/j.1365-2966.2012.21100.x}, \eprint{1202.3206}

\bibitem[{{Takizawa}(2005)}]{2005ApJ...629..791T}
{Takizawa} M (2005) {Hydrodynamic Simulations of a Moving Substructure in a
  Cluster of Galaxies: Cold Fronts and Turbulence Generation}. \apj
  629:791--796, \doi{10.1086/431927}, \eprint{astro-ph/0505274}

\bibitem[{{Taylor} et~al(2015){Taylor}, {Agudo}, {Akahori}, {Beck}, {Gaensler},
  {Heald}, {Johnston-Hollitt}, {Langer}, {Rudnick}, {Scaife}, {Schleicher},
  {Stil}, and {Ryu}}]{2015aska.confE.113T}
{Taylor} R, {Agudo} I, {Akahori} T, {Beck} R, {Gaensler} B, {Heald} G,
  {Johnston-Hollitt} M, {Langer} M, {Rudnick} L, {Scaife} A, {Schleicher} D,
  {Stil} J, {Ryu} D (2015) {SKA Deep Polarization and Cosmic Magnetism}.
  Advancing Astrophysics with the Square Kilometre Array (AASKA14) 113,
  \eprint{1501.02298}

\bibitem[{{Thomas} and {Couchman}(1992)}]{1992MNRAS.257...11T}
{Thomas} PA, {Couchman} HMP (1992) {Simulating the formation of a cluster of
  galaxies}. \mnras 257:11--31, \doi{10.1093/mnras/257.1.11}

\bibitem[{{Tricco} et~al(2016){Tricco}, {Price}, and
  {Federrath}}]{2016MNRAS.461.1260T}
{Tricco} TS, {Price} DJ, {Federrath} C (2016) {A comparison between grid and
  particle methods on the small-scale dynamo in magnetized supersonic
  turbulence}. \mnras 461:1260--1275, \doi{10.1093/mnras/stw1280},
  \eprint{1605.08662}

\bibitem[{{Trivedi} et~al(2014){Trivedi}, {Subramanian}, and
  {Seshadri}}]{2014PhRvD..89d3523T}
{Trivedi} P, {Subramanian} K, {Seshadri} TR (2014) {Primordial magnetic field
  limits from the CMB trispectrum: Scalar modes and constraints}. \prd
  89(4):043523, \doi{10.1103/PhysRevD.89.043523}, \eprint{1312.5308}

\bibitem[{{Turner} and {Widrow}(1988)}]{1988PhRvD..37.2743T}
{Turner} MS, {Widrow} LM (1988) {Inflation-produced, large-scale magnetic
  fields}. \prd 37:2743--2754, \doi{10.1103/PhysRevD.37.2743}

\bibitem[{{V{\" o}lk} and {Atoyan}(2000)}]{Volk&Atoyan..ApJ.2000}
{V{\" o}lk} HJ, {Atoyan} AM (2000) {Early Starbursts and Magnetic Field
  Generation in Galaxy Clusters}. \apj 541:88--94

\bibitem[{{Vacca} et~al(2012){Vacca}, {Murgia}, {Govoni}, {Feretti},
  {Giovannini}, {Perley}, and {Taylor}}]{2012A&A...540A..38V}
{Vacca} V, {Murgia} M, {Govoni} F, {Feretti} L, {Giovannini} G, {Perley} RA,
  {Taylor} GB (2012) {The intracluster magnetic field power spectrum in A2199}.
  \aap 540:A38, \doi{10.1051/0004-6361/201116622}, \eprint{1201.4119}

\bibitem[{{Vacca} et~al(2016){Vacca}, {Oppermann}, {En{\ss}lin}, {Jasche},
  {Selig}, {Greiner}, {Junklewitz}, {Reinecke}, {Br{\"u}ggen}, {Carretti},
  {Feretti}, {Ferrari}, {Hales}, {Horellou}, {Ideguchi}, {Johnston-Hollitt},
  {Pizzo}, {R{\"o}ttgering}, {Shimwell}, and {Takahashi}}]{2016A&A...591A..13V}
{Vacca} V, {Oppermann} N, {En{\ss}lin} T, {Jasche} J, {Selig} M, {Greiner} M,
  {Junklewitz} H, {Reinecke} M, {Br{\"u}ggen} M, {Carretti} E, {Feretti} L,
  {Ferrari} C, {Hales} CA, {Horellou} C, {Ideguchi} S, {Johnston-Hollitt} M,
  {Pizzo} RF, {R{\"o}ttgering} H, {Shimwell} TW, {Takahashi} K (2016) {Using
  rotation measure grids to detect cosmological magnetic fields: A Bayesian
  approach}. \aap 591:A13, \doi{10.1051/0004-6361/201527291},
  \eprint{1509.00747}

\bibitem[{{Valdarnini}(2011)}]{2011A&A...526A.158V}
{Valdarnini} R (2011) {The impact of numerical viscosity in SPH simulations of
  galaxy clusters}. \aap 526:A158, \doi{10.1051/0004-6361/201015340},
  \eprint{1010.3378}

\bibitem[{{Va{\u i}nshte{\u i}n} and {Zel'dovich}(1972)}]{1972SvPhU..15..159V}
{Va{\u i}nshte{\u i}n} SI, {Zel'dovich} YB (1972) {REVIEWS OF TOPICAL PROBLEMS:
  Origin of Magnetic Fields in Astrophysics (Turbulent ``Dynamo'' Mechanisms)}.
  Soviet Physics Uspekhi 15:159--172, \doi{10.1070/PU1972v015n02ABEH004960}

\bibitem[{{Vazza} et~al(2006){Vazza}, {Tormen}, {Cassano}, {Brunetti}, and
  {Dolag}}]{2006MNRAS.369L..14V}
{Vazza} F, {Tormen} G, {Cassano} R, {Brunetti} G, {Dolag} K (2006) {Turbulent
  velocity fields in smoothed particle hydrodymanics simulated galaxy clusters:
  scaling laws for the turbulent energy}. \mnras 369:L14--L18,
  \doi{10.1111/j.1745-3933.2006.00164.x}, \eprint{arXiv:astro-ph/0602247}

\bibitem[{{Vazza} et~al(2009){Vazza}, {Brunetti}, {Kritsuk}, {Wagner},
  {Gheller}, and {Norman}}]{2009A&A...504...33V}
{Vazza} F, {Brunetti} G, {Kritsuk} A, {Wagner} R, {Gheller} C, {Norman} M
  (2009) {Turbulent motions and shocks waves in galaxy clusters simulated with
  adaptive mesh refinement}. \aap 504:33--43,
  \doi{10.1051/0004-6361/200912535}, \eprint{0905.3169}

\bibitem[{{Vazza} et~al(2011){Vazza}, {Brunetti}, {Gheller}, {Brunino}, and
  {Br{\"u}ggen}}]{va11turbo}
{Vazza} F, {Brunetti} G, {Gheller} C, {Brunino} R, {Br{\"u}ggen} M (2011)
  {Massive and refined. II. The statistical properties of turbulent motions in
  massive galaxy clusters with high spatial resolution}. \aap 529:A17+,
  \doi{10.1051/0004-6361/201016015}, \eprint{1010.5950}

\bibitem[{{Vazza} et~al(2012){Vazza}, {Roediger}, and {Brueggen}}]{va12filter}
{Vazza} F, {Roediger} E, {Brueggen} M (2012) {Turbulence in the ICM from
  mergers, cool-core sloshing and jets: results from a new multi-scale
  filtering approach}. ArXiv e-prints 12025882 \eprint{1202.5882}

\bibitem[{{Vazza} et~al(2014){Vazza}, {Br{\"u}ggen}, {Gheller}, and
  {Wang}}]{va14mhd}
{Vazza} F, {Br{\"u}ggen} M, {Gheller} C, {Wang} P (2014) {On the amplification
  of magnetic fields in cosmic filaments and galaxy clusters}. \mnras
  445:3706--3722, \doi{10.1093/mnras/stu1896}, \eprint{1409.2640}

\bibitem[{{Vazza} et~al(2017{\natexlab{a}}){Vazza}, {Br{\"u}ggen}, {Gheller},
  {Hackstein}, {Wittor}, and {Hinz}}]{va17cqg}
{Vazza} F, {Br{\"u}ggen} M, {Gheller} C, {Hackstein} S, {Wittor} D, {Hinz} PM
  (2017{\natexlab{a}}) {Simulations of extragalactic magnetic fields and of
  their observables}. Classical and Quantum Gravity 34(23):234001,
  \doi{10.1088/1361-6382/aa8e60}, \eprint{1711.02669}

\bibitem[{{Vazza} et~al(2017{\natexlab{b}}){Vazza}, {Jones}, {Br{\"u}ggen},
  {Brunetti}, {Gheller}, {Porter}, and {Ryu}}]{va17turb}
{Vazza} F, {Jones} TW, {Br{\"u}ggen} M, {Brunetti} G, {Gheller} C, {Porter} D,
  {Ryu} D (2017{\natexlab{b}}) {Turbulence and vorticity in Galaxy clusters
  generated by structure formation}. \mnras 464:210--230,
  \doi{10.1093/mnras/stw2351}, \eprint{1609.03558}

\bibitem[{{Vazza} et~al(2018{\natexlab{a}}){Vazza}, {Brunetti}, {Br{\"u}ggen},
  and {Bonafede}}]{va18mhd}
{Vazza} F, {Brunetti} G, {Br{\"u}ggen} M, {Bonafede} A (2018{\natexlab{a}})
  {Resolved magnetic dynamo action in the simulated intracluster medium}.
  \mnras 474:1672--1687, \doi{10.1093/mnras/stx2830}, \eprint{1711.02673}

\bibitem[{{Vazza} et~al(2018{\natexlab{b}}){Vazza}, {Brunetti}, {Br{\"u}ggen},
  and {Bonafede}}]{2018MNRAS.474.1672V}
{Vazza} F, {Brunetti} G, {Br{\"u}ggen} M, {Bonafede} A (2018{\natexlab{b}})
  {Resolved magnetic dynamo action in the simulated intracluster medium}.
  \mnras 474:1672--1687, \doi{10.1093/mnras/stx2830}, \eprint{1711.02673}

\bibitem[{{Vikhlinin} et~al(2001){Vikhlinin}, {Markevitch}, and
  {Murray}}]{vik01}
{Vikhlinin} A, {Markevitch} M, {Murray} SS (2001) {Chandra Estimate of the
  Magnetic Field Strength near the Cold Front in A3667}. \apjl 549:L47--L50,
  \doi{10.1086/319126}, \eprint{astro-ph/0008499}

\bibitem[{{Vikhlinin} and {Markevitch}(2002)}]{vik02}
{Vikhlinin} AA, {Markevitch} ML (2002) {A Cold Front in the Galaxy Cluster
  A3667: Hydrodynamics, Heat Conduction and Magnetic Field in the Intergalactic
  Medium}. Astronomy Letters 28:495--508, \doi{10.1134/1.1499173},
  \eprint{astro-ph/0209551}

\bibitem[{{Vladimirov} et~al(2006){Vladimirov}, {Ellison}, and
  {Bykov}}]{2006ApJ...652.1246V}
{Vladimirov} A, {Ellison} DC, {Bykov} A (2006) {Nonlinear Diffusive Shock
  Acceleration with Magnetic Field Amplification}. \apj 652:1246--1258,
  \doi{10.1086/508154}, \eprint{astro-ph/0606433}

\bibitem[{{Vogt} and {En{\ss}lin}(2003)}]{2003A&A...412..373V}
{Vogt} C, {En{\ss}lin} TA (2003) {Measuring the cluster magnetic field power
  spectra from Faraday rotation maps of Abell 400, Abell 2634 and Hydra A}.
  \aap 412:373--385

\bibitem[{{Voit}(2005)}]{2005RvMP...77..207V}
{Voit} GM (2005) {Tracing cosmic evolution with clusters of galaxies}. Reviews
  of Modern Physics 77:207--258, \doi{10.1103/RevModPhys.77.207},
  \eprint{astro-ph/0410173}

\bibitem[{{Waagan} et~al(2011){Waagan}, {Federrath}, and
  {Klingenberg}}]{2011JCoPh.230.3331W}
{Waagan} K, {Federrath} C, {Klingenberg} C (2011) {A robust numerical scheme
  for highly compressible magnetohydrodynamics: Nonlinear stability,
  implementation and tests}. Journal of Computational Physics 230:3331--3351,
  \doi{10.1016/j.jcp.2011.01.026}, \eprint{1101.3007}

\bibitem[{{Walker} et~al(2017){Walker}, {Hlavacek-Larrondo},
  {Gendron-Marsolais}, {Fabian}, {Intema}, {Sanders}, {Bamford}, and {van
  Weeren}}]{swalker2017}
{Walker} SA, {Hlavacek-Larrondo} J, {Gendron-Marsolais} M, {Fabian} AC,
  {Intema} H, {Sanders} JS, {Bamford} JT, {van Weeren} R (2017) {Is there a
  giant Kelvin-Helmholtz instability in the sloshing cold front of the Perseus
  cluster?} \mnras 468:2506--2516, \doi{10.1093/mnras/stx640},
  \eprint{1705.00011}

\bibitem[{{Werner} et~al(2016){Werner}, {ZuHone}, {Zhuravleva}, {Ichinohe},
  {Simionescu}, {Allen}, {Markevitch}, {Fabian}, {Keshet}, {Roediger},
  {Ruszkowski}, and {Sanders}}]{wer16}
{Werner} N, {ZuHone} JA, {Zhuravleva} I, {Ichinohe} Y, {Simionescu} A, {Allen}
  SW, {Markevitch} M, {Fabian} AC, {Keshet} U, {Roediger} E, {Ruszkowski} M,
  {Sanders} JS (2016) {Deep Chandra observation and numerical studies of the
  nearest cluster cold front in the sky}. \mnras 455:846--858,
  \doi{10.1093/mnras/stv2358}, \eprint{1506.06429}

\bibitem[{{Widrow} et~al(2012){Widrow}, {Ryu}, {Schleicher}, {Subramanian},
  {Tsagas}, and {Treumann}}]{wi11}
{Widrow} LM, {Ryu} D, {Schleicher} DRG, {Subramanian} K, {Tsagas} CG,
  {Treumann} RA (2012) {The First Magnetic Fields}. \ssr 166:37--70,
  \doi{10.1007/s11214-011-9833-5}, \eprint{1109.4052}

\bibitem[{{Wiener} et~al(2013){Wiener}, {Oh}, and {Guo}}]{wie13}
{Wiener} J, {Oh} SP, {Guo} F (2013) {Cosmic ray streaming in clusters of
  galaxies}. \mnras 434:2209--2228, \doi{10.1093/mnras/stt1163},
  \eprint{1303.4746}

\bibitem[{{Wiener} et~al(2018){Wiener}, {Zweibel}, and {Oh}}]{wie18}
{Wiener} J, {Zweibel} EG, {Oh} SP (2018) {High {$\beta$} effects on cosmic ray
  streaming in galaxy clusters}. \mnras 473:3095--3103,
  \doi{10.1093/mnras/stx2603}, \eprint{1706.08525}

\bibitem[{{Willson}(1970)}]{1970MNRAS.151....1W}
{Willson} MAG (1970) {Radio observations of the cluster of galaxies in Coma
  Berenices - the 5C4 survey.} \mnras 151:1--44, \doi{10.1093/mnras/151.1.1}

\bibitem[{{Wittor} et~al(2017{\natexlab{a}}){Wittor}, {Jones}, {Vazza}, and
  {Br{\"u}ggen}}]{wi17b}
{Wittor} D, {Jones} T, {Vazza} F, {Br{\"u}ggen} M (2017{\natexlab{a}})
  {Evolution of vorticity and enstrophy in the intracluster medium}. \mnras
  471:3212--3225, \doi{10.1093/mnras/stx1769}, \eprint{1706.02315}

\bibitem[{{Wittor} et~al(2017{\natexlab{b}}){Wittor}, {Vazza}, and
  {Br{\"u}ggen}}]{wi17}
{Wittor} D, {Vazza} F, {Br{\"u}ggen} M (2017{\natexlab{b}}) {Testing cosmic ray
  acceleration with radio relics: a high-resolution study using MHD and
  tracers}. \mnras 464:4448--4462, \doi{10.1093/mnras/stw2631},
  \eprint{1610.05305}

\bibitem[{{Xu} et~al(2009){Xu}, {Li}, {Collins}, {Li}, and
  {Norman}}]{2009ApJ...698L..14X}
{Xu} H, {Li} H, {Collins} DC, {Li} S, {Norman} ML (2009) {Turbulence and Dynamo
  in Galaxy Cluster Medium: Implications on the Origin of Cluster Magnetic
  Fields}. \apjl 698:L14--L17, \doi{10.1088/0004-637X/698/1/L14},
  \eprint{0905.2196}

\bibitem[{{Xu} et~al(2011){Xu}, {Li}, {Collins}, {Li}, and
  {Norman}}]{2011ApJ...739...77X}
{Xu} H, {Li} H, {Collins} DC, {Li} S, {Norman} ML (2011) {Evolution and
  Distribution of Magnetic Fields from Active Galactic Nuclei in Galaxy
  Clusters. II. The Effects of Cluster Size and Dynamical State}. \apj 739:77,
  \doi{10.1088/0004-637X/739/2/77}, \eprint{1107.2599}

\bibitem[{{Yakhot} and {Sreenivasan}(2005)}]{2005JSP...121..823Y}
{Yakhot} V, {Sreenivasan} KR (2005) {Anomalous Scaling of Structure Functions
  and Dynamic Constraints on Turbulence Simulations}. Journal of Statistical
  Physics 121:823--841, \doi{10.1007/s10955-005-8666-6}, \eprint{nlin/0506038}

\bibitem[{{Zandanel} and {Ando}(2014)}]{2014MNRAS.440..663Z}
{Zandanel} F, {Ando} S (2014) {Constraints on diffuse gamma-ray emission from
  structure formation processes in the Coma cluster}. \mnras 440:663--671,
  \doi{10.1093/mnras/stu324}, \eprint{1312.1493}

\bibitem[{{Zeldovich} et~al(1983){Zeldovich}, {Ruzmaikin}, and
  {Sokolov}}]{1983flma....3.....Z}
{Zeldovich} IB, {Ruzmaikin} AA, {Sokolov} DD (eds) (1983) {Magnetic fields in
  astrophysics}, vol~3

\bibitem[{{Zel'dovich}(1970)}]{1970SvA....13..608Z}
{Zel'dovich} YB (1970) {The Hypothesis of Cosmological Magnetic Inhomogeneity.}
  \sovast 13:608

\bibitem[{{Zhuravleva} et~al(2014){Zhuravleva}, {Churazov}, {Schekochihin},
  {Allen}, {Ar{\'e}valo}, {Fabian}, {Forman}, {Sanders}, {Simionescu},
  {Sunyaev}, {Vikhlinin}, and {Werner}}]{2014Natur.515...85Z}
{Zhuravleva} I, {Churazov} E, {Schekochihin} AA, {Allen} SW, {Ar{\'e}valo} P,
  {Fabian} AC, {Forman} WR, {Sanders} JS, {Simionescu} A, {Sunyaev} R,
  {Vikhlinin} A, {Werner} N (2014) {Turbulent heating in galaxy clusters
  brightest in X-rays}. \nat 515:85--87, \doi{10.1038/nature13830},
  \eprint{1410.6485}

\bibitem[{{Zinger} et~al(2018){Zinger}, {Dekel}, {Birnboim}, {Nagai}, {Lau},
  and {Kravtsov}}]{zin18}
{Zinger} E, {Dekel} A, {Birnboim} Y, {Nagai} D, {Lau} E, {Kravtsov} AV (2018)
  {Cold fronts and shocks formed by gas streams in galaxy clusters}. \mnras
  476:56--70, \doi{10.1093/mnras/sty136}, \eprint{1609.05308}

\bibitem[{{Zirakashvili} and {Ptuskin}(2008)}]{2008ApJ...678..939Z}
{Zirakashvili} VN, {Ptuskin} VS (2008) {Diffusive Shock Acceleration with
  Magnetic Amplification by Nonresonant Streaming Instability in Supernova
  Remnants}. \apj 678:939--949, \doi{10.1086/529580}, \eprint{0801.4488}

\bibitem[{{Zuhone} and {Roediger}(2016)}]{zuh16}
{Zuhone} JA, {Roediger} E (2016) {Cold fronts: probes of plasma astrophysics in
  galaxy clusters}. Journal of Plasma Physics 82(3):535820301,
  \doi{10.1017/S0022377816000544}, \eprint{1603.08882}

\bibitem[{{ZuHone} et~al(2011){ZuHone}, {Markevitch}, and {Lee}}]{zuh11}
{ZuHone} JA, {Markevitch} M, {Lee} D (2011) {Sloshing of the Magnetized Cool
  Gas in the Cores of Galaxy Clusters}. \apj 743:16,
  \doi{10.1088/0004-637X/743/1/16}, \eprint{1108.4427}

\bibitem[{{ZuHone} et~al(2013){ZuHone}, {Markevitch}, {Brunetti}, and
  {Giacintucci}}]{zuh13}
{ZuHone} JA, {Markevitch} M, {Brunetti} G, {Giacintucci} S (2013) {Turbulence
  and Radio Mini-halos in the Sloshing Cores of Galaxy Clusters}. \apj 762:78,
  \doi{10.1088/0004-637X/762/2/78}, \eprint{1203.2994}

\bibitem[{{ZuHone} et~al(2015{\natexlab{a}}){ZuHone}, {Brunetti},
  {Giacintucci}, and {Markevitch}}]{zuh15}
{ZuHone} JA, {Brunetti} G, {Giacintucci} S, {Markevitch} M (2015{\natexlab{a}})
  {Testing Secondary Models for the Origin of Radio Mini-Halos in Galaxy
  Clusters}. \apj 801:146, \doi{10.1088/0004-637X/801/2/146},
  \eprint{1403.6743}

\bibitem[{{ZuHone} et~al(2015{\natexlab{b}}){ZuHone}, {Kunz}, {Markevitch},
  {Stone}, and {Biffi}}]{2015ApJ...798...90Z}
{ZuHone} JA, {Kunz} MW, {Markevitch} M, {Stone} JM, {Biffi} V
  (2015{\natexlab{b}}) {The Effect of Anisotropic Viscosity on Cold Fronts in
  Galaxy Clusters}. \apj 798:90, \doi{10.1088/0004-637X/798/2/90},
  \eprint{1406.4031}

\end{thebibliography}

%
%

\end{document}